\documentclass[11pt,a4paper]{article}
\pdfoutput=1

\usepackage[colorlinks=true, linkcolor=black!50!blue, urlcolor=blue, citecolor=blue, anchorcolor=blue]{hyperref}
\usepackage[font=small,labelfont=bf,margin=0mm,labelsep=period,tableposition=top]{caption}
\usepackage[a4paper,top=1.5cm,bottom=2cm,left=1.5cm,right=1.5cm,bindingoffset=0mm]{geometry}

\usepackage{placeins,cite}
\usepackage{graphicx}
\usepackage{float}
\usepackage{afterpage}
\usepackage{epsfig}
\usepackage{amssymb}
\usepackage{amsmath}
\usepackage{bm}
\usepackage{multirow}
\usepackage{url}
\usepackage{xcolor}
\usepackage{booktabs}
\usepackage{textcomp}
\usepackage[shortlabels]{enumitem}
\usepackage{tabularx}
\usepackage{subcaption}
\usepackage[capitalise]{cleveref}
\graphicspath{{../}{./figures/}}

\newcommand{\la}{\left\langle}
\newcommand{\ra}{\right\rangle}
\newcommand{\lp}{\left(}
\newcommand{\rp}{\right)}

\newcolumntype{C}{>{\centering\arraybackslash}X}

\numberwithin{equation}{section}
\numberwithin{figure}{section}
\numberwithin{table}{section}

\begin{document}

\vspace{-2.0cm}
\begin{flushright}
CERN-TH-2025-054
\end{flushright}
\vspace{0.6cm}

\begin{center}
  {\Large \bf {\sc NNPDFpol2.0}: a global determination of polarized PDFs
    \\[0.25cm]
   and their uncertainties at next-to-next-to-leading order
     }\\
  \vspace{1.1cm}
  {\small
    Juan Cruz-Martinez$^1$,
    Toon Hasenack$^{2,3}$,
    Felix Hekhorn$^{4,5}$,
    Giacomo Magni$^{3,6}$,
    Emanuele R. Nocera$^7$,\\
    Tanjona R. Rabemananjara$^{3,6}$,
    Juan Rojo$^{3,6}$,
    Tanishq Sharma$^7$,
    and Gijs van Seeventer$^{2,3}$
  }\\
  
\vspace{0.7cm}

{\it \small
  ~$^1$CERN, Theoretical Physics Department,
  CH-1211 Geneva 23, Switzerland\\[0.1cm]
  ~$^2$Institute for Theoretical Physics, Utrecht University,
  Leuvenlaan 4, 3584 CE Utrecht\\[0.1cm]
  ~$^3$Nikhef Theory Group,
  Science Park 105, 1098 XG Amsterdam, The Netherlands\\[0.1cm]
  ~$^4$University of Jyvaskyla, Department of Physics,
  P.O. Box 35, FI-40014 University of Jyvaskyla, Finland\\[0.1cm]
  ~$^5$Helsinki Institute of Physics,
  P.O. Box 64, FI-00014 University of Helsinki, Finland\\[0.1cm]
  ~$^6$Department of Physics and Astronomy, Vrije Universiteit,
  NL-1081 HV Amsterdam\\[0.1cm]
  ~$^7$Dipartimento di Fisica, Universit\`a degli Studi di Torino and\\
  INFN, Sezione di Torino, Via Pietro Giuria 1, 10125 Torino, Italy\\[0.1cm]
 }

\vspace{1.0cm}

{\bf \large Abstract}

\end{center}

We present {\sc NNPDFpol2.0}, a new set of collinear helicity parton
distribution functions (PDFs) of the proton based on legacy measurements of
structure functions in inclusive neutral-current longitudinally polarised
deep-inelastic scattering (DIS), and of $W$-boson, single-inclusive, and di-jet
production asymmetries in longitudinally polarised proton-proton collisions.
The determination is accurate to next-to-next-to-leading order in the strong
coupling, and includes heavy quark mass corrections in the analysis of DIS
data. Uncertainties due to missing higher-order corrections are systematically
incorporated by means of a covariance matrix determined by scale variations.
{\sc NNPDFpol2.0} is based on a machine learning methodology, that makes use of
Monte Carlo sampling for the representation of uncertainties into PDFs, of a
neural network for the parametrisation of PDFs, of stochastic gradient descent
for the optimisation of PDF parameters, and of hyperoptimisation for the
selection of the best fitting model. We study the impact on PDFs of
higher-order corrections, of the positivity constraint, and of the data.
We demonstrate two phenomenological applications of {\sc NNPDFpol2.0},
specifically the determination of the proton spin fraction carried by gluons
and quarks, and of theoretical predictions for single-hadron production in
longitudinally polarised DIS and proton-proton collisions.

\clearpage

\tableofcontents

\section{Introduction}
\label{sec:introduction}

Helicity-dependent (polarised, henceforth) parton distribution functions
(PDFs)~\cite{Ethier:2020way} are defined as differences between the density of
partons with spin  aligned parallel or anti-parallel to the spin of the proton
to which they belong. The interest in their determination is mainly related to
the fact that their lowest moments are proportional to the proton axial
currents, which express the fraction of proton spin carried by gluons and
quarks~\cite{Anselmino:1994gn}. In spite of impressive experimental and
theoretical investigations over the past thirty years, knowledge of polarised
PDFs remains limited in comparison to their unpolarised counterparts, in
particular concerning the distributions of sea quarks and gluons. This fact
hinders the fundamental understanding of proton spin decomposition in the
framework of Quantum Chromodynamics (QCD)~\cite{Ji:2020ena}.

The Electron-Ion Collider (EIC)~\cite{AbdulKhalek:2021gbh,AbdulKhalek:2022hcn},
expected to start its operations in the early 2030s, is designed to
revolutionise this state of affairs. The EIC will have the possibility to
collide polarised proton and lepton beams, so to measure the polarised
inclusive and semi-inclusive deep-inelastic scattering (DIS) structure
functions, to which polarised PDFs are related through factorisation
theorems~\cite{Collins:1989gx}. These measurements are forecast to cover an
unprecedented range of proton momentum fraction $x$ and virtuality $Q^2$ (see
{\it e.g.}~Fig.~1 in~\cite{Ethier:2020way}) while attaining percent-level
precision (see {\it e.g.}~Sect.~II in~\cite{AbdulKhalek:2021gbh}). 
These prospects call for a matching accuracy of the corresponding theoretical
predictions, which require in turn an improvement in the accuracy of the
perturbative computations involving polarised protons in the initial state,
and of the polarised PDF determinations extracted from these.

Concerning the accuracy of perturbative computations, significant progress has
been made to incorporate next-to-next-to-leading order (NNLO) QCD corrections
into the matrix elements of a range of polarised observables. For polarised
inclusive DIS, the massless polarised structure function $g_1$ has been known at
NNLO for a long time~\cite{Zijlstra:1993sh}. Very recently, zero-mass N$^3$LO
corrections have been computed~\cite{Blumlein:2022gpp}, as have NNLO massive
contributions~\cite{Hekhorn:2018ywm}, their asymptotic
limit~\cite{Behring:2015zaa,Ablinger:2019etw,Behring:2021asx,Blumlein:2021xlc,
  Bierenbaum:2022biv,Ablinger:2022wbb,Ablinger:2023ahe,Ablinger:2024xtt},
  and NNLO parity-violating massless
polarised structure functions~\cite{Borsa:2022irn}. For polarised
semi-inclusive DIS, the massless polarised structure function $g_1^h$ has been
determined up to NNLO lately: first using an approximation based on
the threshold resummation formalism~\cite{Abele:2021nyo}; then exactly
using various analytical methods~\cite{Bonino:2024wgg,Goyal:2024tmo,
  Goyal:2024emo}. Finally, NNLO corrections have also been obtained for
$W$-boson production in polarised proton-proton
collisions~\cite{Boughezal:2021wjw}. In the last years,
a similar effort was put to compute NNLO corrections to polarised splitting
functions entering DGLAP evolution equations~\cite{Moch:2014sna,
  Moch:2015usa,Blumlein:2021enk,Blumlein:2021ryt}, and their matching conditions
for heavy quarks~\cite{Bierenbaum:2022biv}.

Concerning the accuracy of PDF determinations, a first polarised NNLO set was
determined some time ago by analysing only inclusive DIS
data~\cite{Taghavi-Shahri:2016idw}. This data does not allow for a
separation of polarised quark and antiquark PDFs and has a limited sensitivity
to the polarised gluon PDF. Therefore, other polarised PDF
sets, {\it e.g.} {\sc DSSV14}~\cite{deFlorian:2014yva} and
{\sc NNPDFpol1.1}~\cite{Nocera:2014gqa}, despite being accurate only to
next-to-leading order (NLO), have been more widely used so far. The NNLO
analysis presented in~\cite{Taghavi-Shahri:2016idw} has been followed by two
newer NNLO determinations last year: {\sc MAPPDFpol1.0}~\cite{Bertone:2024taw}
and {BDSSV24}~\cite{Borsa:2024mss}. These two determinations incorporate
experimental data other than inclusive DIS in a NNLO analysis. Both of them,
however, are approximated. The former~\cite{Bertone:2024taw}, which is based on
polarised inclusive and semi-inclusive DIS measurements, relies on the
approximation of~\cite{Abele:2021nyo} to analyse the latter. The necessary
fragmentation functions (FFs) are determined consistently using the same
approximation. The latter~\cite{Borsa:2024mss}, which is based on polarised
inclusive and semi-inclusive DIS measurements and on single-inclusive jet,
gauge boson, and hadron production measurements in polarised proton-proton
collisions, is based on the same approximation. The necessary FF sets are used
at NLO. Unknown NNLO corrections in the matrix elements of single-inclusive jet
and hadron production in polarised proton-proton collisions are included by
means of $K$-factors determined from soft-gluon
approximation~\cite{deFlorian:2007tye,Hinderer:2018nkb}.
Whereas the more extended data set provides a better sensitivity to polarised
PDFs, these approximations introduce additional theory uncertainties which are
not accounted for.

As a further step towards the understanding of the polarised partonic 
structure of the nucleon, in this paper we present {\sc NNPDFpol2.0},
a new determination of the proton polarised PDFs based on the NNPDF
methodology~\cite{NNPDF:2021njg}. This determination
improves our previous one, {\sc NNPDFpol1.1}~\cite{Nocera:2014gqa}, in three
main respects, which set it apart from the other two recent NNLO fits of
polarised PDFs~\cite{Bertone:2024taw,Borsa:2024mss} described above.

\begin{enumerate}

\item We significantly extend the range of fitted data sets. We specifically
  consider measurements of polarised inclusive lepton-nucleon DIS, including
  legacy measurements from HERMES and COMPASS, and measurements of $W$-boson,
  single-inclusive jet, and di-jet production from STAR. Many of these data sets
  have become available after the publication of {\sc NNPDFpol1.1}, and
  represent the legacy of experimental programs that are now completed.
  In contrast to {\sc NNPDFpol1.1}, we drop measurements of open charm
  production, which were demonstrated to be
  immaterial~\cite{Nocera:2013yia,Nocera:2014gqa}. As in {\sc NNPDFpol1.1},
  we do not consider measurements of single-hadron production in polarised DIS
  and polarised proton-proton collisions, the analysis of which requires
  explicit knowledge of the FFs. 

\item We incorporate higher-order corrections to both the DGLAP evolution and
  to the hard cross sections, whenever available, up to NNLO in the strong
  coupling. We likewise incorporate heavy quark mass
  corrections in the analysis of polarised inclusive DIS structure functions
  according to the FONLL general-mass scheme~\cite{Forte:2010ta} as implemented
  in~\cite{Hekhorn:2024tqm,Barontini:2024xgu}. We account for the uncertainties
  associated to missing higher-order QCD corrections (MHOUs) by means of
  the methodology developed in~\cite{NNPDF:2019vjt,NNPDF:2019ubu,NNPDF:2024dpb},
  whereby MHOUs are treated through a theory covariance matrix determined by
  scale variations. Theoretical predictions are computed with the
  {\sc pineline}~\cite{Barontini:2023vmr} framework, which combines various
  pieces of open-source software specifically designed for PDF
  determination: {\sc EKO}~\cite{Candido:2022tld,candido_2022_6340153},
  for the evolution of PDFs; {\sc yadism}~\cite{Candido:2024rkr,candido_2023_8066034}, for the
  computation of inclusive DIS structure functions; and
  {\sc PineAPPL}~\cite{Carrazza:2020gss,christopher_schwan_2024_12795745} for
  the delivery of theoretical predictions as PDF-independent fast interpolation
  grids. Here this computational framework has been extended to deal
  with polarised observables, including in the case in which unpolarised and
  polarised PDFs ought to be used simultaneously, such as in the computation of
  spin asymmetries.

\item We deploy the machine-learning methodology developed
  in~\cite{NNPDF:2021njg}. Whereas the core ingredients of the fitting
  methodology are the same as in {\sc NNPDFpol1.1} (a Monte Carlo sampling for
  uncertainty representation and a neural network for PDF parametrisation), all
  aspects of the parametrisation and optimisation (such as the neural network
  architecture or the choice of minimisation algorithm) are now selected
  through a hyperparameter optimisation
  procedure~\cite{Carrazza:2019mzf,Cruz-Martinez:2024wiu}, which consists in an
  automated scan of the space of models. To ensure that the optimal model does
  not lead to overfitted PDFs, a $k$-folding partition is used, which verifies
  the effectiveness of any given model on sets of data excluded in turn from
  the fit. In contrast to {\sc NNPDFpol1.1}, all data sets are fitted, instead
  of being partly fitted and partly being incorporated by means of Bayesian
  reweighting~\cite{Ball:2010gb,Ball:2011gg}. Furthermore, model optimisation
  is realised through a gradient descent algorithm instead of a genetic
  algorithm.
  
\end{enumerate}

\begin{figure}[!t]
  \includegraphics[width=0.49\textwidth]{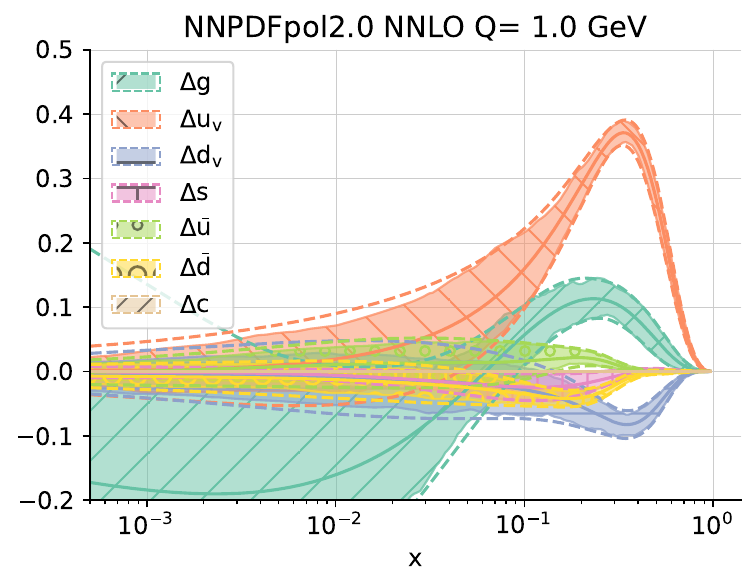}
  \includegraphics[width=0.49\textwidth]{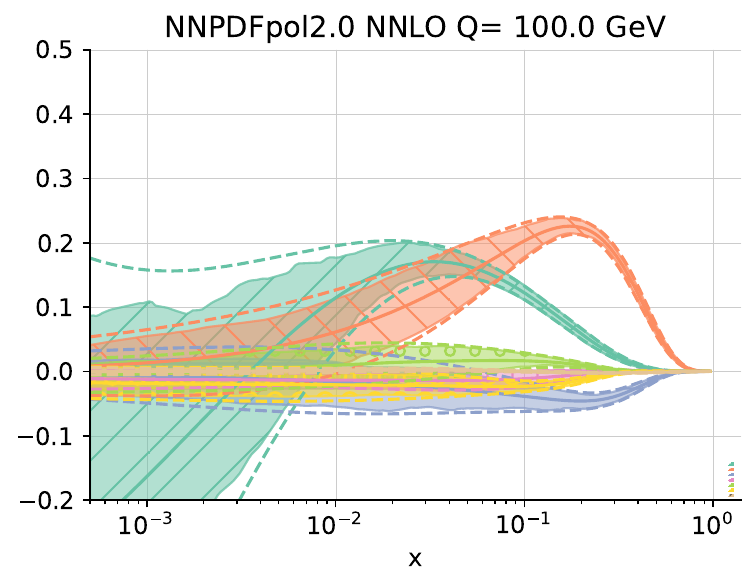}\\
  \caption{The {\sc NNPDFpol2.0} NNLO PDFs at $Q=1$~GeV (left) and
    $Q=100$~GeV (right). Bands with dashed contours correspond to one-sigma
    uncertainties, whereas bands with continuous contours correspond to 68\%
    confidence-level uncertainties.}
  \label{fig:NNPDFpol2.0}
\end{figure}

The {\sc NNPDFpol2.0} polarised PDF sets are released at LO, NLO and NNLO
without and with MHOUs for a single value of the strong coupling,
$\alpha_s(m_Z)=0.118$. The NNLO set remains approximate, insofar as NNLO
corrections to single-inclusive jet and di-jet production asymmetries in
polarised proton-proton collision are unknown. The corresponding PDFs, which
represent the main outcome of this paper, are displayed in
\cref{fig:NNPDFpol2.0}. All parton sets are made available in the
{\sc LHAPDF} format~\cite{Buckley:2014ana,LHAPDFurl} as ensembles of both 1000
and 100 Monte Carlo replicas, the latter being obtained from compression of the
former with the algorithm developed in~\cite{Carrazza:2021hny,pycompressor}.
The open-source NNPDF software~\cite{NNPDF:2021uiq} has been extended to
include the input and tools needed to reproduce the {\sc NNPDFpol2.0} sets
presented here.

The structure of this paper is as follows. In \cref{sec:data-theory} we
present the experimental and theoretical input entering the {\sc NNPDFpol2.0}
determination. In \cref{sec:methodology} we discuss the methodology
deployed to determine it, focusing on how aspects of PDF parametrisation,
optimisation, and hyperoptimisation are adapted to the polarised case.
In \cref{sec:results} we present the {\sc NNPDFpol2.0} parton set,
study its fit quality and perturbative stability, compare it to
{\sc NNPDFpol1.1} and to other polarised PDF sets, and assess its dependence
on the positivity constraint and on the data. In \cref{sec:pheno} we
demonstrate two applications of {\sc NNPDFpol2.0}, specifically on
the determination of the spin content of the proton and on the computation of
theoretical predictions for longitudinal spin asymmetries of single-inclusive
hadron production in DIS and polarised proton-proton collisions. The conclusion
and an outlook are provided in \cref{sec:summary}. The paper is completed
by \cref{app:pinebench}, in which we present a benchmark of the
numerical accuracy of the {\sc PineAPPL} grids used in this work, specifically
to analyse $W$-boson production spin asymmetries.

\section{Experimental and theoretical input}
\label{sec:data-theory}

In this section, we present the experimental and theoretical input entering
the {\sc NNPDFpol2.0} determination. We first introduce the data set,
describing the details of each measurement, and the computational tools used to
obtain the corresponding theoretical predictions. We then discuss their
perturbative accuracy, and specifically the way in which we account for MHOUs.

\subsection{The {\sc NNPDFpol2.0} data set}
\label{subsec:data}

The {\sc NNPDFpol2.0} determination is based on measurements of three different
polarised observables: the structure function $g_1$ in longitudinally polarised
inclusive lepton-nucleon DIS; the longitudinal single-spin asymmetry
$A_L^{W^\pm}$ for $W^\pm$-boson production in polarised proton-proton collisions;
and the longitudinal double-spin asymmetry $A_{LL}^{1-;2-{\rm jet}}$ for
single-inclusive jet and di-jet production in polarised proton-proton
collisions. The definition of these observables can be found, {\it e.g.},
in Sect.~3 of~\cite{Ball:2013lla}, and in Sect.~6.2.2 of~\cite{Nocera:2014vla}.
We review the measurements that we include in
{\sc NNPDFpol2.0} for each of these observables in turn.

\begin{description}

\item[Polarised inclusive DIS structure function.] We include measurements
  performed by the EMC~\cite{EuropeanMuon:1989yki},
  SMC~\cite{SpinMuon:1998eqa,SpinMuon:1999udj},
  and COMPASS~\cite{COMPASS:2015mhb,COMPASS:2016jwv}
  experiments at CERN, by the E142~\cite{E142:1996thl},
  E143~\cite{E143:1998hbs}, E154~\cite{E154:1997xfa},
  and E155~\cite{E155:2000qdr} experiments at SLAC, by the HERMES experiment
  at DESY~\cite{HERMES:1997hjr,HERMES:2006jyl}, and by the
  Hall~A~\cite{JeffersonLabHallA:2016neg,Kramer:2002tt,
    JeffersonLabHallA:2004tea}
  and CLAS~\cite{CLAS:2014qtg,CLAS:2006ozz} experiments at JLab. All of these
  experiments provide data for the polarised inclusive DIS structure function
  $g_1$, reconstructed from the longitudinal double-spin asymmetry (see,
  {\it e.g.},~Sect.~2.1 in~\cite{Ball:2013lla} for details), except SMC low-$x$,
  E155, Hall A, and CLAS, which instead provide data for $g_1$ normalised to the
  unpolarised inclusive structure function $F_1$. The details of the
  measurements, including their kinematic coverage in the proton momentum
  fraction $x$ and virtuality $Q^2$, are summarised in Table~\ref{tab:DIS_data}.
  In comparison to {\sc NNPDFpol1.1}, we update the COMPASS data with the legacy
  measurements of~\cite{COMPASS:2015mhb,COMPASS:2016jwv}, which combine events
  recorded in 2007 and 2011, and we add the JLab
  Hall~A~\cite{JeffersonLabHallA:2016neg,Kramer:2002tt,
    JeffersonLabHallA:2004tea}
  and CLAS~\cite{CLAS:2014qtg,CLAS:2006ozz} data. Note also that
  in {\sc NNPDFpol1.1} we reconstructed $g_1$ from the measured longitudinal and
  transverse spin asymmetries, assuming either a vanishing structure function
  $g_2$ or relating $g_2$ to $g_1$ with the Wandzura-Wilczek
  relation~\cite{Wandzura:1977qf}. Here we use instead the values of $g_1$
  reconstructed in the corresponding experimental papers. The two procedures
  lead to differences that are smaller than the uncertainty propagated by us
  or estimated in the experimental measurements, therefore we deem them to be
  equivalent.

\begin{table}[!t]
  \scriptsize
  \centering
  \renewcommand{\arraystretch}{1.4}
  \begin{tabularx}{\textwidth}{Xccccc}
  \toprule
  Data set
  & Ref.
  & $N_{\rm {dat}}$
  & $x$
  & $Q^2$~[GeV$^2$]
  & Theory \\
  \midrule
  EMC $g_1^p$
  & \cite{EuropeanMuon:1989yki}
  & $10 \ (10)$
  & [0.015, 0.466]
  & [3.5, 29.5]
  & {\sc Yadism} \\
  SMC $g_1^p$
  & \cite{SpinMuon:1998eqa}
  & $13 \ (12)$
  & [0.002, 0.48]
  & [0.50, 54.8]
  & {\sc Yadism} \\
  SMC $g_1^d$
  & \cite{SpinMuon:1998eqa}
  & $13 \ (12)$
  & [0.002, 0.48]
  & [0.50, 54.80]
  & {\sc Yadism} \\
  SMC low-$x$ $g_1^p/F_1^p$
  & \cite{SpinMuon:1999udj}
  & $15 \ (8)$
  & [0.00011, 0.121]
  & [0.03, 23.1]
  & {\sc Yadism} \\
  SMC low-$x$ $g_1^d/F_1^d$
  & \cite{SpinMuon:1999udj}
  & $15 \ (8)$
  & [0.00011, 0.121]
  & [0.03, 22.9]
  & {\sc Yadism} \\
  COMPASS $g_1^p$
  & \cite{COMPASS:2015mhb}
  & $17 \ (17)$
  & [0.0036, 0.57]
  & [1.1, 67.4]
  & {\sc Yadism} \\
  COMPASS $g_1^d$
  & \cite{COMPASS:2016jwv}
  & $15 \ (15)$
  & [0.0046, 0.567]
  & [1.1, 60.8]
  & {\sc Yadism}\\
  \midrule
  E142 $g_1^n$
  & \cite{E142:1996thl}
  & $8 \ (8)$
  & [0.035, 0.466]
  & [1.1, 5.5]
  & {\sc Yadism} \\
  E143 $g_1^p$
  & \cite{E143:1998hbs}
  & $28 \ (27)$
  & [0.035, 0.466]
  & [1.27, 9.52]
  & {\sc Yadism} \\
  E143 $g_1^d$
  & \cite{E143:1998hbs}
  & $28 \ (27)$
  & [0.031, 0.749]
  & [1.27, 9.52]
  & {\sc Yadism} \\       
  E154 $g_1^n$
  & \cite{E154:1997xfa}
  & $11 \ (11)$
  & [0.017, 0.024]
  & [1.2, 15.0]
  & {\sc Yadism} \\
  E155 $g_1^p/F_1^p$
  & \cite{E155:2000qdr}
  & $24 \ (24)$
  & [0.015, 0.750]
  & [1.22, 34.72]
  & {\sc Yadism} \\
  E155 $g_1^n/F_1^n$
  & \cite{E155:2000qdr}
  & $24 \ (24)$
  & [0.015, 0.750]
  & [1.22, 34.72]
  & {\sc Yadism} \\
  \midrule 
  HERMES $g_1^n$
  & \cite{HERMES:1997hjr}
  & $9 \ (9)$
  & [0.033, 0.464]
  & [1.22, 5.25]
  & {\sc Yadism} \\
  HERMES $g_1^p$
  & \cite{HERMES:2006jyl}
  & $15 \ (15)$
  & [0.0264, 0.7248]
  & [1.12, 12.21]
  & {\sc Yadism} \\
  HERMES $g_1^d$
  & \cite{HERMES:2006jyl}
  & $15 \ (15)$
  & [0.0264, 0.7248]
  & [1.12, 12.21]
  & {\sc Yadism} \\
  \midrule
  JLab E06 014 $g_1^n/F_1^n$
  & \cite{JeffersonLabHallA:2016neg}
  & $6 \ (4)$
  & [0.277, 0.548]
  & [3.078, 3.078]
  & {\sc Yadism} \\
  JLab E97 103 $g_1^n$
  & \cite{Kramer:2002tt}
  & $5 \ (2)$
  & [0.160, 0.200]
  & [0.57, 1.34]
  & {\sc Yadism}\\
  JLab E99 117 $g_1^n/F_1^n$
  & \cite{JeffersonLabHallA:2004tea}
  & $3 \ (1)$
  & [0.33, 0.60]
  & [2.71, 4.83]
  & {\sc Yadism}\\
  JLab EG1 DVCS $g_1^p/F_1^p$
  & \cite{CLAS:2014qtg}
  & $47 \ (21)$
  & [0.154, 0.578]
  & [1.064, 4.115]
  & {\sc Yadism} \\
  JLab EG1 DVCS $g_1^d/F_1^d$
  & \cite{CLAS:2014qtg}
  & $44 \ (19)$
  & [0.158, 0.574]
  & [1.078, 4.666]
  & {\sc Yadism}\\      
  JLab EG1B $g_1^p/F_1^p$
  & \cite{CLAS:2006ozz}
  & $787 \ (114)$
  & [0.0262, 0.9155]
  & [0.0496, 4.96]
  & {\sc Yadism} \\
  JLab EG1B $g_1^d/F_1^d$
  & \cite{CLAS:2006ozz}
  & $2465 \ (301)$
  & [0.0295, 0.9337]
  & [0.0496, 4.16]
  & {\sc Yadism} \\
  \bottomrule
\end{tabularx}

  \vspace{0.3cm}
  \caption{The polarised inclusive DIS measurements included in
    {\sc NNPDFpol2.0}. We denote each data set with the name used throughout
    this paper, and we indicate its reference, number of data points before
    (after) applying kinematic cuts, kinematic coverage (before cuts), and the
    piece of software used to compute the corresponding theoretical
    predictions.}
  \label{tab:DIS_data}
\end{table}

We compute the corresponding theoretical predictions with
{\sc yadism}~\cite{Candido:2024rkr}, which we have developed to handle the
computation of the polarised structure function $g_1$ and interfaced to
{\sc PineAPPL}. Predictions are accurate up to NNLO and include charm-quark
mass corrections through the FONLL general-mass variable-flavour-number
scheme~\cite{Forte:2010ta}, recently extended to the case of polarised 
structure functions~\cite{Hekhorn:2024tqm}. Target mass corrections are also
included as explained in Appendix~B of~\cite{Hekhorn:2024tqm}.
We apply kinematic cuts on the virtuality $Q^2$ and on the
invariant mass of the final state $W^2$, by requiring
$Q^2\geq Q^2_{\rm min}=1.0~{\rm GeV}^2$ and $W^2\geq W^2_{\rm min}=4.0~{\rm GeV}^2$.
These cuts remove, respectively, the region where perturbative QCD becomes
unreliable due to the growth of the strong coupling, and the region where
dynamical higher-twist corrections in the factorisation of $g_1$ (which we do
not include) may become sizeable. The cut on $W^2$ also suppresses contributions
from the resonance production region. The choice of the value of $Q^2_{\rm cut}$
is a common choice in several determinations of polarised PDFs, including
{\sc NNPDFpol1.0} and {\sc NNPDFpol1.1}, and results from a tradeoff between
incorporating as much experimental information as possible and preserving the
reliability of perturbative computations. The choice of the value of
$W^2_{\rm cut}$ is slightly smaller than that used in {\sc NNPDFpol1.0} and
{\sc NNPDFpol1.1}. There, the choice was based on the study performed
in~\cite{Simolo:2006iw}, where higher-twist terms were added to the polarised
structure functions, with a cofficient fitted to the data; it was shown
that these additional terms became compatible with zero for a value of
$W^2_{\rm cut}$ around 6.25~GeV$^2$. We have repeated our determination with this
more conservative cut, and have found very small differences of fit quality
and of PDFs in comparison to our baseline fit. We therefore conclude that this
is robust upon lack of higher-twist corrections. The less restrictive cut,
however, allows us to incorporate in the fit a larger fraction of JLab data,
hence the reason why we prefer it. Nuclear corrections affecting experiments
that utilise a deuterium target are neglected. Whereas, in principle, they
could be accounted for as described in~\cite{Ball:2020xqw}, we consider that
they be negligible in comparison to the precision of the experimental
measurements. We therefore model the deuteron as the average of a proton and a
neutron, and relate the PDFs of the latter to the PDFs of the former assuming
exact isospin symmetry.

\item[$W$-boson longitudinal single-spin asymmetry.] We include the
  measurement of the longitudinal single-spin asymmetry for $W^\pm$-boson
  production in polarised proton-proton collisions, $A_L^{W^\pm}$, performed by
  STAR at a centre-of-mass-energy $\sqrt{s}=510$~GeV~\cite{STAR:2018fty}.
  The measurement combines events recorded during the 2011-2012 and 2013 runs,
  and it supersedes the previous one~\cite{STAR:2014afm} used in
  {\sc NNPDFpol1.1}~\cite{Nocera:2014gqa}. It is given as a differential
  distribution in the lepton pseudorapidity $\eta_{\ell^\pm}$, which is
  strongly correlated to the $W^\pm$-boson rapidity, and it covers the interval
  $-1.25\leq\eta_{\ell^\pm}\leq +1.25$. The details of the measurement are
  summarised in Table~\ref{tab:DY_data}. We do not include measurements of
  the longitudinal single-spin asymmetry for $W^\pm+Z^0$-boson production,
  $A_L^{W^\pm +Z^0}$, measured by PHENIX~\cite{PHENIX:2015ade}, nor of the
  longitudinal double-spin asymmetry for $W^\pm$-boson production,
  $A_{LL}^{W^\pm}$, measured by STAR~\cite{STAR:2018fty}. The reason being that
  these are expected to provide little constraints on polarised
  PDFs~\cite{Nocera:2014gqa}.

\begin{table}[!t]
  \scriptsize
  \centering
  \renewcommand{\arraystretch}{1.4}
  \begin{tabularx}{\textwidth}{Xccccc}
  \toprule
  Data set
  & Ref.
  & $N_{\rm dat}$
  & $\eta_{\ell}$
  & $\sqrt{s}$~[GeV]
  & Theory \\
  \midrule
  STAR $A_L^{W^+}$
  & \cite{STAR:2018fty}
  & 6
  & [$-$1.25, +1.25]
  & $510$
  & \cite{Boughezal:2021wjw} \\
  STAR $A_L^{W^-}$
  & \cite{STAR:2018fty}
  & 6
  & [$-$1.25, +1.25]
  & $510$
  & \cite{Boughezal:2021wjw} \\
  \bottomrule
\end{tabularx}

  \vspace{0.3cm}
  \caption{Same as Table~\ref{tab:DIS_data} for $W^\pm$-boson production data.
    The piece of software used for the computations is the version of
    {\sc\small MCFM} developed in~\cite{Boughezal:2021wjw} and modified to
    produce {\sc PineAPPL} grids, see also Appendix~\ref{app:pinebench}.}
    \label{tab:DY_data}
\end{table}
  
  We compute the corresponding theoretical predictions with the modified
  version of {\sc MCFM}~\cite{Boughezal:2021wjw}, which we interfaced to
  {\sc PineAPPL} up to NLO, see Appendix~\ref{app:pinebench} for a numerical
  benchmark. NNLO corrections to the partonic matrix elements
  are included, for both the unpolarised and polarised cross sections entering
  the asymmetry, by means of a bin-by-bin $K$-factor, which we compute with the
  same version of {\sc MCFM}. We observe that NNLO corrections are generally
  small ($K$-factors are at most of $\mathcal{O}(3\%)$) and that they are
  relatively independent from the lepton rapidity, consistently with Fig.~2
  in~\cite{Boughezal:2021wjw}. This feature can be explained from the fact that
  cancellations occur between the polarised numerator and the unpolarised
  denominator in the asymmetry.

\item[Single-inclusive jet and di-jet longitudinal double-spin asymmetry.] We
  include measurements of the longitudinal double-spin asymmetry for
  single-inclusive jet and di-jet production in polarised proton-proton
  collisions, $A_{\rm LL}^{\text{1-;2-jet}}$, performed by
  PHENIX~\cite{PHENIX:2010aru} at a centre-of-mass energy $\sqrt{s}=200$~GeV,
  and by STAR at centre-of-mass energies $\sqrt{s}=200$~GeV~\cite{STAR:2012hth,
    STAR:2014wox,STAR:2021mfd} and
  $\sqrt{s}=510$~GeV~\cite{STAR:2019yqm,STAR:2021mqa}. The STAR measurements
  correspond, respectively, to the 2005-2006~\cite{STAR:2012hth},
  2009~\cite{STAR:2014wox,STAR:2016kpm,STAR:2018yxi},
  2012~\cite{STAR:2019yqm}, 2013~\cite{STAR:2021mqa}, and
  2015~\cite{STAR:2021mfd} runs. The measurements are given as distributions
  differential in the transverse momentum of the leading jet, $p_T$, in the
  case of single-inclusive jet production, and in the invariant mass of the
  di-jet system, $m_{jj}$, in the case of di-jet production. For di-jet
  production, we consider all the topologies provided. The details of these
  measurements, which include their kinematic coverage, are summarised in
  Table~\ref{tab:JET_data}. The single-inclusive jet measurements presented
  in~\cite{PHENIX:2010aru,STAR:2012hth,STAR:2014wox} were
  already included in {\sc NNPDFpol1.1}~\cite{Nocera:2014gqa}.

  We compute the corresponding theoretical predictions with the code
  presented in~\cite{deFlorian:1998qp,Jager:2004jh}, which we interfaced to
  {\sc PineAPPL}, and modified to handle the necessary kinematic cuts that
  define different di-jet topologies. For each data set, we use a jet
  algorithm consistent with that used in the corresponding experimental
  analysis. It was shown in~\cite{Mukherjee:2012uz} that single-inclusive jet
  and di-jet longitudinal double spin asymmetries are however rather
  insensistive to the jet algorithm, be it of cone- or $k_t$-type. Predictions
  are accurate up to NLO, given that NNLO corrections are not known. We
  therefore supplement them with a theory uncertainty, accounting for missing
  higher orders, estimated by varying the renormalisation scale, as we explain
  in Sect.~\ref{subsec:th_covmat}.

\begin{table}[!t]
  \scriptsize
  \centering
  \renewcommand{\arraystretch}{1.4}
  \begin{tabularx}{\textwidth}{Xcccccc}
   \toprule
   Data set
   & Ref.
   & $N_{\rm {dat}}$
   & $p_{T}$ or $m_{jj}$~[GeV]
   & $\sqrt{s}$~[GeV]
   & Theory \\
   \midrule
   PHENIX $A_{LL}^{\text{1-jet}}$
   & \cite{PHENIX:2010aru}
   & $6$
   & [2.4,10.]
   & 200
   & \cite{deFlorian:1998qp,Jager:2004jh} \\
   \midrule
   STAR $A_{LL}^{\text{1-jet}}$ (2005)
   & \cite{STAR:2012hth}
   & $10$
   & [2.4, 11.]
   & 200
   & \cite{deFlorian:1998qp,Jager:2004jh}\\
   STAR $A_{LL}^{\text{1-jet}}$ (2006)
   & \cite{STAR:2012hth}
   & $9$
   & [8.5, 35.]
   & 200
   & \cite{deFlorian:1998qp,Jager:2004jh}\\
   STAR $A_{LL}^{\text{1-jet}}$ (2009)
   & \cite{STAR:2014wox} 
   & $22$
   & [5.5, 32.]
   & 200
   & \cite{deFlorian:1998qp,Jager:2004jh} \\
   STAR $A_{LL}^{\text{2-jet}}$ (2009)
   & \cite{STAR:2016kpm,STAR:2018yxi}
   & $33$
   & [17., 68.]
   & 200
   & \cite{deFlorian:1998qp,Jager:2004jh} \\
   STAR $A_{LL}^{\text{1-jet}}$ (2012)
   & \cite{STAR:2019yqm}
   & $14$
   & [6.8, 55.]
   & 510
   & \cite{deFlorian:1998qp,Jager:2004jh} \\
   STAR $A_{LL}^{\text{2-jet}}$ (2012)
   & \cite{STAR:2019yqm}
   & $42$
   & [20., 110.]
   & 510
   & \cite{deFlorian:1998qp,Jager:2004jh} \\
   STAR $A_{LL}^{\text{1-jet}}$ (2013)
   & \cite{STAR:2021mqa}
   & $14$
   & [8.7, 63.]
   & 510
   & \cite{deFlorian:1998qp,Jager:2004jh} \\
   STAR $A_{LL}^{\text{2-jet}}$ (2013)
   & \cite{STAR:2021mqa}
   & $49$
   & [14., 133.]
   & 510
   & \cite{deFlorian:1998qp,Jager:2004jh} \\
   STAR $A_{LL}^{\text{1-jet}}$ (2015)
   & \cite{STAR:2021mfd} 
   & $22$
   & [5.8, 34.]
   & 200
   & \cite{deFlorian:1998qp,Jager:2004jh} \\
   STAR $A_{LL}^{\text{2-jet}}$ (2015)
   & \cite{STAR:2021mfd}
   & $14$
   & [20., 71.]
   & 200
   & \cite{deFlorian:1998qp,Jager:2004jh} \\
  \bottomrule
\end{tabularx}

  \vspace{0.3cm}
  \caption{Same as Table~\ref{tab:DIS_data} for single-inclusive jet and
    di-jet production data. The pieces of software used to compute the
    corresponding theoretical predictions~\cite{deFlorian:1998qp,Jager:2004jh}
    have been extended to allow for the generation of {\sc PineAPPL} grids.}
  \label{tab:JET_data}
\end{table}

\end{description}
 
\begin{figure}[!t]
  \centering
  \includegraphics[width=0.8\textwidth]{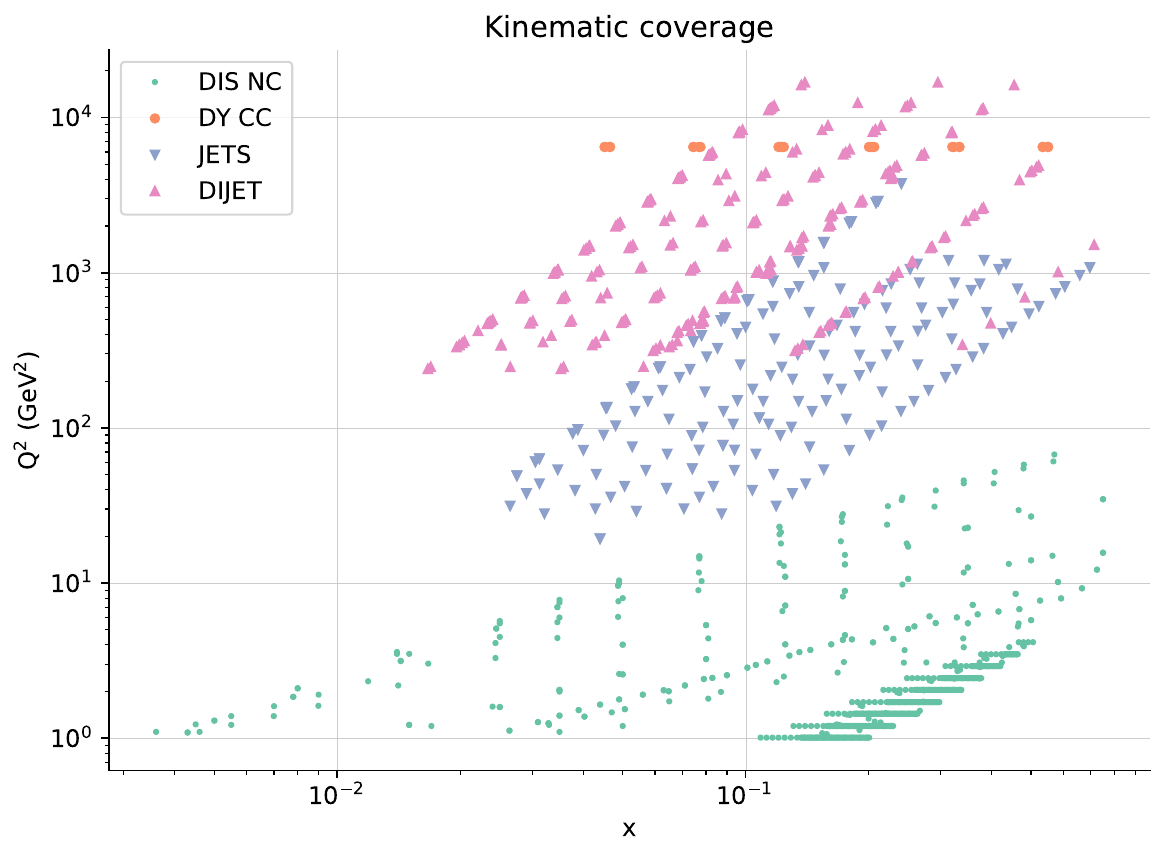}
  \caption{The kinematic coverage of the {\sc NNPDFpol2.0} data set in the
    $(x, Q^2)$ plane after applying kinematic cuts. The data sets are
    categorised as explained in Sect.~\ref{subsec:th_covmat}.}
  \label{fig:kinplot}
\end{figure}

The total number of data points included in {\sc NNPDFpol2.0}, after applying
the aforementioned kinematic cuts, is $N_{\rm dat}=951$, irrespective of the
perturbative accuracy of the determination. The corresponding kinematic
coverage in the $(x,Q^2)$ plane is displayed in Fig.~\ref{fig:kinplot}.
The data sets are categorised as explained in Sect.~\ref{subsec:th_covmat}.
For $W^\pm$-boson, single-inclusive jet, and di-jet production in polarised
proton-proton collisions, LO kinematic relations have been used to determine
$x$ and $Q^2$ from the relevant hadronic variables.

As can be seen from Fig.~\ref{fig:kinplot}, the largest number of data points
correspond to polarised inclusive DIS measurements. Given the available
coverage in the virtuality $Q^2$, the scattering is mediated by a virtual
photon, hence it is neutral current. At LO, these measurements are therefore
sensitive only to the singlet PDF flavour combination $\Delta \Sigma$ and,
thanks to the fact that deuteron and proton targets are used, also to the
non-singlet triplet $\Delta T_3$. The sensitivity to the gluon PDF,
which enters only at higher orders, is suppressed by powers of the strong
coupling. Valence-like PDF flavour combinations are probed thanks to
$W^\pm$-boson production measurements in polarised proton-proton collisions,
which is a parity-violating process. Complementary to this are measurements of
single-hadron production in DIS, that however we do not consider because they
require the simultaneous knowledge of FFs. Sensitivity to the gluon PDF is
achieved thanks to the single-inclusive jet and di-jet production measurements
in polarised proton-proton collisions, which account for the rest of our data
set. Additional constraints on the gluon PDF may be provided by measurements of
two other processes: single-hadron production in polarised proton-proton
collisions, which we do not consider because of the need for the simultaneous
knowledge of FFs; and open-charm production in DIS, which we do not consider
because the available data sets were demonstrated to bring in a negligible
amount of information~\cite{Nocera:2014gqa}.

The complete information on experimental uncertainties, including on their
correlations, is taken into account whenever available from the
{\sc HEPdata} repository~\cite{Maguire:2017ypu} or from the corresponding
publications. Specifically, full covariance matrices are provided only for the
HERMES measurement of~\cite{HERMES:2006jyl} and for
the STAR jet measurements of~\cite{STAR:2014wox,STAR:2016kpm,STAR:2018yxi,
  STAR:2019yqm,STAR:2021mqa,STAR:2021mfd}. Most notably, the latter include
correlations between all the single-inclusive jet and di-jet bins, a fact that
allows us to include all the measurements at the same time in the fit without
incurring in double counting. Information on correlations is generally not
provided by other experiments, except for the highlight of
a multiplicative, fully correlated, uncertainty due to the beam polarisation.
We use experimental uncertainties to construct the so-called experimental
covariance matrix, see Eq.~(8) in~\cite{Ball:2012wy} for its definition.
This covariance matrix will be used to present the fit quality in
Sect.~\ref{sec:results}. We also construct the $t_0$ covariance
matrix, according to Eq.~(9) in~\cite{Ball:2012wy}, whereby relative
multiplicative uncertainties are multiplied by the corresponding theoretical
predictions instead of experimental data. This prescription, which was
designed to avoid D'Agostini bias~\cite{Ball:2009qv}, will be used for
parameter optimisation.

\subsection{Perturbative accuracy}
\label{subsec:th_covmat}

The perturbative accuracy of the theoretical predictions corresponding to the
measurements described in Sect.~\ref{subsec:data} depends on the perturbative
accuracy of the partonic matrix elements and of the DGLAP splitting functions,
which are both expanded as a series in the strong coupling.
Concerning this perturbative accuracy, this work pursues two goals: first, to
include corrections up to NNLO in both; and, second, to include MHOUs arising
from the truncation of the expansion series to a finite accuracy.

The first goal is achieved by deploying a set of open-source computational
tools specifically designed for PDF fitting. As already mentioned, these
include: {\sc Yadism}~\cite{Candido:2024rkr} for the computation of the
polarised inclusive structure function $g_1$;
{\sc PineAPPL}~\cite{Carrazza:2020gss,christopher_schwan_2024_12795745}
(interfaced with the codes presented
in~\cite{Boughezal:2021wjw,deFlorian:1998qp,Jager:2004jh} and
used to compute the polarised proton-proton collision spin asymmetries)
for the construction of PDF-independent fast interpolation grids;
{\sc EKO}~\cite{Candido:2022tld,candido_2022_6340153} for PDF evolution;
and the {\sc pineline} framework~\cite{Barontini:2023vmr} to combine them all.
Each of these pieces of software have been extended to handle the computation
of the polarised observables at the desired accuracy. This amounted to the
following. We implemented in {\sc Yadism} the FONLL general-mass
variable-flavour-number scheme up to
NNLO~\cite{Hekhorn:2024tqm,Barontini:2024xgu}, which combines
the massless computation~\cite{Zijlstra:1993sh} with the recent massive
one~\cite{Hekhorn:2018ywm} and its asymptotic
limit~\cite{Bierenbaum:2022biv}. We implemented in {\sc EKO} the polarised
splitting functions, including the known corrections up to
NNLO~\cite{Moch:2014sna,Moch:2015usa,Blumlein:2021enk,Blumlein:2021ryt} and
their matching conditions~\cite{Bierenbaum:2022biv} for heavy quarks. We have
finally extended {\sc PineAPPL} and {\sc pineko} to deal with polarised
observables, including in the case in which unpolarised and polarised PDFs
ought to be used simultaneously, such as in the computation of spin asymmetries.
In all the computations, we use the same values of the physical parameters as
in {\sc NNPDF4.0}~\cite{NNPDF:2021njg}.

The second goal is achieved following the methodology developed
in~\cite{NNPDF:2019vjt,NNPDF:2019ubu,NNPDF:2024dpb}. Specifically,
we supplement the experimental covariance matrix, constructed from knowledge
of experimental uncertainties, with a MHOU covariance matrix, constructed from
renormalisation and factorisation scale variations
\begin{equation}
  \text{cov}^{(\text{tot})}_{ij}
  = \text{cov}^{(\text{exp})}_{ij}
  + \text{cov}^{(\text{mhou})}_{ij} \,, \quad i,j = 1 \dots N_{\text{dat}}\, .
  \label{eq:tot_cov}
\end{equation}
Renormalisation scale variations govern the scale dependence of matrix elements,
while factorisation scale variations govern the scale dependence of DGLAP
evolution equations. The former are correlated only across data points that
belong to the same physical processes; the latter are correlated across all
data points. To properly correlate renormalisation scale variations, we
therefore define four process categories: neutral-current DIS (DIS NC),
corresponding to measurements of $g_1$; charged-current Drell-Yan (DY CC),
corresponding to measurements of $A_L^{W^\pm}$; single-inclusive jet production
(JETS), corresponding to measurements of $A_{LL}^{\text{1-jet}}$; and di-jet
production (DIJET), corresponding to measurements of $A_{LL}^{\text{2-jet}}$.
We thus assume four independent renormalisation scales $\mu_{r,i}$ and one
common factorisation scale $\mu_{f}$. For the renormalisation $\mu_r$ and
factorisation $\mu_f$ scales we define the normalised ratios
$\rho_{r,k}=\mu_{r,k}/Q$ and $\rho_{f,k}=\mu_{f,k}/Q$, respectively, where $Q$
denotes the central scale of the process ({\it e.g.}~$Q=m_{jj}$ for di-jet
production).

The computation of $\text{cov}^{(\text{mhou})}_{ij}$ then follows the scheme B
prescription detailed in~\cite{NNPDF:2019ubu}. Using an approach similar to
that developed in~\cite{NNPDF:2024nan}, we can distinguish two different scale
variation procedures, depending on which MHOU component we want to estimate.

\begin{figure}[!t]
  \centering
  \includegraphics[width=0.49\textwidth]{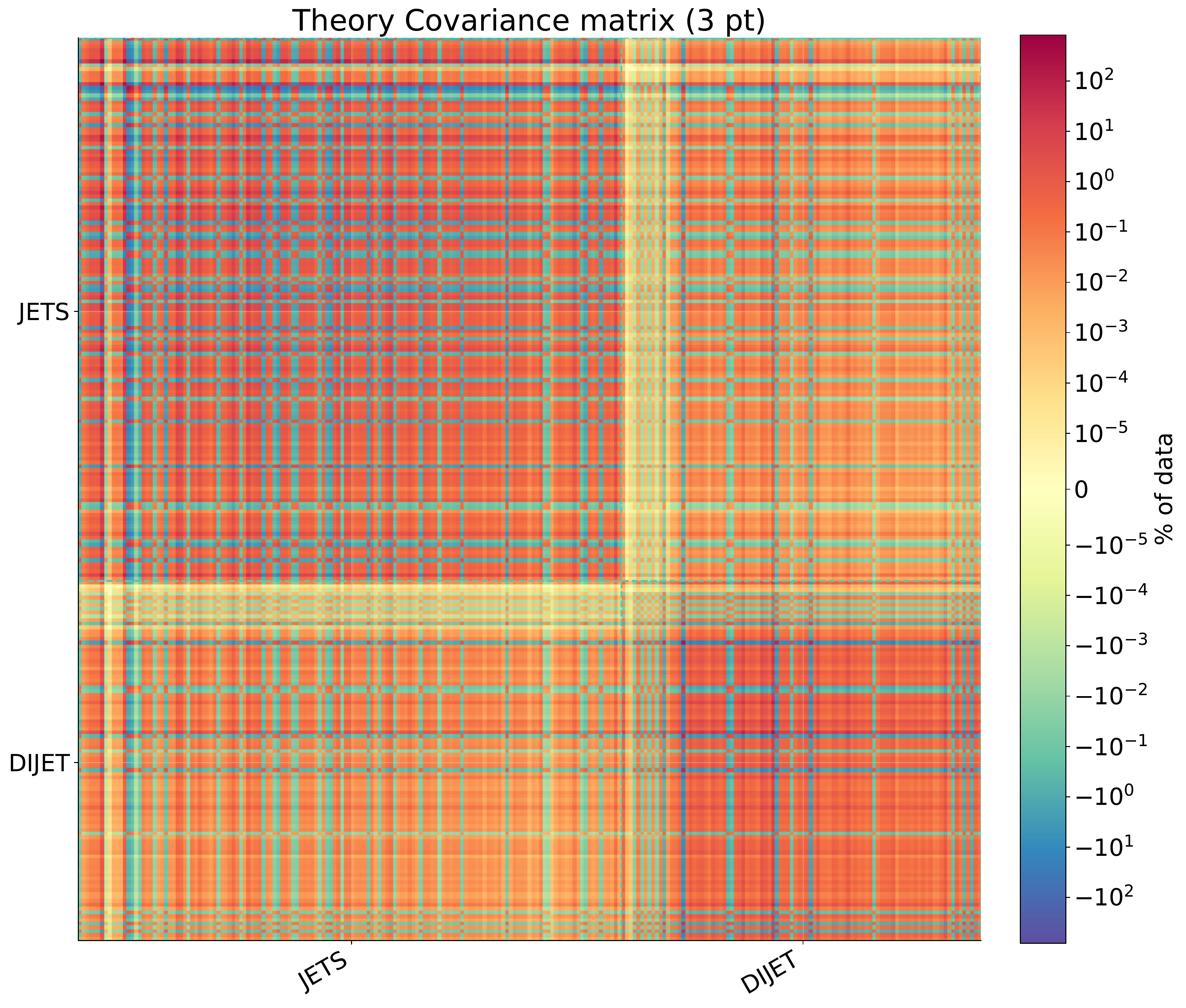}
  \includegraphics[width=0.49\textwidth]{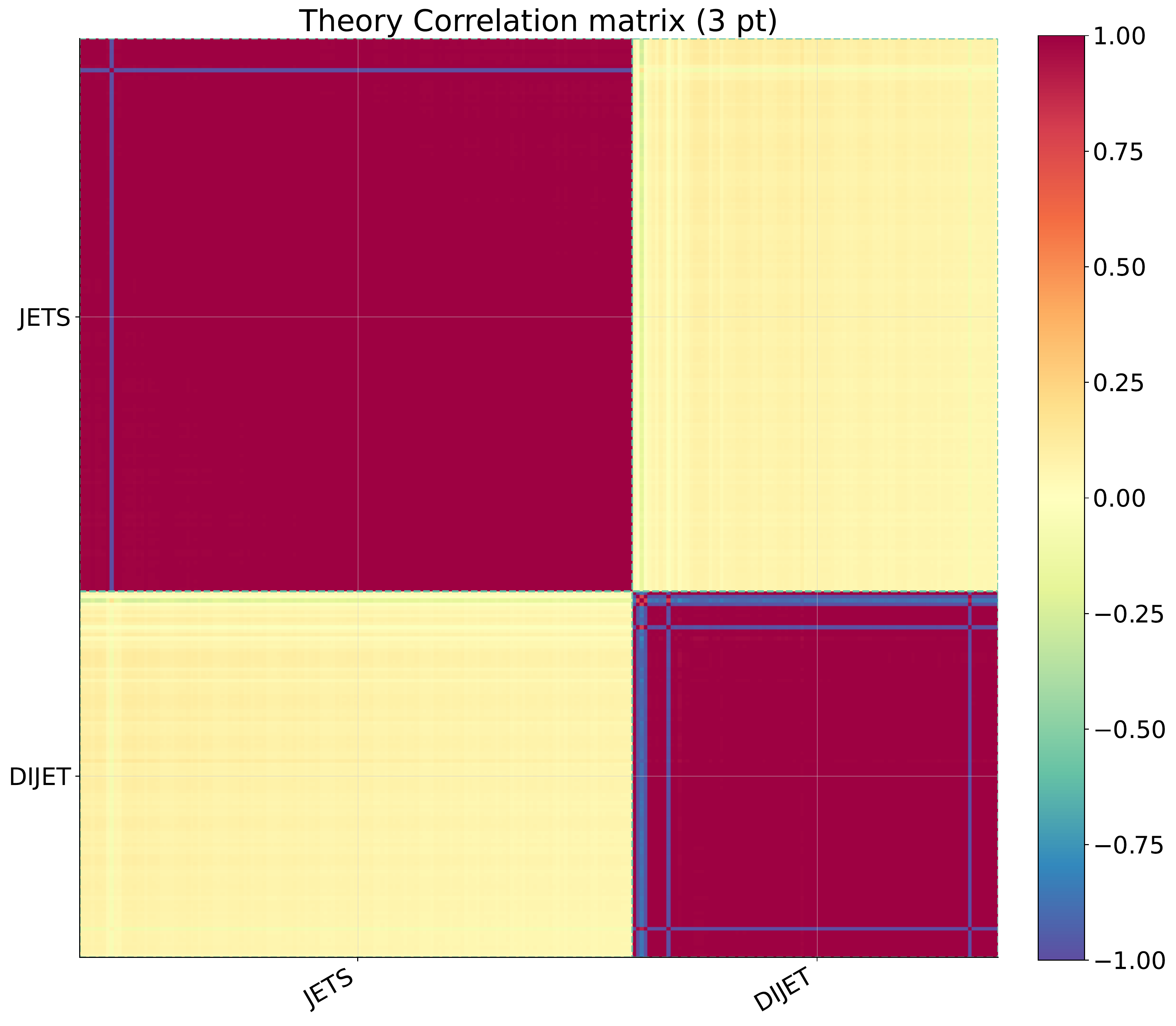}
  \caption{The JETS and DIJETS theory covariance (left) and correlation (right)
    matrices used to estimate unknown NNLO corrections to the corresponding
    matrix elements. The matrix is computed with renormalisation scale 
    variations corresponding to prescription (i) and normalised for each row
    to the central value of the data point.
    Note the logarithmic (linear) scale used in the left (right) panel. 
    }
  \label{fig:3pt_jets_covmat}
\end{figure}

\begin{figure}[!t]
  \centering
  \begin{subfigure}[!t]{0.49\textwidth}
    \includegraphics[width=\textwidth]{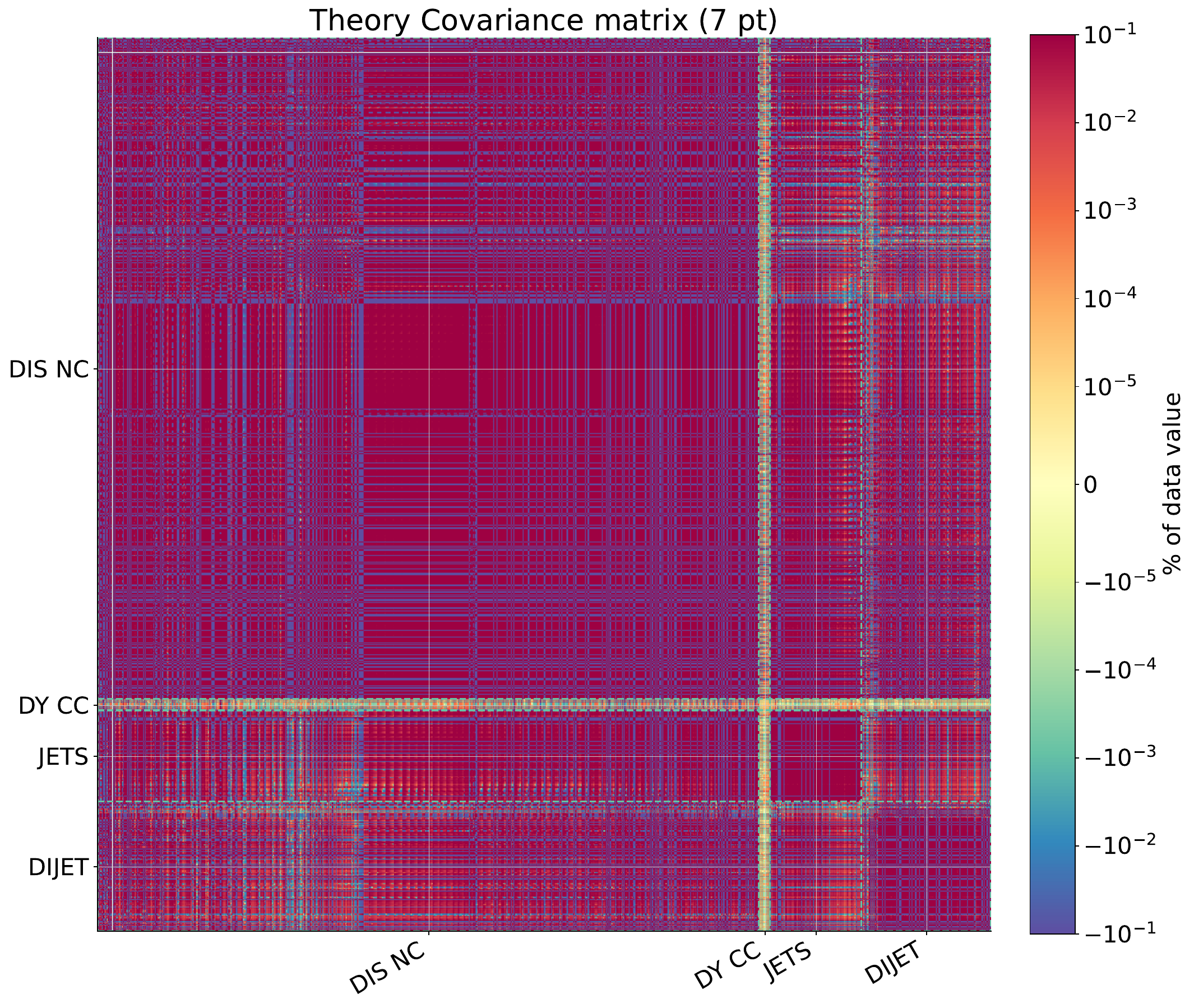}
    \caption{NLO}
  \end{subfigure}
  \begin{subfigure}[!t]{0.49\textwidth}
    \includegraphics[width=\textwidth]{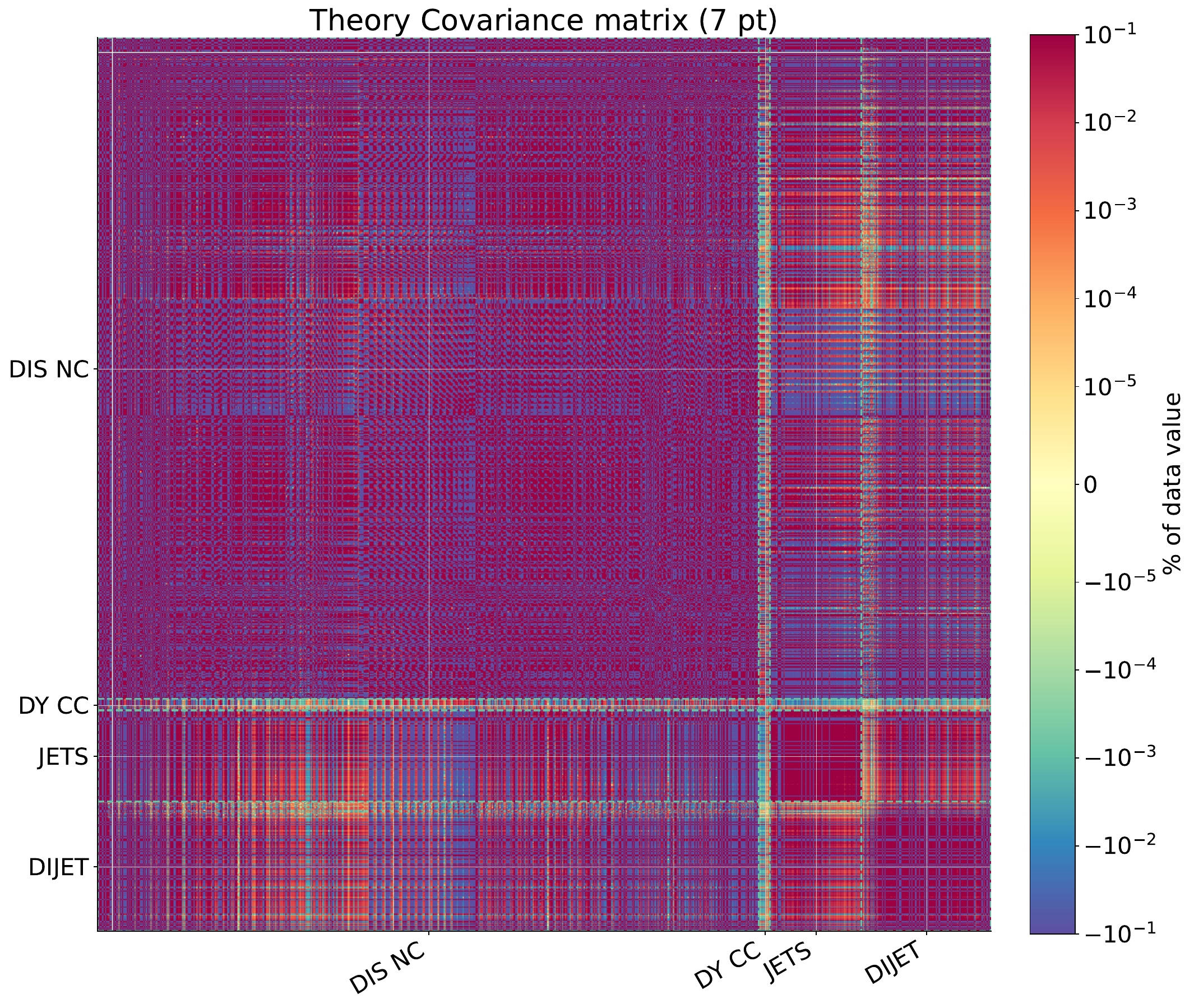}
    \caption{NNLO}
  \end{subfigure}
  \caption{The MHOU covariance matrix for all the data sets entering
    {\sc NNPDFpol2.0}, computed with the 7-point prescription (ii) at NLO
    (left) and NNLO (right).}
  \label{fig:7pt_correlation}
\end{figure}

\begin{enumerate}[label=(\roman*)]

\item We adopt a 3-point renormalisation scale variation prescription to
  estimate missing NNLO corrections in the matrix elements of those processes
  for which they are unknown, that is, single-inclusive jet and di-jet
  production in polarised proton-proton collisions. We therefore vary the
  ratios $\rho_{r,i}$ of the JETS and DIJET processes in the range
  $\rho_{r,i} \in \{0.5, 1.0, 2.0\}$. The resulting covariance matrix is shown
  in Fig.~\ref{fig:3pt_jets_covmat} (left), where each entry is expressed as
  a percentage of the experimental central value; there we also show the
  corresponding correlation matrix (right).
  
\item We adopt a 7-point renormalisation and factorisation scale variation
  prescription to estimate the MHOU for the complete data set. This
  prescription can be applied both at NLO and at NNLO. We therefore consider
  simultaneous variations of the factorisation and renormalisation scales in
  the range $\rho_k=\{0.5, 1.0, 2.0\}$, and discard the two outermost
  combinations, namely $(\rho_{r,i},\rho_f)=( 0.5,2.0)$ and $( 2.0,0.5)$.
  The resulting covariance matrix is displayed in
  Fig.~\ref{fig:7pt_correlation}, both at NLO (left) and at NNLO (right).
\end{enumerate}

The two prescriptions are exclusive. Prescription (i) will be adopted only in
the NNLO fits that we will call {\it without MHOUs} in Sect.~\ref{sec:results}.
This nomenclature puts the emphasis on the fact that MHOUs are
included only partially, and specifically only to account for unknown NNLO
corrections. All the other fits called {\it without MHOUs}, either LO or NLO,
will not use either prescription. Prescription (ii) will be adopted
in all the fits, whether LO, NLO or NNLO, that we will call {\it with MHOUs}
in Sect.~\ref{sec:results}. This nomenclature puts the emphasis on the fact
that MHOUs, beyond the nominal accuracy of the fit, are included on all data
points. Be that as it may, we have checked that MHOUs are generally smaller
than experimental uncertainties, and that they are significantly more
correlated. From Fig.~\ref{fig:7pt_correlation} we can also appreciate that
covariances are relatively independent from the perturbative order. This
behaviour is different from that observed in the unpolarised case, see
Fig.~3.1 in~\cite{NNPDF:2024dpb}. The reason being that here physical
observables are spin asymmetries, therefore their perturbative dependence
mostly cancels out in the ratio. We therefore expect a fit of polarised PDFs
to depend on the perturbative accuracy only mildly. This is indeed what we will
explicitly find in Sect.~\ref{subsec:pert_stability}.

\begin{figure}[!t]
  \centering
  \includegraphics[width=\textwidth]{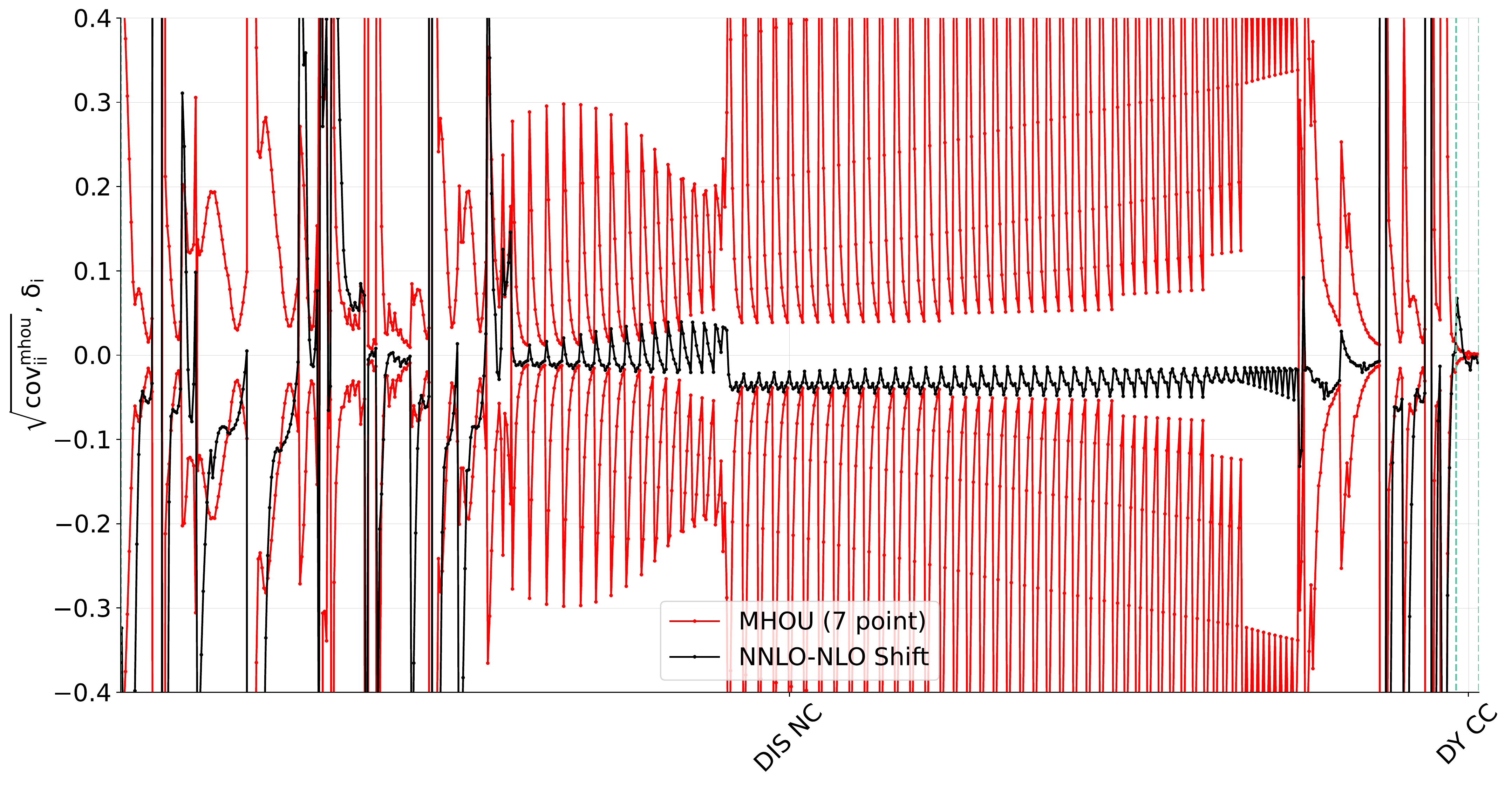}
  \caption{Comparison between the shift $\delta_i$ between known NLO and NNLO
    predictions and the square root of the diagonal elements of the NLO MHOU
    covariance matrix, computed as explained in the text. The two quantities
    are shown for each data point $i$ that belongs to the NC DIS and CC DY data
    sets listed in Tables~\ref{tab:DIS_data} and~\ref{tab:DY_data},
    respectively. Both quantities are normalised to the central value of the
    experimental data.}
  \label{fig:shift_nlo_nnlo}
\end{figure}

To verify whether the procedure is able to correctly estimate the magnitude of
higher order corrections, for each data point $i$, we can compare the shift
$\delta_i$ between known NLO and NNLO predictions to the square root of the
diagonal elements of the NLO MHOU covariance matrix,
$\sqrt{{\rm cov}_{ii}^{\rm mhou}}$, computed using the 7-point prescription and
scheme B (see~\cite{NNPDF:2019ubu}). Both quantities are normalised to the
central value of the experimental data. If the former is encompassed by the
latter, we conclude that the way in which we estimate the theoretical
uncertainty is sufficiently good to grasp the missing higher order correction.
This comparison is shown in Fig.~\ref{fig:shift_nlo_nnlo} for the DIS NC and
the DY CC data points, ordered according to the list of experiments displayed
in Table~\ref{tab:DIS_data}; the JETS and DIJET data are not shown because NNLO
corrections are unknown. We observe that MHOUs and the shift display a very
similar pattern, and, more importantly, that for most of the data points the
NLO MHOU overestimates the NLO-NNLO shift. We should however consider that the
somewhat conservative estimate of MHOUs will be compensated by correlations
once the corresponding covariance matrix is included in a PDF fit, as we will
see in Sect.~\ref{sec:results}.

\section{Fitting methodology}
\label{sec:methodology}

In this section, we present the methodology deployed to determine
{\sc NNPDFpol2.0} based on parametric regression with machine learning.
This methodology closely follows the one used in~\cite{NNPDF:2021njg,
  NNPDF:2021uiq} for the determination of the {\sc NNPDF4.0} set of
unpolarised PDFs. We review in turn which aspects of the PDF parametrisation,
optimisation, and hyperoptimisation are upgraded and adapted from the
unpolarised to the polarised case. 

\subsection{Parametrisation}
\label{subsec:parametrisation}

Parton distribution parametrisation entails two choices: first, a choice of
parametrisation basis, that is, the set of linearly independent distributions
that are parametrised; second, a choice of parametrisation form, that is, the
function that maps the PDF parameters into the elements of the basis.

Concerning the parametrisation basis, we choose the set
\begin{equation}
  \Delta f(x,Q_0^2)
  =
  \{
  \Delta g, \Delta\Sigma, \Delta T_3, \Delta T_8, \Delta V, \Delta V_3
  \}(x,Q_0^2)\,,
  \label{eq:ev_basis}
\end{equation}
made of the gluon PDF $\Delta g$ and of five independent quark flavour PDF
combinations: the singlet $\Delta\Sigma$, the non-singlet sea triplet
$\Delta T_3$ and octet $\Delta T_8$, the total valence $\Delta V$, and the
non-singlet valence $\Delta V_3$. These PDF combinations are defined as
\begin{align}
  \Delta \Sigma (x, Q_0^2)
    &=
    \Delta u^+(x, Q_0^2)
    + \Delta d^+(x, Q_0^2)
    + \Delta s^+ (x, Q_0^2)\,,
    \nonumber \\
  \Delta T_3 (x, Q_0^2)
    &=
    \Delta u^+(x, Q_0^2)
    - \Delta d^+(x, Q_0^2)\,,
    \nonumber \\
  \Delta T_8 (x, Q_0^2)
    &=
    \Delta u^+(x, Q_0^2)
    + \Delta d^+(x, Q_0^2)
    - 2 \Delta s^+ (x, Q_0^2)\,, \label{eq:param_basis}
    \\
      \Delta V (x, Q_0^2)
    &=
    \Delta u^-(x, Q_0^2)
    + \Delta d^-(x, Q_0^2)
  + \Delta s^- (x, Q_0^2)\,,
    \nonumber \\
  \Delta V_3 (x, Q_0^2)
    &=
    \Delta u^-(x, Q_0^2)
    - \Delta d^-(x, Q_0^2)\,,
    \nonumber 
  \label{eq:param-basis}
\end{align}
where $\Delta q^\pm = \Delta q \pm \Delta \bar{q}$, with $q=u, d, s$. The basis
defined in Eq.~(\ref{eq:param_basis}) corresponds to the eigenbasis of the
DGLAP evolution equations. The parametrisation scale is taken to be
$Q_0^2=1.0$~GeV$^2$; PDFs are then evolved to the scale of the physical
processes by means of DGLAP equations, see Sect.~\ref{sec:data-theory}. 

Because the available experimental data is essentially insensitive to a
possible asymmetry between $\Delta s$ and $\Delta\bar s$, we assume
$\Delta s(x,Q_0^2)=\Delta\bar s(x,Q_0^2)$. At the parametrisation scale, the
non-singlet distribution $\Delta V_8(x,Q_0^2)=\Delta u^-(x, Q_0^2)
+ \Delta d^-(x, Q_0^2) - 2 \Delta s^- (x, Q_0^2)$ is therefore equal to the
total valence distribution, $\Delta V_8(x,Q_0^2) = \Delta V(x,Q_0^2)$. 
Small differences between $\Delta s$ and $\Delta \bar s$ occur for
$Q^2>Q_0^2$ at NNLO and beyond, because higher-order QCD corrections induce a
different evolution for the two distributions~\cite{Catani:2004nc}.
We assume that charm is generated from gluon splitting through parton
evolution, therefore we set to zero, and do not parametrise, a possible
polarised intrinsic charm component at the parametrisation scale $Q_0^2$.

Concerning the parametrisation form, we choose a feed-forward neural network
with six output nodes, each of which corresponds to an element of the fitting
basis defined in Eq.~\eqref{eq:ev_basis}. The neural network has two input
nodes that correspond to values of $x$ and $\ln x$. The neural network
architecture and activation functions are determined according to the
hyperparameter optimisation procedure delineated in
Sect.~\ref{subsec:hyperoptimisation}. The output of the neural network is then
related to the polarised PDFs as
\begin{equation}
  x\Delta f(x,Q_0^2,\boldsymbol\theta)
  =
  A_{\Delta f}
  x^{1-\alpha_{\Delta f}}
  (1-x)^{\beta_{\Delta f}}
  {\rm NN}_{\Delta f}(x,\boldsymbol\theta)\,,
  \label{eq:parametrisation}
\end{equation}
where $\Delta f$ denotes each element of the chosen basis, $A_{\Delta f}$ is a
normalisation factor, $\alpha_{\Delta f}$ and $\beta_{\Delta f}$ are preprocessing
exponents, and ${\rm NN}_{\Delta f}(x,\boldsymbol\theta)$ is the output of the
neural network, which depends on weights and biases, collectively denoted as
$\boldsymbol\theta$.

The normalisation factor $A_{\Delta f}$ is equal to one for all PDFs but the
non-singlet triplet and octet PDF combinations, $\Delta T_3$ and $\Delta T_8$,
for which we define
\begin{equation}
  A_{\Delta T_3}
  =
  a_3 \left[ \int_{x_{\rm min}}^{1} dx \, \Delta T_3 (x, Q_0^2) \right]^{-1}
  \qquad{\rm and}\qquad
  A_{\Delta T_8}
  =
  a_8 \left[ \int_{x_{\rm min}}^{1} dx \, \Delta T_8 (x, Q_0^2) \right]^{-1}\,.
  \label{eq:normalisation}
\end{equation}
Here $a_3$ and $a_8$ are the (scale-independent) baryon octet decay constants,
whose experimental values are taken from the PDG
review~\cite{ParticleDataGroup:2024cfk}
\begin{equation}
  a_3 = 1.2756\pm 0.0013
  \qquad{\rm and}\qquad
  a_8 = 0.585\pm 0.025\,.
  \label{eq:decay_constants}
\end{equation}
The integrals in Eq.~\eqref{eq:normalisation} are computed each time the
parameters $\boldsymbol\theta$ change, taking $x_{\rm min}=10^{-4}$. For each
replica, the values of $a_3$ and $a_8$ are random numbers sampled from a
Gaussian distribution with mean value and standard deviation equal to the
corresponding experimental central value and uncertainty,
Eq.~\eqref{eq:decay_constants}. Enforcing Eq.~\eqref{eq:normalisation}
therefore corresponds to requiring that SU(2) and SU(3) flavour symmetries are
exact up to the experimental uncertainties quoted in
Eq.~\eqref{eq:decay_constants}.  A sizeable breaking of SU(3) is
advocated in the literature~\cite{Flores-Mendieta:1998tfv}, which
may result in an increase of the uncertainty on $a_8$ up to $30\%$. We take
as default this more conservative estimate, and set $a_8 = 0.585\pm 0.176$.

Finally, the preprocessing exponents $\alpha_{\Delta f}$ and $\beta_{\Delta f}$,
which are needed to speed up the neural network training, are determined by
means of an iterative procedure, firstly introduced in~\cite{Ball:2013lla}.
Specifically, their values are initially random numbers sampled from a flat
distribution in the ranges summarised in Table~\ref{tab:preprocessing}.
These ranges are re-determined from PDFs after these are fitted to the data;
PDFs are then refitted using values of the effective exponents that are sampled
within the newly determined ranges. The procedure is repeated until PDFs do not
change, a result that is typically reached in a couple of iterations. We use
the ranges of Table~\ref{tab:preprocessing} for all our LO, NLO, and NNLO fits,
without and with MHOUs.

\begin{table}[!t]
  \scriptsize
  \centering
  \renewcommand{\arraystretch}{1.7}
\begin{tabularx}{\textwidth}{Xcccccc}
  \toprule
  $\Delta f$
  & $\Delta \Sigma$
  & $\Delta g$
  & $\Delta T_3$
  & $\Delta T_8$
  & $\Delta V_3$
  & $\Delta V$ \\
  \midrule
  $\alpha_{\Delta f}$
  & [+1.09, +1.12]
  & [+0.82, +1.84]
  & [$-$0.44, +0.93]
  & [+0.59, +0.85]
  & [+0.47, +0.96]
  & [+0.08, +0.95] \\
  $\beta_{\Delta f}$
  & [+1.46, +3.00]
  & [+2.59, +5.70]
  & [+1.77, +3.33]
  & [+1.53, +3.44]
  & [+1.57, +3.56]
  & [+1.51, +3.45] \\
  \bottomrule
\end{tabularx}

  \vspace{0.3cm}
  \caption{The initial ranges from which the small- and large-$x$ preprocessing
    exponents are sampled for each PDF replica. The ranges are the same for the
    LO, NLO, and NNLO fits, without and with MHOUs.}
  \label{tab:preprocessing}
\end{table}

\subsection{Parameter optimisation}
\label{subsec:optimisation}

Optimisation of the neural network parameters $\boldsymbol\theta$ requires
a choice of loss function and of optimisation algorithm, including a stopping
criterion. We discuss each of these two choices in turn.

Concerning the loss function, we make considerations that are peculiar 
to the determination of polarised PDFs. For each Monte Carlo PDF replica $k$,
we define the loss function $L^{(k)}(\boldsymbol\theta)$ as the sum of three
contributions:
\begin{equation}
  L^{(k)}(\boldsymbol\theta)
  \equiv
  \chi^{2(k)}_{t_0}(\boldsymbol\theta)
  + \Lambda_{\rm int} R_{\rm int}^{(k)}(\boldsymbol\theta)
  + \Lambda_{\rm pos} R_{\rm pos}^{(k)}(\boldsymbol\theta)\,.
  \label{eq:loss_function}
\end{equation}

The first term in Eq.~\eqref{eq:loss_function} is the usual Gaussian likelihood
evaluated with the $t_0$ prescription
\begin{equation}
  \chi^{2(k)}_{t_0}(\boldsymbol\theta)
  =
  \frac{1}{N_{\rm dat}}\sum_{i,j=1}^{N_{\rm dat}}
  \left[
  T_i\left(\Delta f^{(k)}(x,Q_0^2,\boldsymbol\theta)\right) - D_i^{(k)}
  \right]
  \lp {\rm cov}_{t_0}\rp_{ij}^{-1}
  \left[
  T_j\left(\Delta f^{(k)}(x,Q_0^2,\boldsymbol\theta)\right) - D_j^{(k)}
  \right]\,,
  \label{eq:chi2}
\end{equation}
where $i,j$ are indices that run on the number of data points $N_{\rm dat}$,
${\rm cov}_{ij,t_0}$ is the $t_0$ covariance matrix, $D_{i}^{(k)}$ are the $k$-th
experimental pseudodata replicas,
and $T_{i}\left(\Delta f^{(k)}(x,Q_0^2,\boldsymbol\theta)\right)$ are the
corresponding theoretical predictions. The covariance matrix is computed as
explained in Sect.~\ref{sec:data-theory}. Specifically, the $t_0$
prescription~\cite{Ball:2009qv} is used to determine the contribution due to
experimental uncertainties in Eq.~(\ref{eq:tot_cov}), whereas point
prescriptions are used to determine the MHOU contribution, when these are
taken into account.

Theoretical predictions are matched to experimental data, which consists of
measurements of asymmetries, see Sect.~\ref{subsec:data}. As such, they are
given by the ratio between cross sections that depend on polarised PDFs and
cross sections that depend on unpolarised PDFs, where the latter are not
fitted to the data:
\begin{equation}
  T_i\left(\Delta f(x,Q_0^2,\boldsymbol\theta)\right)
  =
  \frac{T_i^{\rm (pol)}\left(\Delta f(x,Q_0^2,\boldsymbol\theta)\right)}{T_i^{\rm (unp)}\left(f(x,Q_0^2)\right)}\,.
  \label{eq:asy_theo}
\end{equation}
Therefore, parameter optimisation enters theoretical predictions only through
the numerator of Eq.~\eqref{eq:asy_theo}, which is evaluated by
convolving the parametrised PDFs with fast-kernel (FK) interpolating tables
that are pre-computed in the {\sc PineAPPL} format. The FK tables are in turn a
convolution of partonic matrix elements and evolution kernel operators (EKOs),
that evolve PDFs from the parametrisation scale $Q_0^2$ to the scale $Q^2$ of
the physical process. 

The numerical implementation of these theory predictions proceeds as follows.
Given a grid of $x$ values $\{x\}_{n=1}^{N_{\rm grid}}$, a theoretical prediction
for $T^{\rm (pol)}$ is obtained as an interpolation from grid values 
\begin{align}
  T^{\rm (pol)}\left(\Delta f(x,Q_0^2,\boldsymbol\theta)\right)
  & =
  \sum_a\sum_n \Delta f_a(x_n,Q_0^2,\boldsymbol\theta)
  \Delta{\rm FK}_a\left(x_n, Q^2\leftarrow Q_0^2 \right)\,,
  \label{eq:FK_convolution_1}    
  \\
  T^{\rm (pol)}\left(\Delta f(x,Q_0^2,\boldsymbol\theta)\right)
  & =
  \sum_{a,b}\sum_{m,n} \Delta\mathcal{L}_{ab}^{\rm L}(x_m, x_n, Q_0^2,\boldsymbol\theta)
  \Delta{\rm FK}^{\rm L}_{ab}\left(x_m, x_n,Q^2\leftarrow Q_0^2 \right)\,,
  \label{eq:FK_convolution_2} 
  \\
  T^{\rm (pol)}\left(\Delta f(x,Q_0^2,\boldsymbol\theta)\right)
  & =
  \sum_{a,b}\sum_{m,n} \Delta\mathcal{L}_{ab}^{\rm LL}(x_m, x_n, Q_0^2,\boldsymbol\theta)
  \Delta{\rm FK}^{\rm LL}_{ab}\left(x_m, x_n,Q^2\leftarrow Q_0^2 \right)\,,
  \label{eq:FK_convolution_3}
\end{align}
for processes with one, Eq.~\eqref{eq:FK_convolution_1}, or two,
Eqs.~\eqref{eq:FK_convolution_2}-\eqref{eq:FK_convolution_3}, protons in the
initial state. The indexes $a$ and $b$ run over the active partons, whereas the
indexes $m$ and $n$ run over the $N_{\rm grid}$ values of the $x$ interpolating
grid. The partonic luminosities $\Delta\mathcal{L}_{ab}^{\rm L}$ in
Eq.~\eqref{eq:FK_convolution_2} and $\Delta\mathcal{L}_{ab}^{\rm LL}$ in
Eq.~\eqref{eq:FK_convolution_3} are the product of two PDFs, in particular,
depending on the kind of theoretical prediction, of an unpolarised and a
polarised PDF or of two polarised PDFs:
\begin{align}
  \Delta\mathcal{L}_{ab}^{\rm L}(x_m,x_n,Q_0^2,\boldsymbol\theta)
  & =
  f_a(x_m,Q_0^2) \Delta f_b(x_n,Q_0^2,\boldsymbol\theta)\,,
  \label{eq:lumi_singlepol}
  \\
  \Delta\mathcal{L}_{ab}^{\rm LL}(x_m,x_n,Q_0^2,\boldsymbol\theta)
  & =
  \Delta f_a(x_m,Q_0^2,\boldsymbol\theta) \Delta f_b(x_n,Q_0^2,\boldsymbol\theta)\,.
  \label{eq:lumi_doublepol}
\end{align}
The FK tables in Eqs.~\eqref{eq:FK_convolution_1}-\eqref{eq:FK_convolution_3}
are
\begin{align}
  \Delta{\rm FK}_a\left(x_n, Q^2\leftarrow Q_0^2 \right)
  & =
  \sum_{c,j,k,\ell}\alpha_s^{p+k}
  \Delta{\rm EKO}_{a,n}^{c,j,\ell}\
  \Delta\sigma_c^{(k)}(x_\ell,\mu_j^2)\,,
  \label{eq:FKtables_1}
  \\
  \Delta{\rm FK}_{ab}^{\rm L}\left(x_m, x_n,Q^2\leftarrow Q_0^2 \right)
  & =
  \sum_{c,d,j,k,\ell,o}\alpha_s^{p+k}
  \Delta{\rm EKO}_{a,m}^{c,j,\ell}\
  {\rm EKO}_{b,n}^{d,j,o}\
  \Delta\sigma_{cd}^{(k),{\rm L}}(x_\ell,x_o,\mu_j^2)\,,
  \label{eq:FKtables_2}
  \\
  \Delta{\rm FK}_{ab}^{\rm LL}\left(x_m, x_n,Q^2\leftarrow Q_0^2 \right)
  & =
  \sum_{c,d,j,k,\ell,o}\alpha_s^{p+k}
  \Delta{\rm EKO}_{a,m}^{c,j,\ell}\
  \Delta{\rm EKO}_{b,n}^{d,j,o}\
  \Delta\sigma_{cd}^{(k),{\rm LL}}(x_\ell,x_o,\mu_j^2)\,,
  \label{eq:FKtables_3}
\end{align}
where the indexes $a,b,c$, and $d$ denote the active partons, the
indexes $m,n,\ell$, and $o$ denote points in the $x$ interpolation grids, the
index $p$ denotes the leading power of the strong coupling for a given process,
the index $k$ denotes the perturbative order, and the index $j$ labels
 the renormalisation and factorisation scale variations. Depending
on the process, a single polarised EKO ($\Delta{\rm EKO}_{a,n}^{c,j,\ell}$) is
required, a polarised and an unpolarised EKOs are required
($\Delta{\rm EKO}_{a,m}^{c,j,\ell}\ {\rm EKO}_{b,n}^{d,j,o}$), or two polarised EKOs
are required ($\Delta{\rm EKO}_{a,m}^{c,j,\ell}\ \Delta{\rm EKO}_{b,n}^{d,j,o}$).

Note that, in Eq.~\eqref{eq:asy_theo} and in Eq.~\eqref{eq:lumi_singlepol},
the unpolarised PDF does not depend on the parameters $\boldsymbol\theta$,
because this is not determined from the data in the optimisation process.
It is instead kept fixed to an external PDF set, which is chosen to be
{\sc NNPDF4.0}. Specifically, we use PDF sets determined assuming perturbative
charm, at a perturbative order consistent with that of the accuracy of the
determination of the polarised PDF. At LO, we therefore use
{\tt NNPDF40\_lo\_pch\_as\_01180}, at NLO {\tt NNPDF40\_nlo\_pch\_as\_01180},
and at NNLO {\tt NNPDF40\_nnlo\_pch\_as\_01180}. These PDF sets were determined
from an independent data set, therefore we expect little interplay between the
unpolarised and polarised PDFs, also in light of the fact that the former are
comparatively better constrained. The unpolarised and polarised PDFs are
determined with the same values of the physical parameters and with the same
methodology. Finally, the denominator in Eq.~\eqref{eq:asy_theo} is computed
using FK tables as the numerator. The only difference being that the polarised
PDF, $\Delta f(x,Q_0^2,\boldsymbol{\theta})$, in
Eqs.~\eqref{eq:FK_convolution_1}, \eqref{eq:lumi_singlepol},
and~\eqref{eq:lumi_doublepol}, and the
polarised EKOs, $\Delta{\rm EKO}$, and the polarised cross sections,
$\Delta\sigma$, in Eqs.~\eqref{eq:FKtables_1}-\eqref{eq:FKtables_3} are
replaced with their unpolarised counterparts.

The second term in Eq.~\eqref{eq:loss_function},
$\Lambda_{\rm int}R_{\rm int}(\boldsymbol\theta)$, is a regularisation term that
enforces the lowest moments of polarised PDFs to be finite. This requirement
follows from the assumption that the nucleon matrix element of the axial
current be finite for each parton. Therefore, the small-$x$ behaviour of
polarised PDFs must obey
\begin{align}
  \lim _{x \rightarrow 0} x \Delta f(x, Q^2) = 0
  & \qquad\mbox{for $\Delta f=\Delta g,\Delta\Sigma,\Delta T_3,\Delta T_8$}\,,\\
  \lim _{x \rightarrow 0} x^2 \Delta f(x, Q^2) = 0
  & \qquad\mbox{for $\Delta f=\Delta V_3,\Delta V_8$}\,.
  \label{eq:integrablity}
\end{align}
The first of these two conditions is fulfilled by construction for the polarised
quark triplet and octet PDF combinations, given the choice of normalisation
made in their parametrisation, see Eq.~\eqref{eq:normalisation}. The
regularisation term is defined as
\begin{align}
  \Lambda_{\rm int}R_{\rm int}(\boldsymbol\theta)
  =
  \Lambda_{\rm int} \sum_{f} 
  \left[x \Delta f\left(x_{\rm{int}}, Q_{\rm int}^2,\boldsymbol\theta\right)\right]^2
  & \qquad\mbox{for $\Delta f = \Delta g,\Delta\Sigma$}\,,
  \label{eq:int_reg_1}
  \\
  \Lambda_{\rm int}R_{\rm int}(\boldsymbol\theta)
  =
  \Lambda_{\rm int} \sum_{f} 
  \left[x^2 \Delta f\left(x_{\rm{int}}, Q_{\rm int}^2,\boldsymbol\theta\right)\right]^2
  & \qquad\mbox{for $\Delta f=\Delta V, \Delta V_3$}\,,
  \label{eq:int_reg_2}  
\end{align}
where $Q_{\rm int}^2=1$~GeV$^2$ and $x_{\rm{int}}=10^{-5}$. The Lagrange multiplier
$\Lambda_{\rm int}$ grows exponentially during the fit and reaches the maximum
value $\Lambda_{\rm int}=100$ at maximum training length.

The third term in Eq.~\eqref{eq:loss_function},
$\Lambda_{\rm pos} R_{\rm pos}^{(k)}(\boldsymbol\theta)$, is a regularisation term
that enforces PDFs to lead to positive cross sections. At LO, this implies
that polarised PDFs are bound by their unpolarised counterparts for each
parton $f$, for each $x$, and for each $Q^2$~\cite{Altarelli:1998gn}
\begin{equation}
  \vert \Delta f (x, Q^2) \vert \leq f(x, Q^2)\,.
  \label{eq:positivity_pdfs}
\end{equation}
Beyond LO, bounds on polarised PDFs can be derived using physical cross
sections. In the case of the gluon PDF, for instance, by studying Higgs boson
production in polarised proton-proton collisions~\cite{deFlorian:2024utd}.
Be that as it may, however, NLO corrections distort the LO positivity
condition, Eq.~\eqref{eq:positivity_pdfs}, only mildly at large $x$, where it
can actually affect polarised PDFs~\cite{Altarelli:1998gn,Hekhorn:2024foj}.
We therefore conclude that Eq.~\eqref{eq:positivity_pdfs} can be effectively
used to enforce positivity, without introducing a bias in our determination.
This conclusion is supported by the results that we get if the positivity
constraint, Eq.~\eqref{eq:positivity_pdfs}, is lifted from the fit, see
Sect.~\ref{subsec:PDF_dependence}. The regularisation term is defined as
\begin{equation}
  \Lambda_{\rm{pos}} R_{\rm pos}^{(k)}(\boldsymbol\theta)
  =
  \Lambda_{\rm{pos}} \sum_f \sum_{i=1}^{n} 
  \operatorname{ReLU}\left(-\mathcal{C}_f\left(x^i_{\rm pos}, Q^2_{\rm pos},\boldsymbol\theta\right)\right), \qquad \operatorname{ReLU}(t)=\left\{\begin{array}{ll}
  t & \text { if } t>0 \\
  0 & \text { if } t \leq 0
  \end{array}\right.\,,
  \label{eq:positivity-constraints}
\end{equation}
where the function
\begin{align}
  \mathcal{C}_f \left(x^i_{\rm pos}, Q^2_{\rm pos},\boldsymbol\theta\right)
  = f\left(x^i_{\rm pos},Q^2_{\rm pos}\right)
  -\left|\Delta f\left(x^i_{\rm pos},Q^2_{\rm pos},\boldsymbol\theta\right)\right|
  +\sigma_f\left(x^i_{\rm pos},Q^2_{\rm pos}\right),
  \label{eq:c-function}
\end{align}
encodes the positivity condition of Eq.~\eqref{eq:positivity_pdfs}, including
the uncertainty on the unpolarised PDF.
In Eqs.~\eqref{eq:positivity-constraints}-\eqref{eq:c-function},
$f$ denotes the parton, and $i$ denotes the point at which the function
$\mathcal{C}_f$ is evaluated. In particular, $n=20$ points are sampled
in the range $\left[5\cdot 10^{-7}, 9\cdot 10^{-1}\right]$, half of which are
logarithmically spaced below $10^{-1}$ and half of which are linearly spaced
above. The unpolarised PDF $f$ and its one-sigma uncertainty $\sigma_f$
are taken from NNPDF4.0, specifically from the same PDF set that enters the
computation of theoretical predictions, see above. Finally,
$Q^2_{\rm pos}=5$~GeV$^2$ and the Lagrange multiplier
$\Lambda_{\rm pos}$ grows exponentially during the fit, reaching the maximum
value $\Lambda_{\rm int}=10^{10}$ at maximum training length.

Optimisation of the parameters $\boldsymbol\theta$ is achieved through
stochastic gradient descent applied to the loss function,
Eq.~\eqref{eq:loss_function}, as in NNPDF4.0, see in particular Sect.~3.2
of~\cite{NNPDF:2021njg}. The specific optimisation algorithm is selected
from those that are available in the {\sc Keras}
library~\cite{chollet2015keras,tensorflow2015-whitepaper} through
hyperparameter optimisation, as discussed in
Sect.~\ref{subsec:hyperoptimisation}. As in NNPDF4.0, cross-validation
is used to avoid overfitting. To this purpose, for each pseudodata replica,
the data points are split into a training and a validation sets, in a
proportion of 60\% and 40\%. The loss function, Eq.~\eqref{eq:loss_function},
is then optimised on the training set and monitored on the validation set. The
stopping criterion is as in NNPDF4.0, specifically, the fit stops when the
validation loss does not improve for a number of epochs given by the product of
the maximum number of epochs and the stopping patience. Both the maximum number
of epochs and the stopping patience are hyperparameters whose values are
obtained as explained in Sect.~\ref{subsec:hyperoptimisation}.
Post-fit checks are enforced to exclude parameter configurations that violate
the positivity constraint, or, as in {\sc NNPDF4.0}, that have values of the
$\chi^2$ or the arc-length that are outside the four-sigma interval of their
distribution.

\subsection{Hyperparameter optimisation}
\label{subsec:hyperoptimisation}

Hyperparamter optimisation consists in an automated scan of the space of
models, that are appraised according to a properly defined figure of merit,
see, {\it e.g.}, Sect.~3.3 in~\cite{NNPDF:2021njg}. To ensure that the optimal
model does not lead to overfitted PDFs, a $k$-folding partition of the data set
is used, which verifies the effectiveness of any given model on sets of data
excluded in turn from the fit. A methodology to perform hyperparameter
optimisation therefore requires three choices: first, a selection of the
hyperparameters to optimise; second, a partition of the data sets into folds;
and third, a definition of the figure of merit to optimise.

Concerning the hyperparameters to optimise, similarly to what was done in
{\sc NNPDF4.0}, we scan the neural network architecture (specifically the
number of layers, the number of nodes, and the form of the activation function),
the optimiser, the value of the clipnorm, and the learning rate. We keep fixed
the maximum number of training epochs, the stopping patience, and the initial
values of the Lagrange multipliers in the integrability and positivity
regularisation terms, $\Lambda_{\rm int}$ and $\Lambda_{\rm pos}$, in
the loss function, Eq.~\eqref{eq:loss_function}.

Concerning the $k$-folding partition, as in {\sc NNPDF4.0}, we construct four
different folds, by selecting individual measurements in such a way that each
fold is representative of the global data set for both the kinematic coverage
and process types. Our selection is reported in Table~\ref{tab:kfold-datasets}.
The JLab data sets are included in all four folds, hence they are not
explicitly listed in Table~\ref{tab:kfold-datasets}. This choice is motivated
by the fact that they provide little constraints on polarised PDFs, despite
being composed of a large number of data points.

\begin{table}[!t]
  \scriptsize
  \centering
  \renewcommand{\arraystretch}{1.4}
  \begin{tabularx}{\textwidth}{CCCC}
  \toprule
  Fold 1 & Fold 2 & Fold 3 & Fold 4 \\
  \midrule
    COMPASS $g_1^p$
  & COMPASS $g_1^d$
  & E143 $g_1^p$
  & E143 $g_1^d$ \\
    E142 $g_1^n$
  & E155 $g_1^p/F_1^p$
  & HERMES $g_1^n$
  & E154 $g_1^n$\\
    E155 $g_1^n/F_1^n$
  & EMC $g_1^p$
  & SMC $g_1^d$
  & SMC $g_1^p$ \\
    HERMES $g_1^d$
  & HERMES $g_1^p$
  & SMC low-$x$ $g_1^p/F_1^p$
  & SMC low-$x$ $g_1^d/F_1^d$ \\
  \midrule
    STAR $A_L^{W^-}$
  & STAR $A_L^{W^+}$
  & STAR $A_L^{W^-}$
  & STAR $A_L^{W^+}$\\  
  \midrule
    STAR $A_{LL}^{\rm 1-jet}$ (2005)
  & STAR $A_{LL}^{\rm 1-jet}$ (2006)
  & STAR $A_{LL}^{\rm 1-jet}$ [CC] (2009)
  & STAR $A_{LL}^{\rm 1-jet}$ [CF] (2009) \\
    STAR $A_{LL}^{\rm 1-jet}$ (2012)
  & STAR $A_{LL}^{\rm 1-jet}$ (2013)
  & STAR $A_{LL}^{\rm 1-jet}$ [CC] (2015)
  & STAR $A_{LL}^{\rm 1-jet}$ [CF] (2015) \\
    STAR $A_{LL}^{\rm 2-jet}$ [A] (2009)
  & STAR $A_{LL}^{\rm 2-jet}$ [B] (2009)
  & STAR $A_{LL}^{\rm 2-jet}$ [C] (2009)
  & PHENIX $A_{LL}^{\rm 1-jet}$ 200 GeV \\
    STAR $A_{LL}^{\rm 2-jet}$ [A] (2012)
  & STAR $A_{LL}^{\rm 2-jet}$ [B] (2012)
  & STAR $A_{LL}^{\rm 2-jet}$ [C] (2012)
  & STAR $A_{LL}^{\rm 2-jet}$ [D] (2012) \\
    STAR $A_{LL}^{\rm 2-jet}$ [D] (2013)
  & STAR $A_{LL}^{\rm 2-jet}$ [C] (2013)
  & STAR $A_{LL}^{\rm 2-jet}$ [B] (2013)
  & STAR $A_{LL}^{\rm 2-jet}$ [A] (2013) \\
    STAR $A_{LL}^{\rm 2-jet}$ [SS] (2019)
  & STAR $A_{LL}^{\rm 2-jet}$ [OS] (2015)
  & STAR $A_{LL}^{\rm 2-jet}$ [SS] (2015)
  & STAR $A_{LL}^{\rm 2-jet}$ [OS] (2009) \\
  \bottomrule
\end{tabularx}

  \vspace{0.3cm}
  \caption{The partition of the {\sc NNPDFpol2.0} data set into the four folds
    used in the $k$-folding hyperoptimisation procedure. The letters in brackets
    denote different topologies of the single-inclusive jet and di-jet
    measurements. The JLab measurements, not listed explicitly in the table,
    are included in each of the four folds.}
  \label{tab:kfold-datasets}
\end{table}

Concerning the figure of merit to optimise, we make a choice that differs from
{\sc NNPDF4.0}. Specifically, instead of making our model selection based on
fits to the central data, we now hyperoptimise at the level of the PDF
distributions resulting from fits to a Monte Carlo PDF replica ensemble.
This was not possible before due to several limitations, the main one being the
inability to perform simultaneous fits of multiple replicas at once. Thanks to
various technical improvements~\cite{Cruz-Martinez:2024wiu},
it is now possible to perform simultaneous multiple replicas fits using
graphics processing units (GPUs). Such improvements also allow us to distribute
the hyperoptimisation scans across multiple GPUs, allowing for an asynchronous
search of the parameter space. This development implies that the larger the
number of GPUs utilised, the faster the scan of the hyperparameter space.

The algorithm starts each trial with a selected set of hyperparameter
configurations from which $n_{\rm folds}$ folds are constructed. For each subset
of folds, the $p$-th fold is left-out and the remaining folds are combined into
a data set from which the neural network is optimised according to the procedure
described in Sect.~\ref{subsec:optimisation} by fitting simultaneously
$N_{\rm rep}$ replicas. The figure of merit is then defined as the
$\chi^2_{\rm PDF}$ of the non-fitted folds averaged over the folds,
\begin{equation}
  L_{\rm hopt}^{(\chi^2_{\rm PDF})} \lp \boldsymbol{\hat{\theta}}\rp
  =
  \frac{1}{n_\text{folds}}\sum_{p=1}^{n_{\rm folds}}
  \underset{\boldsymbol{\theta} \in {\boldsymbol{\Theta}}}{\text{ min}^*} \lp \ \la \chi^2_{{\rm PDF},p}
  \lp \boldsymbol{\theta},\boldsymbol{\hat{\theta}} \rp \ra_{\rm rep} \rp\,,
  \label{eq:hyperopt-loss}
\end{equation}
where we distinguish between the model parameters $\boldsymbol{\theta}$
(the neural network weights and biases) and hyperparameters
$\boldsymbol{\hat{\theta}}$. The $\chi^{2(k)}_{p}$ of the $p$-th fold now
includes contributions  from the PDF uncertainties, added in quadrature to the
experimental covariance matrix
\begin{equation}
  \chi^{2(k)}_{{\rm PDF},p}(\boldsymbol{\theta})
  =
  \frac{1}{n_p} \displaystyle\sum_{i,j \in p} \lp D^{(0)}_{i}-T^{(k)}_{i}
  (\boldsymbol{\theta}) \rp\lp {\rm cov}^{\rm (exp)}
  +
  {\rm cov}^{\rm (PDF)}  \rp^{-1}_{ij} \lp D^{(0)}_{j}-T^{(k)}_{j}
  (\boldsymbol{\theta})\rp 
  \,,
  \label{eq:chi2definition}
\end{equation}
following the theory covariance matrix formalism presented
in~\cite{NNPDF:2024dpb,Cruz-Martinez:2021rgy}, and where $n_p$ indicates the
number of data points in the $p$-th fold.

For all the $n_{\rm trials}$ hyperparameter configurations explored in the
parameter space, one obtains $n_{\rm trials}$ losses computed with
Eq.~\eqref{eq:hyperopt-loss}. Additionally, for each point in the hyperparameter
space, we also evaluate the standard deviation over the replica sample in units
of the data uncertainty on the left-out folds. A suitable hyperoptimisation
loss that displays such a property can be defined
as~\cite{Cruz-Martinez:2024wiu}
\begin{equation}
  L_{\rm hopt}^{(\varphi^2)} \lp \boldsymbol{\hat{\theta}}\rp  \equiv \lp  \frac{1}{n_\text{folds}}
  \displaystyle\sum^{n_\text{folds}}_{p=1} \varphi_{\chi^2_p}^2 \lp \boldsymbol{\hat{\theta}}\rp\rp^{-1} \, ,
  \label{eq:hyperoptloss_phi2}
\end{equation}
where the metric that probes the second moment of the PDF distribution
is given by
\begin{equation}
  \varphi^2_{\chi^2_p} = \langle \chi^2_p \left[ T \left(\Delta f_{\rm fit}\right), D \right] \rangle_{\rm rep} 
  - \chi^2_p \left[  \langle T \left( \Delta f_{\rm fit} \right) \rangle_{\rm rep},D \right]\,.
  \label{eq:phi2_metric}
\end{equation} 
This equation measures the PDF uncertainty on the scale of the data
uncertainties, therefore given a successful fit to the data excluding the
$k$-th fold, the preferred extrapolation to the non-fitted $k$-th fold is the
one with the largest uncertainties, {\it i.e.}~with small values of
Eq.~\eqref{eq:hyperoptloss_phi2}.

\begin{figure}[!t]
  \centering
  \includegraphics[width=\textwidth]{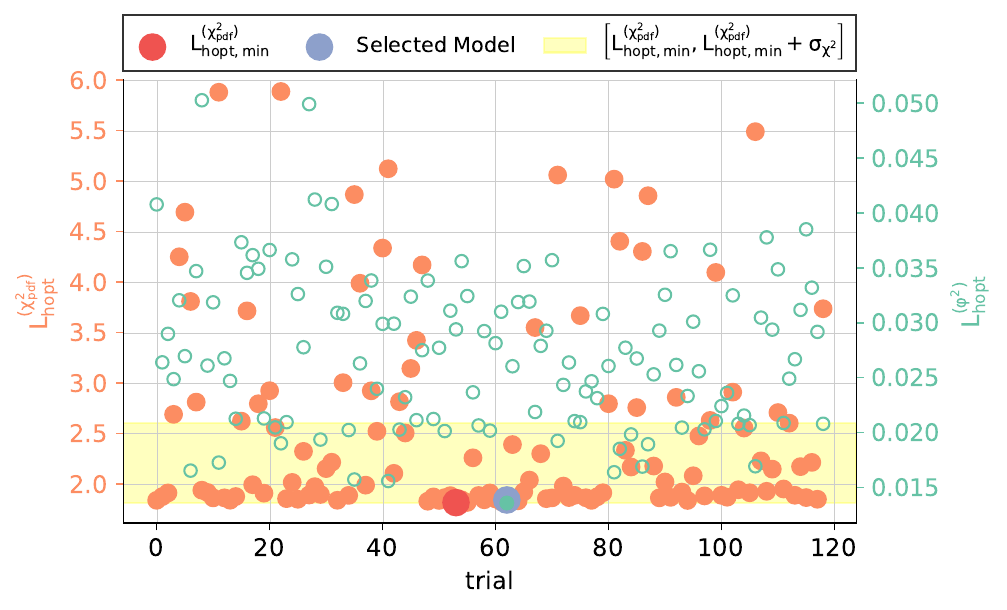}
  \caption{The values of the two hyperoptimisation metrics considered in this
    work, Eq.~\eqref{eq:hyperopt-loss} and Eq.~\eqref{eq:hyperoptloss_phi2},
    for the $n_{\rm trials}$ considered configurations of model hyperparameters.
    Specifically, for each trial, we display a pair of symbols: a point,
    corresponding to the value of the figure of merit displayed on the left
    axis, Eq.~\eqref{eq:hyperopt-loss}, and a circle, corresponding to the
    figure of merit displayed on the right axis,
    Eq.~\eqref{eq:hyperoptloss_phi2}. The band indicates the selection range
    defined in Eq.~\eqref{eq:selection_range}. The optimal model, selected to
    be the one that yields the lowest value of Eq.~\eqref{eq:hyperoptloss_phi2}
    within the range defined by Eq.~\eqref{eq:selection_range}, is highlighted
    in colour.}
  \label{fig:hyperopt-dist}
\end{figure}

Given a set of successful models, for which
$L_{\rm hopt}^{(\chi^2_{\rm pdf})} \leq L_{\rm hopt, threshold}^{(\chi^2_{\rm pdf})}$,
we then select the best one as follows. We evaluate the standard deviation
$\sigma_{\chi^2}$ from the spread of $\chi_{{\rm PDF},k}^{2(k)}$,
Eq.~\eqref{eq:chi2definition}, among the $N_{\rm rep}$ replicas of the selected
fit given by
$\hat{\boldsymbol{\theta}}^\star \equiv L_{\rm hopt, min}^{(\chi^2_{\rm pdf})}$.
We then use this value to define a selection range
\begin{equation}
\label{eq:selection_range}
  \mathcal{R}: \left[ \hat{\boldsymbol{\theta}}^\star, \hat{\boldsymbol{\theta}}^\star + \sigma_{\chi^2} \right], \qquad
  \text{with} \qquad \hat{\boldsymbol{\theta}}^\star = \underset{\hat{\boldsymbol{\theta}} \in \hat{\boldsymbol{\Theta}}}{\arg \min }
  \left( L_{\rm hopt}^{(\chi^2_{\rm pdf})} \lp \boldsymbol{\hat{\theta}}\rp \right)\,.
\end{equation}
The optimal set of hyperparameters is then selected to be those that yield the
lowest value of Eq.~(\ref{eq:hyperoptloss_phi2}) within the range $\mathcal{R}$.

\begin{table}[!t]
  \scriptsize
  \centering
  \renewcommand{\arraystretch}{1.4}
  \begin{tabularx}{\textwidth}{Xlll}
\toprule
Parameter                            & \multicolumn{2}{c}{Sampled range}                 & \multicolumn{1}{c}{Optimal model} \\
\cmidrule{2-3}
                                     &  min.                  & max.                     &               \\
\midrule
NN architecture                      & $n_1,n_2,n_3=10$       & $n_1,n_2,n_3=40$         & $n_1=29$, $n_2=12$, $n_3=6$ \\
Number of layers                     & $2$                    & $3$                      & $3$ \\ 
NN initializer                       & \textsc{Glorot\_normal}& \textsc{Glorot\_uniform} & \textsc{Glorot\_uniform} \\
Activation functions                 & \textsc{Tanh}          & \textsc{Sigmoid}         & \textsc{Tanh}  \\
Optimizer                            & \textsc{Nadam}         & \textsc{Adam}            & \textsc{Nadam} \\
Clipnorm                             & $10^{-7}$              & $10^{-4}$                & $2.95 \times 10^{-5}$ \\
Learning rate                        & $10^{-4}$              & $10^{-2}$                & $1.40 \times 10^{-3}$ \\
\midrule
Maximum \# training epochs           & \multicolumn{2}{c}{$17000$}                       & $17000$ \\
Stopping patience                    & \multicolumn{2}{c}{$0.1$}                         & $0.1$   \\
Initial positivity  multiplier       & \multicolumn{2}{c}{$185$}                         & $185$   \\   
Initial integrability multiplier     & \multicolumn{2}{c}{$10$}                          & $10$    \\
\bottomrule
\end{tabularx}

  \vspace{0.3cm}
  \caption{The hyperparameters considered in this study and the values selected
    for the optimal model. Hyperparameters that are kept fixed are reported
    in the bottom part of the table.}
  \label{tab:hyper_param}
\end{table}

\begin{figure}[!t]
  \centering
  \includegraphics[width=0.495\textwidth]{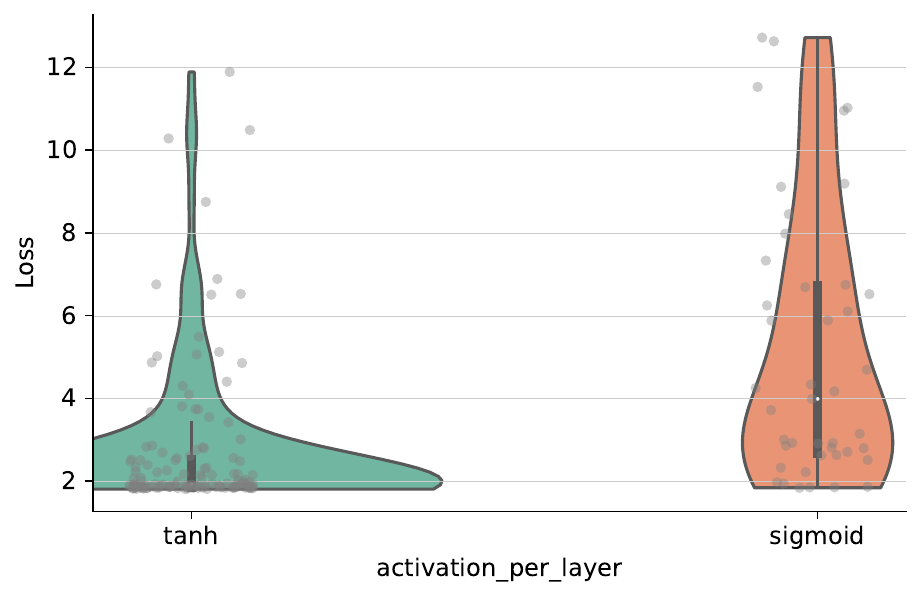}
  \includegraphics[width=0.495\textwidth]{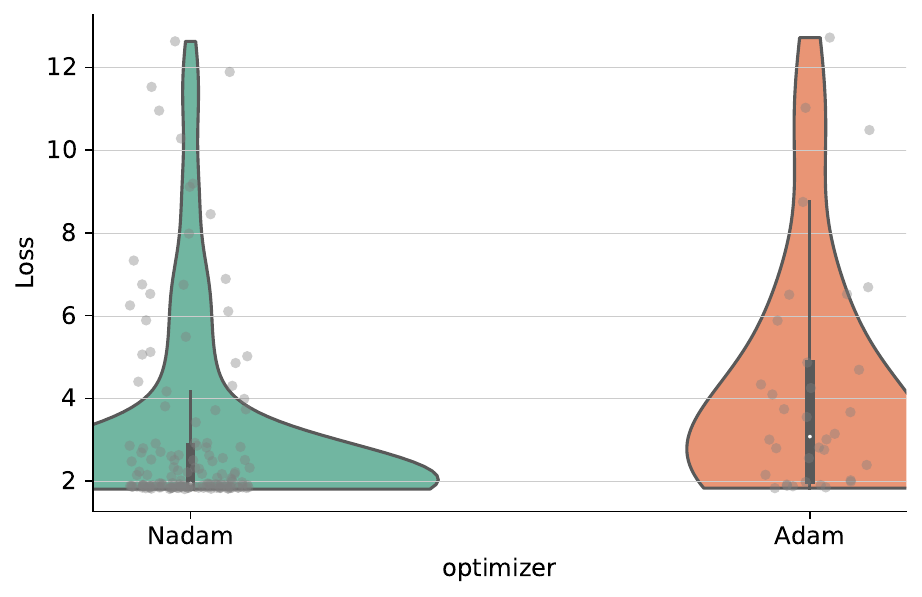}
  \caption{The values taken by the hyperoptimisation loss function
    Eq.~\eqref{eq:hyperopt-loss} for the type of activation function (left)
    and the type of optimiser (right). Results are based on a scan of
    $n_{\rm trials}=200$ configurations. The filled regions indicate the
    reconstructed probability distributions using the kernel density estimate
    method.}
  \label{fig:hyperparameters}
\end{figure}

To determine the best set of hyperparameters in {\sc NNPDFpol2.0}, we performed
a scan of $n_{\rm trials}=200$ possible configurations, distributed across four
A100 Nvidia GPUs, with $N_{\rm rep} = 60$ replicas and $n_{\rm folds} = 4$ each
as in Table~\ref{tab:kfold-datasets}. The values of the two hyperoptimisation
metrics considered here, Eq.~\eqref{eq:hyperopt-loss} and
Eq.~\eqref{eq:hyperoptloss_phi2}, have been monitored for each trial, and are
displayed in Fig.~\ref{fig:hyperopt-dist}. Specifically, for each trial, we
display a pair of symbols: a point, corresponding to the value of the figure of
merit Eq.~\eqref{eq:hyperopt-loss}; and a circle, corresponding to the figure
of merit Eq.~\eqref{eq:hyperoptloss_phi2}. The band indicates the selection
range defined in Eq.~\eqref{eq:selection_range}. The optimal model, selected
to be the one that yields the lowest value of Eq.~\eqref{eq:hyperoptloss_phi2}
within the range defined by Eq.~\eqref{eq:selection_range}, is highlighted in
colour. As reported in the unpolarised case~\cite{Cruz-Martinez:2024wiu}, while
a large number of models displays a very similar quality of agreement with the
central value of the non-fitted folds, Eq.~\eqref{eq:hyperopt-loss}, a much
wider spread is obtained for the metric that measures the PDF uncertainties,
Eq.~\eqref{eq:hyperoptloss_phi2}, which is eventually used to select the
optimal model. The selected values of the hyperparameters are reported in
Table~\ref{tab:hyper_param}. These will be adopted in the rest of this paper.
Finally, Fig.~\ref{fig:hyperparameters} displays the loss function,
Eq.~\eqref{eq:hyperopt-loss}, for the type of activation function and the type
of optimiser. The filled regions are the reconstructed probability
distributions using the kernel density estimate method.

\section{Results}
\label{sec:results}

In this section, we present the {\sc NNPDFpol2.0} parton set. We first discuss
its perturbative stability, in terms of fit quality and of PDFs. We then
compare the {\sc NNPDFpol2.0} PDFs to the previous {\sc NNPDFpol1.1} PDFs and
to those in other recent NNLO determinations. We finally study the dependence
of the {\sc NNPDFpol2.0} parton set upon variations of the positivity
constraint and on the data set.

\subsection{Fit quality, parton distributions, and perturbative stability}
\label{subsec:pert_stability}

In Table~\ref{tab:chi2tot}, we display the number of data points $N_{\rm dat}$
and the $\chi^2$ per data point corresponding to the LO, NLO, and NNLO
{\sc NNPDFpol2.0} baseline determinations. For each perturbative order, we
display a pair of values, corresponding to fits without and with inclusion of
MHOUs, as discussed in Sect.~\ref{subsec:th_covmat}. Specifically, the values
labelled as ``no MHOU'' correspond to the $\chi^2$ computed using the
experimental covariance matrix (reconstructed only from knowledge of
experimental uncertainties and their correlations), whereas the values labelled
as ``MHOU'' correspond to the $\chi^2$ computed using the sum of the
experimental and theory covariance matrices (estimated by means of the scale
variation prescriptions outlined in Sect.~\ref{subsec:th_covmat}). In neither
case the $\chi^2$ corresponds to the figure of merit utilised for parameter
optimisation, Eq.~\eqref{eq:chi2}. Values are displayed for each data set
included in the fits, for data sets aggregated according to the process
categorisation introduced in Sect.~\ref{subsec:th_covmat}, and for the total
data set.

\begin{table}[!p]
  \renewcommand{\arraystretch}{1.60}
  \scriptsize
  \centering
  \begin{tabularx}{\textwidth}{Xrcccccc}
  \toprule
  & 
  & \multicolumn{2}{c}{LO}
  & \multicolumn{2}{c}{NLO}
  & \multicolumn{2}{c}{NNLO} \\
  Dataset
  & $N_{\rm dat}$
  & no MHOU
  & MHOU
  & no MHOU
  & MHOU 
  & no MHOU
  & MHOU \\
  \midrule
  EMC $g_1^p$
  & 10 & 0.86 & 0.75 & 0.64 & 0.93 & 0.71 & 0.72 \\
  SMC $g_1^p$
  & 12 & 0.88 & 0.70 & 0.44 & 0.56 & 0.53 & 0.49 \\
  SMC $g_1^d$
  & 12 & 1.25 & 1.38 & 1.32 & 1.24 & 1.29 & 1.04 \\
  SMC low-$x$ $g_1^p/F_1^p$
  &  8 & 1.80 &	1.42 & 1.51 & 1.21 & 1.48 & 1.22 \\
  SMC low-$x$ $g_1^d/F_1^d$
  &  8 & 0.52 &	0.56 & 0.54 & 0.56 & 0.52 & 0.49 \\
  COMPASS $g_1^p$ 
  & 17 & 1.50 & 1.04 & 0.75 & 0.53 & 0.89 & 0.63 \\
  COMPASS $g_1^d$
  & 15 & 0.86 & 1.07 & 0.57 & 0.76 & 0.62 & 0.57 \\
  E142 $g_1^n$
  &  8 & 6.40 & 4.83 & 3.34 & 1.81 & 3.24 & 1.00 \\
  E143 $g_1^p$
  & 27 & 2.81 & 1.57 & 0.99 & 0.84 & 1.00 & 0.98 \\
  E143 $g_1^d$43 $g_1^d$
  & 27 & 1.40 & 1.73 & 1.31 & 1.46 & 1.36 & 1.32 \\
  E154 $g_1^n$
  & 11 & 1.20 & 0.63 & 0.30 & 0.43 & 0.40 & 0.33 \\
  E155 $g_1^p/F_1^p$
  & 24 & 0.49 & 0.53 & 0.64 & 1.00 & 0.86 & 0.75 \\
  E155 $g_1^n/F_1^n$
  & 24 & 0.64 & 0.58 & 0.62 & 0.56 & 0.75 & 0.49 \\
  HERMES $g_1^n$
  &  9 & 0.26 & 0.23 & 0.24 & 0.20 & 0.22 & 0.19 \\
  HERMES $g_1^p$
  & 15 & 3.71 & 1.85 & 1.07 & 1.03 & 1.46 & 1.29 \\
  HERMES $g_1^d$
  & 15 & 2.04 & 2.22 & 1.10 & 1.30 & 1.33 & 1.20 \\
  JLAB E06 014 $g_1^n/F_1^n$
  &  4 & 4.58 & 3.49 & 4.57 & 3.35 & 4.73 & 4.01 \\
  JLAB E97 103 $g_1^n$
  &  2 & 13.4 & 0.91 & 4.29 & 1.02 & 4.44 & 0.68 \\
  JLAB E99 117 $g_1^n/F_1^n$
  &  1 & 0.36 & 0.01 & 0.53 & 0.05 & 0.75 & 0.32 \\
  JLAB EG1 DVCS $g_1^p/F_1^p$
  & 21 & 0.15 & 0.32 & 0.21 & 0.31 & 0.17 & 0.19 \\
  JLAB EG1 DVCS $g_1^d/F_1^d$
  & 19 & 0.37 & 0.26 & 0.30 & 0.38 & 0.27 & 0.23 \\
  JLAB EG1B $g_1^p/F_1^p$
  &114 & 0.78 & 0.79 & 0.80 & 0.80 & 0.89 & 0.83 \\
  JLAB EG1B $g_1^d/F_1^d$
  &301 & 0.94 & 0.93 & 0.93 & 0.94 & 0.93 & 0.92 \\
  \midrule
  DIS NC
  &704 & 1.15 & 1.03 & 0.97 & 0.86 & 0.95 & 0.93 \\
  \midrule
  STAR $A_L^{W^+}$
  &  6 & 0.90 & 0.77 & 0.33 & 0.45 & 0.94 & 1.00 \\
  STAR $A_L^{W^-}$
  &  6 & 1.56 & 1.44 & 0.80 & 1.05 & 1.05 & 1.06 \\
  \midrule
  DY CC
  & 12 & 1.34 & 1.36 & 1.08 & 0.82 & 1.13 & 1.17 \\
  \midrule
  PHENIX $A_{LL}^{\text{1-jet}}$
  &  6 & 0.22 & 0.21 & 0.22 & 0.21 & 0.21 & 0.21 \\
  STAR $A_{LL}^{\text{1-jet}}$ (2005)
  & 10 & 1.12 & 1.13 & 1.12 & 1.11 & 1.11 & 1.10 \\
  STAR $A_{LL}^{\text{1-jet}}$ (2006)
  &  9 & 0.53 & 0.53 & 0.51 & 0.53 & 0.54 & 0.53 \\
  STAR $A_{LL}^{\text{1-jet}}$ (2009)
  & 22 & 0.76 & 0.77 & 0.73 & 0.81 & 0.80 & 0.78 \\
  STAR $A_{LL}^{\text{1-jet}}$ (2012)
  & 14 & 1.49 & 1.52 & 1.54 & 1.54 & 1.48 & 1.50 \\
  STAR $A_{LL}^{\text{1-jet}}$ (2013)
  & 14 & 1.38 & 1.28 & 1.25 & 1.32 & 1.38 & 1.37 \\
  STAR $A_{LL}^{\text{1-jet}}$ (2015)
  & 22 & 1.23 & 1.07 & 1.14 & 1.07 & 1.21 & 1.12 \\
  \midrule
  JETS
  & 97 & 1.04 & 1.04 & 1.03 & 1.03 & 1.05 & 1.03 \\
  \midrule
  STAR $A_{LL}^{\text{2-jet}}$ (2009)
  & 33 & 1.20 & 1.19 & 1.31 & 1.29 & 1.27 & 1.28 \\
  STAR $A_{LL}^{\text{2-jet}}$ (2012)
  & 42 & 1.15 & 1.15 & 1.20 & 1.19 & 1.19 & 1.17 \\
  STAR $A_{LL}^{\text{2-jet}}$ (2013)
  & 49 & 0.83 & 0.82 & 0.82 & 0.81 & 0.82 & 0.82 \\
  STAR $A_{LL}^{\text{2-jet}}$ (2015)
  & 14 & 1.24 & 1.14 & 1.10 & 1.13 & 1.27 & 1.14 \\
  \midrule
  DIJETS
  &138 & 1.06 & 1.05 & 1.08 & 1.08 & 1.09 & 1.07 \\
  \midrule
  Total
  &951 & 1.12 & 1.03 & 0.98 & 0.90 & 0.97 & 0.95 \\
  \bottomrule
\end{tabularx}

  \vspace{0.3cm}
  \caption{The number of data points $N_{\rm dat}$ and the $\chi^2$ per data
    point corresponding to the LO, NLO, and aNNLO {\sc NNPDFpol2.0} baseline
    determinations, without and with inclusion of MHOUs (see text for details).
    Values are quoted for each data set included in the fits, for data sets
    aggregated according to the process categorisation introduced in
    Sect.~\ref{subsec:th_covmat}, and for the total data set.}
  \label{tab:chi2tot}
\end{table}

From inspection of Table~\ref{tab:chi2tot}, we can assess the impact of the
perturbative accuracy and of MHOUs on the fit quality, as quantified by the
value of the $\chi^2$ per data point. Without MHOUs, the $\chi^2$ decreases when
moving from LO to NLO, while it remains essentially unchanged when moving from
NLO to NNLO. With MHOUs, it again decreases when moving from LO to NLO,
whereas it slightly increases when moving from NLO to NNLO. The observed LO to
NLO decrease amounts to 0.14 and 0.13, in the cases of fits without and with
MHOUs, respectively. This corresponds to about three sigma in units of the
distribution of the $\chi^2$ per data point, with $N_{\rm dat}=951$ data points.
Comparatively, the NLO to NNLO increase observed in the fit with MHOUs amounts
to 0.05, which is about one sigma. It can therefore be regarded as a
statistical fluctuation. We also note that the difference between the $\chi^2$
computed without and with MHOUs decreases as the perturbative order increases.

This trend, observed on the global data set, is mostly driven by the inclusive
DIS data sets, which represent about 74\% of the entire data set. In the case
of inclusive DY, the $\chi^2$ fluctuates a little more, again with a minimum
at NLO, however we do not deem these fluctuations significant, given the
smallness of the data set. In the case of single-inclusive jet and di-jet
production, the fit quality is essentially insensitive to the perturbative
order or the inclusion of MHOUs, and always remains of order one.

All of these facts lead us to three conclusions. First, that the impact of NNLO
perturbative corrections on the fit quality is almost immaterial. Second,
and consistently, that the impact of MHOUs is moderate at LO and NLO and
very limited at NNLO. Third, and consequently, that, with the current data set,
theoretical framework, and methodology, the perturbative expansion has
converged. It is therefore plausible to expect that, had perturbative
corrections to splitting functions and matrix elements been known beyond NNLO,
one would have not observed a further improvement of the fit quality.

\begin{figure}[!t]
  \centering
  \includegraphics[width=0.49\textwidth]{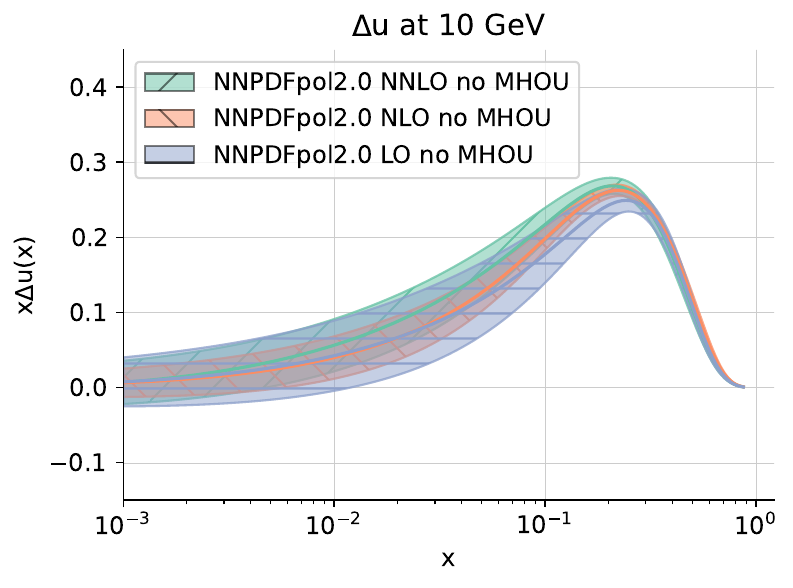}
  \includegraphics[width=0.49\textwidth]{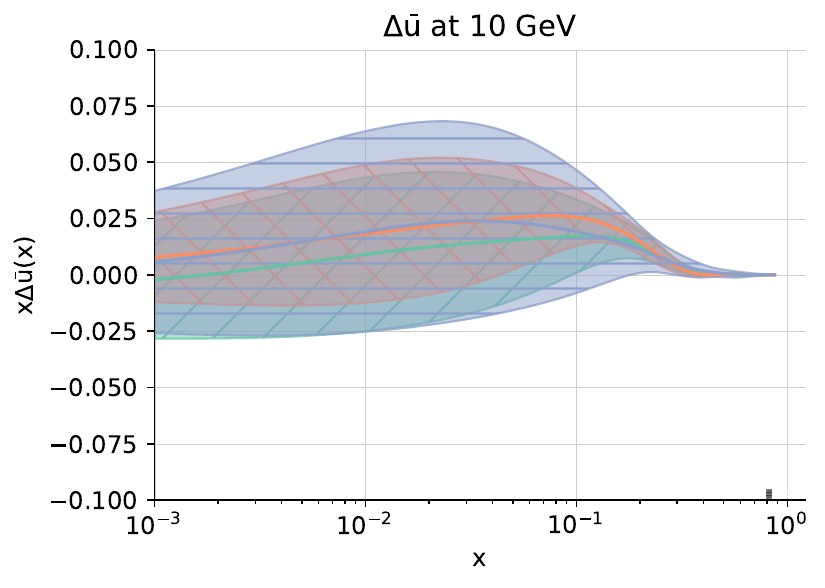}\\
  \includegraphics[width=0.49\textwidth]{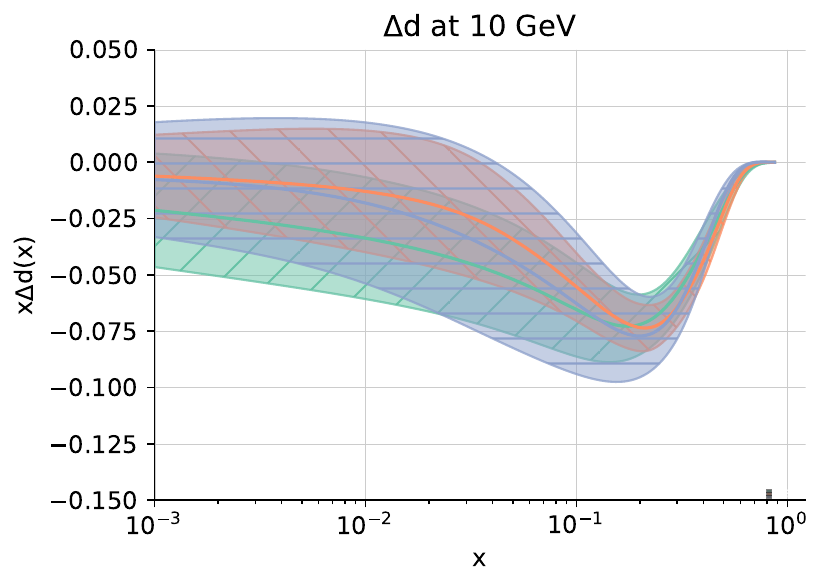}
  \includegraphics[width=0.49\textwidth]{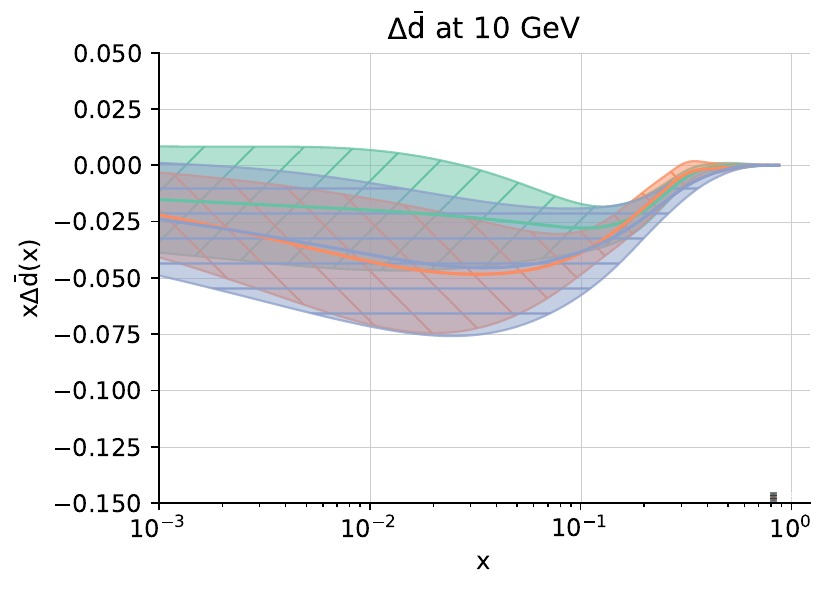}\\
  \includegraphics[width=0.49\textwidth]{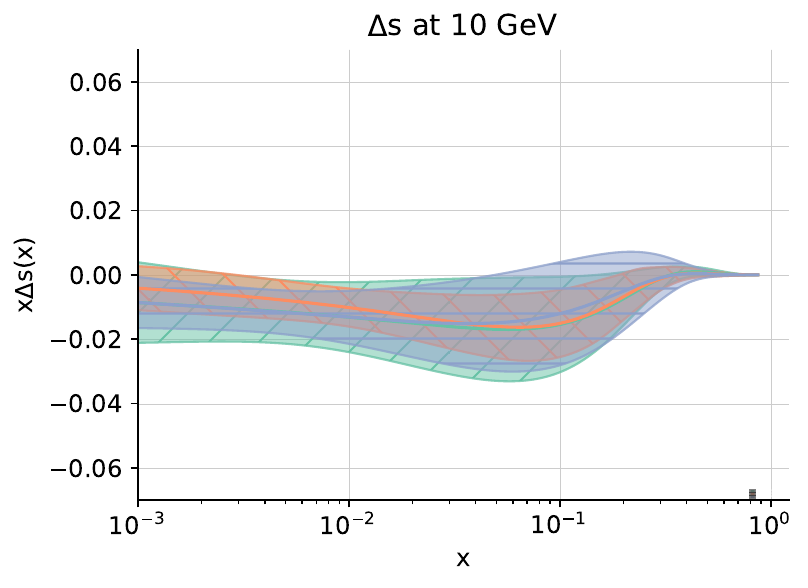}
  \includegraphics[width=0.49\textwidth]{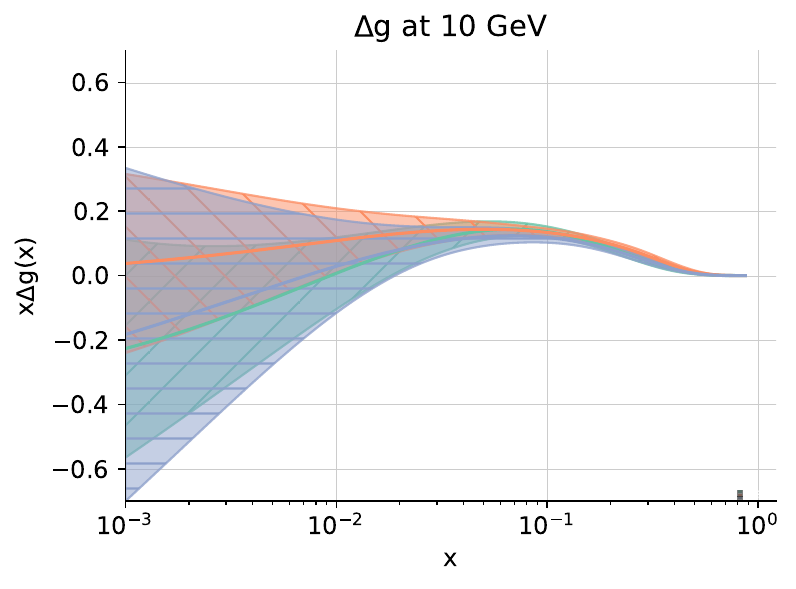}\\
  \caption{The LO, NLO, and NNLO {\sc NNPDFpol2.0} $\Delta u$, $\Delta\bar u$,
    $\Delta d$, $\Delta\bar d$, $\Delta s$, and $\Delta g$ PDFs as a function
    of $x$ in logarithmic scale at $Q=10$~GeV. Error bands correspond to one
    sigma PDF uncertainties, not including MHOUs on the theory predictions
    used in the fit.}
  \label{fig:NNPDFpol20_pert_conv}
\end{figure}

\begin{figure}[!t]
  \centering
  \includegraphics[width=0.49\textwidth]{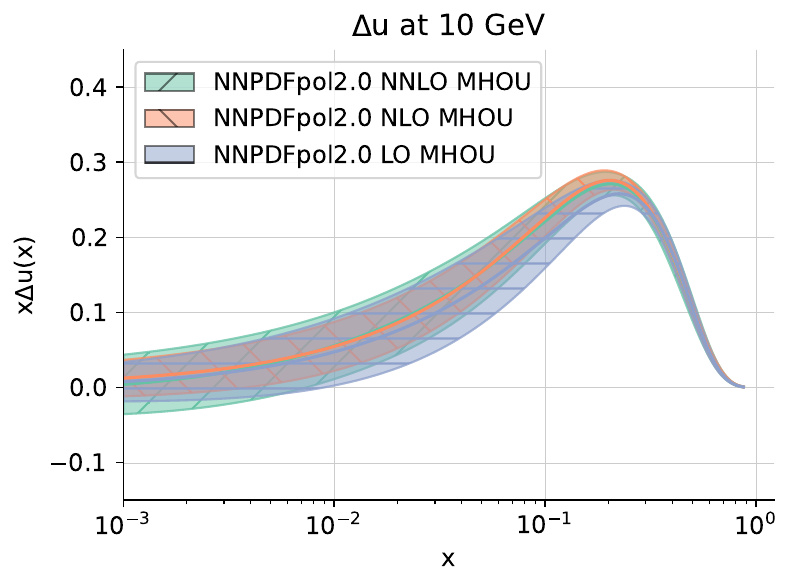}
  \includegraphics[width=0.49\textwidth]{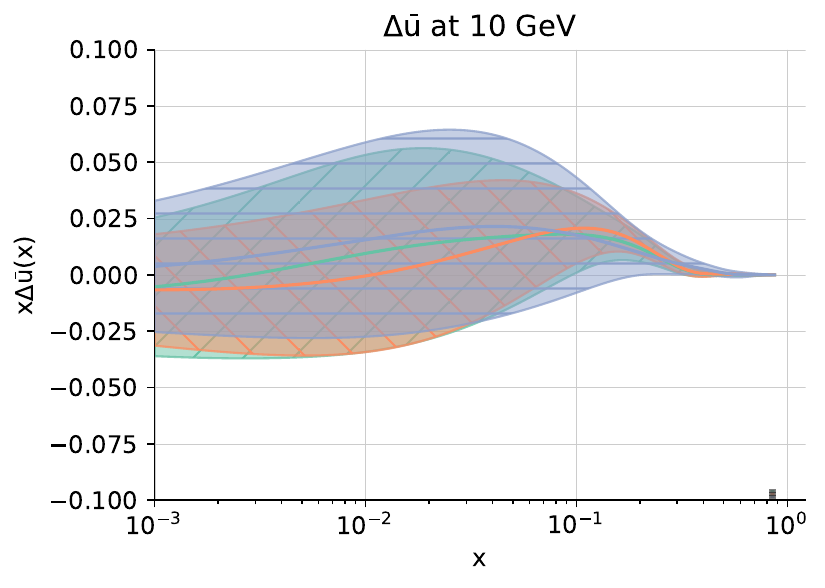}\\
  \includegraphics[width=0.49\textwidth]{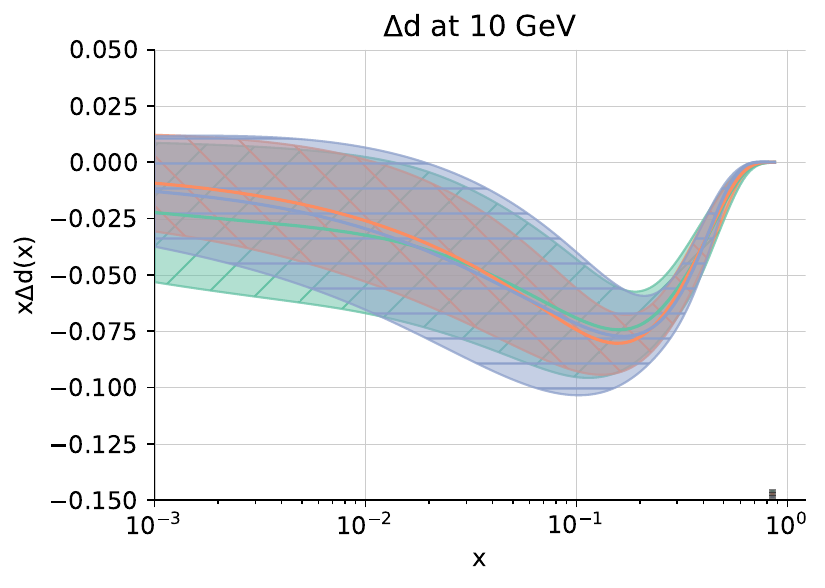}
  \includegraphics[width=0.49\textwidth]{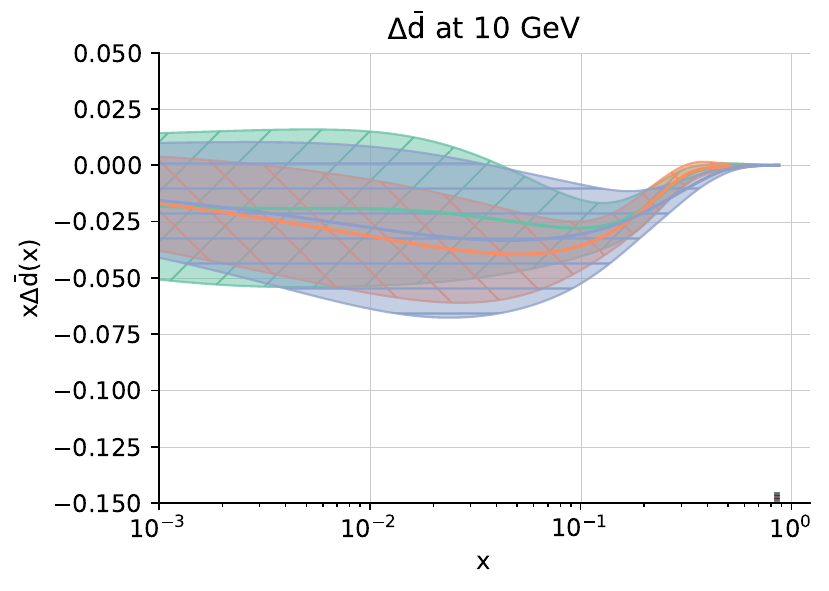}\\
  \includegraphics[width=0.49\textwidth]{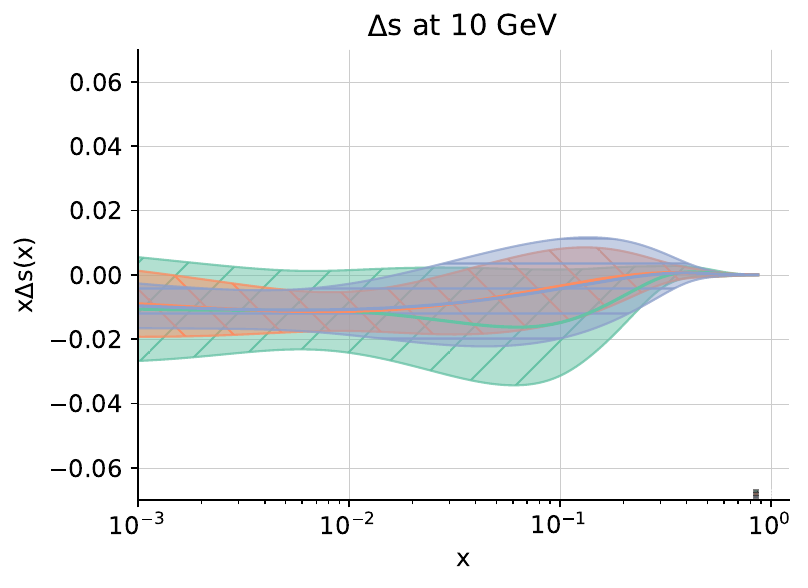}
  \includegraphics[width=0.49\textwidth]{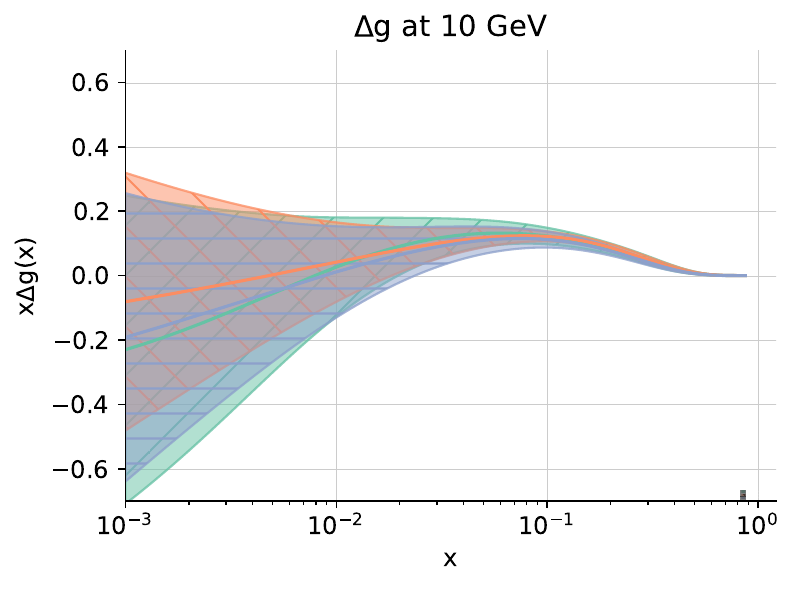}\\
  \caption{Same as Fig.~\ref{fig:NNPDFpol20_pert_conv} for the LO, NLO and
  NNLO {\sc NNPDFpol2.0} PDFs that include MHOUs.}
  \label{fig:NNPDFpol20_pert_conv_mhou}
\end{figure}

It is interesting to note that a similar behaviour was also seen in unpolarised
PDFs~\cite{NNPDF:2024nan}, although one order higher in the perturbative
expansion. In this respect, fits of polarised PDFs look significantly more
insensitive to the perturbative accuracy than their unpolarised counterparts.
We ascribe this evidence to the fact that unpolarised and polarised PDF fits
rely on measurements for different physical observables: the former on cross
sections; the latter on spin asymmetries. By definition, an asymmetry is a
ratio of cross sections, whereby most of the theoretical corrections, including
those of perturbative origin, cancel out.

We then examine the polarised PDFs. We compare the LO, NLO, and NNLO
{\sc NNPDFpol2.0} PDFs, obtained without and with inclusion of MHOUs, in
Figs.~\ref{fig:NNPDFpol20_pert_conv} and~\ref{fig:NNPDFpol20_pert_conv_mhou},
respectively. Specifically, we show the polarised up, anti-up, down, anti-down,
strange and gluon PDFs, as a function of $x$ in logarithmic scale at $Q=10$~GeV.
Error bands correspond to one sigma PDF uncertainties, which do not (no MHOU
sets) or do (MHOU sets) include MHOUs on theoretical predictions as specified
in Sect.~\ref{subsec:th_covmat}.

The features observed in the fit quality are consistently manifest on PDFs.
Without MHOUs, we see that the PDF dependence on the perturbative order is
generally mild, and milder when moving from NLO to NNLO than from LO to NLO.
This can be noticed, in particular, in the polarised gluon PDF around
$x\sim 0.3$, where the difference between the LO and NLO central values is
about one sigma, whereas the same difference between the NLO and NNLO is
almost null. With MHOUs, differences become even milder. The shift in the PDF
central value due to inclusion of higher order corrections (both NLO and NNLO)
is comfortably encompassed by the uncertainty of the LO PDF. This fact confirms
that the MHOU covariance matrix estimated through scale variation is correctly
reproducing the effect due to higher perturbative orders.

We therefore conclude that the LO PDFs with MHOUs are almost as accurate as
their NLO and NNLO counterparts. They are however a little less precise, as we
can see from their uncertainties that are larger than those of their NLO and
NNLO counterparts. This is particularly visible in the case of the polarised up
and anti-up quarks at intermediate-to-large values of $x$. Inclusion of MHOUs
does not inflate uncertainties significantly, as one can realise by
comparing Fig.~\ref{fig:NNPDFpol20_pert_conv} with
Fig.~\ref{fig:NNPDFpol20_pert_conv_mhou}. This means that correlations
across factorisation and renormalisation scale variations
(see Fig.~\ref{fig:7pt_correlation}) counteract the inflation of the diagonal
elements of the covariance matrix (see Fig.~\ref{fig:shift_nlo_nnlo})
due to the inclusion of the MHOU covariance matrix.

\begin{figure}[!t]
  \centering
  \includegraphics[width=0.49\textwidth]{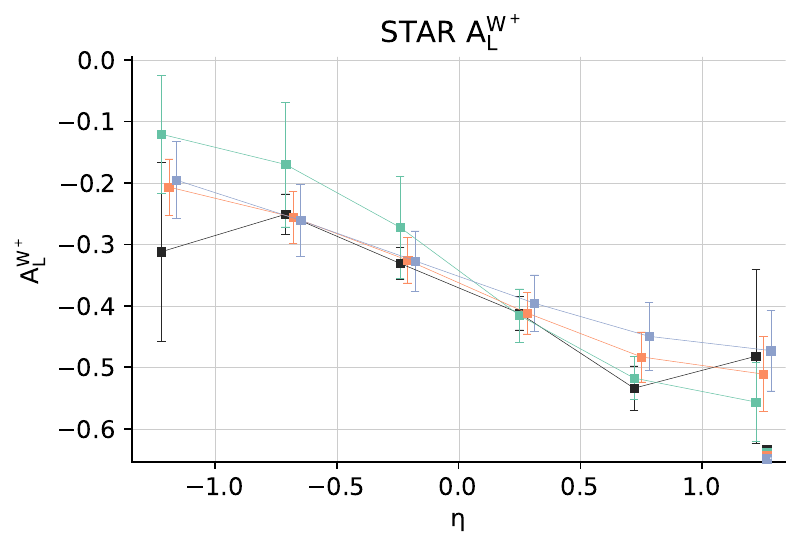}
  \includegraphics[width=0.49\textwidth]{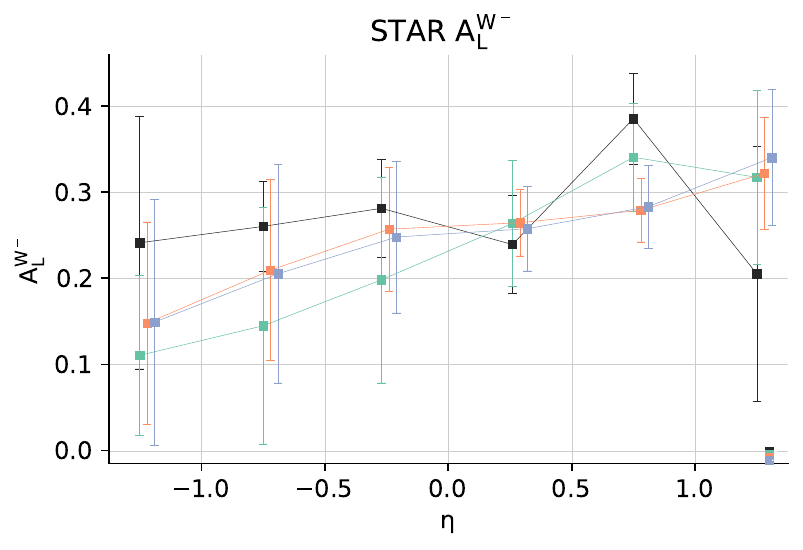}\\
  \includegraphics[width=0.49\textwidth]{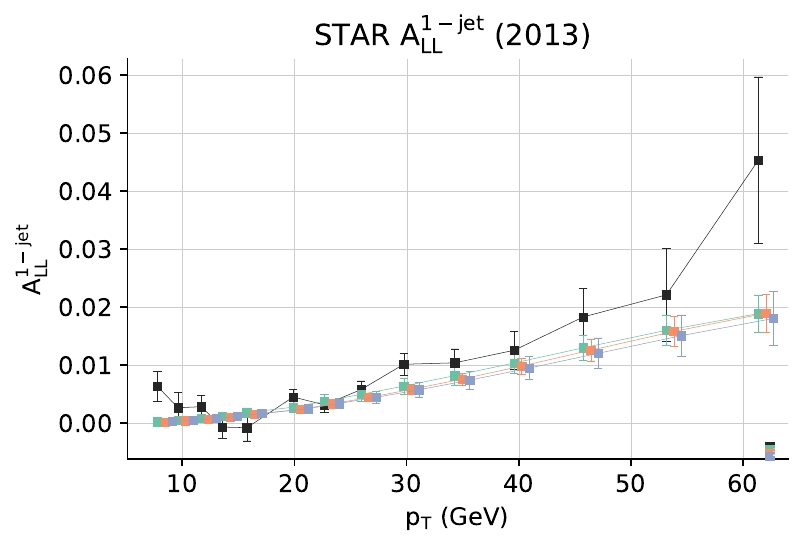}
  \includegraphics[width=0.49\textwidth]{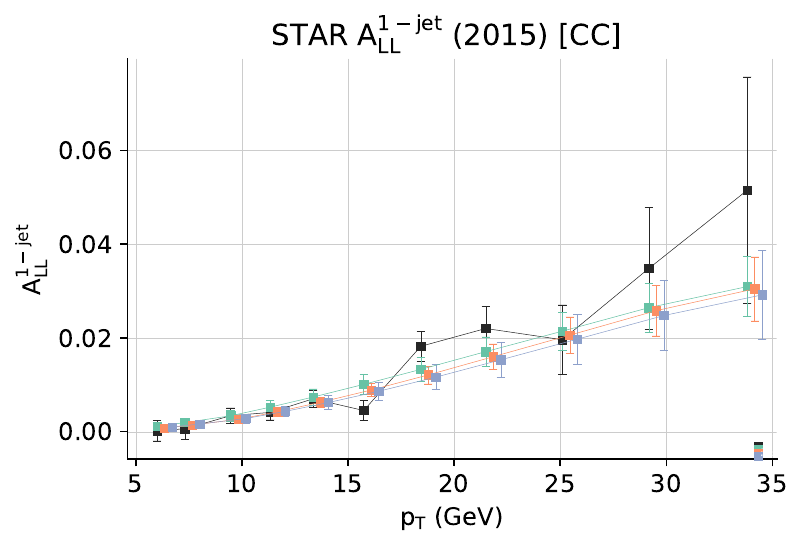}\\
  \includegraphics[width=0.49\textwidth]{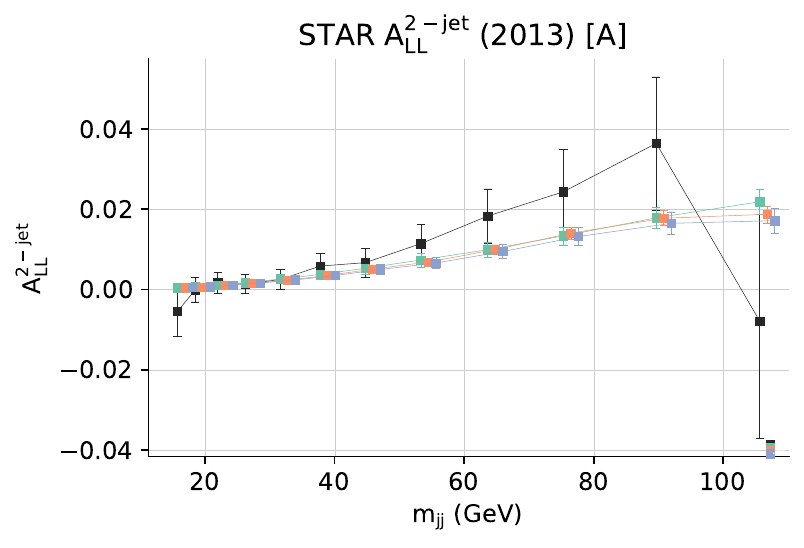}
  \includegraphics[width=0.49\textwidth]{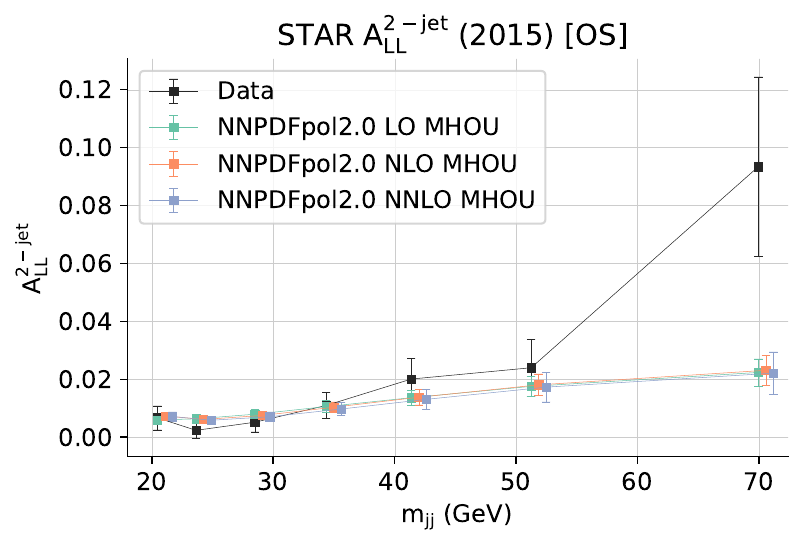}\\
  \caption{Comparison between a representative selection of the data sets
    included in {\sc NNDPFpol2.0} and the corresponding theoretical
    predictions. Specifically, from top to bottom, we display: the longitudinal
    single-spin asymmetry for $W^\pm$-boson production measured by the STAR
    experiment at a centre-of-mass energy of 510~GeV~\cite{STAR:2018fty};
    the longitudinal double-spin asymmetry for single-inclusive jet production
    from the 2013~\cite{STAR:2021mqa} and 2015 (in the central rapidity
    region)~\cite{STAR:2021mfd} runs measured by the STAR experiment at
    centre-of-mass energies of 510~GeV and 200~GeV, respectively; and the
    longitudinal double-spin asymmetry for di-jet production from the same
    2013 (topology [A])~\cite{STAR:2021mqa} and 2015 (opposite-sign [OS]
    topology)~\cite{STAR:2021mfd} runs. Theoretical predictions are computed
    at LO, NLO, and NNLO with the corresponding MHOU {\sc NNDPFpol2.0} PDF
    sets. The unpolarised PDF is taken consistently from the LO, NLO, and
    NNLO {\sc NNPDF4.0} central set. For single-inclusive jet and di-jet
    asymmetries, NNLO predictions are approximate, in that splitting functions
    used for parton evolution are accurate to NNLO, whereas matrix elements are
    accurate only to NLO. Experimental uncertainties are the sum in quadrature
    of all statistical and systematic uncertainties. Uncertainties on
    theoretical predictions correspond to one-sigma polarised PDF uncertainties
    only.}
  \label{fig:data_theory}
\end{figure}

A final evidence of the excellent perturbative stability of the
{\sc NNPDFpol2.0} sets is provided by the comparison of theoretical predictions
and experimental data. In Fig.~\ref{fig:data_theory}, we display such a
comparison for a representative selection of the data sets included in
{\sc NNDPFpol2.0}, specifically: the longitudinal single-spin asymmetry
for $W^\pm$-boson production measured by the STAR experiment at a centre-of-mass
energy of 510~GeV~\cite{STAR:2018fty}; the longitudinal double-spin asymmetry
for single-inclusive jet production from the 2013~\cite{STAR:2021mqa}
and 2015 (in the central rapidity region)~\cite{STAR:2021mfd} runs measured by
the STAR experiment at centre-of-mass energies of 510~GeV and 200~GeV,
respectively; and the longitudinal double-spin asymmetry for di-jet
production from the same 2013 (topology [A])~\cite{STAR:2021mqa}
and 2015 (opposite-sign [OS] topology)~\cite{STAR:2021mfd} runs. Theoretical
predictions are computed at LO, NLO, and NNLO with the corresponding MHOU
{\sc NNDPFpol2.0} PDF sets. The unpolarised PDF is taken consistently
from the LO, NLO, and NNLO {\sc NNPDF4.0} central set. For single-inclusive
jet and di-jet asymmetries, NNLO predictions are approximate, in that
splitting functions used for parton evolution are accurate to NNLO,
whereas matrix elements are accurate only to NLO. Experimental uncertainties
are the sum in quadrature of all statistical and systematic uncertainties.
Uncertainties on theoretical predictions correspond to one-sigma polarised PDF
uncertainties only. As we can see from Fig.~\ref{fig:data_theory}, theoretical
predictions are in good agreement with experimental data and vary only mildly
with the perturbative accuracy.

\subsection{Comparison to {\sc NNPDFpol1.1} and to other PDF sets}
\label{subsec:comparison}

\begin{figure}[!t]
  \centering
  \includegraphics[width=0.49\textwidth]{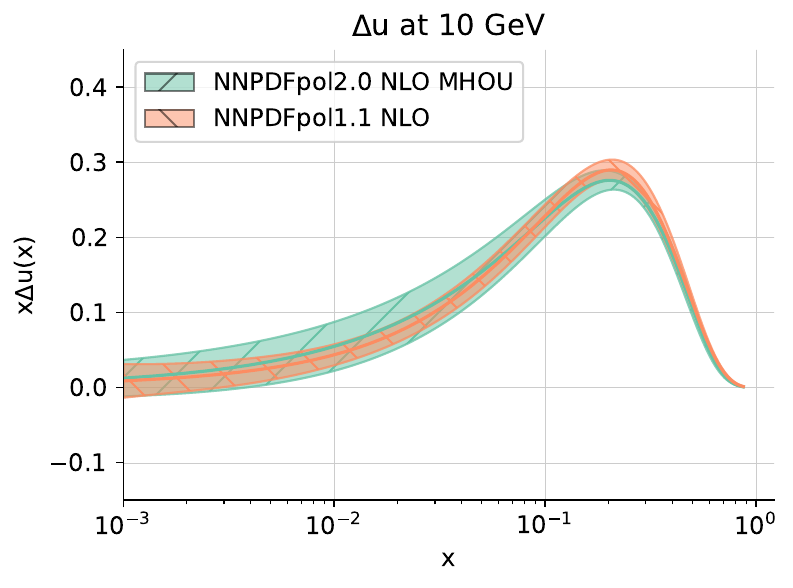}
  \includegraphics[width=0.49\textwidth]{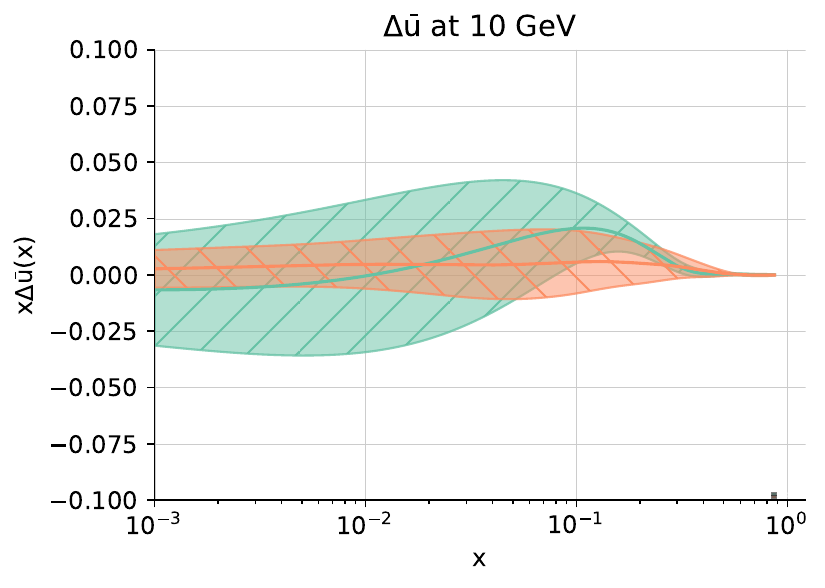}\\
  \includegraphics[width=0.49\textwidth]{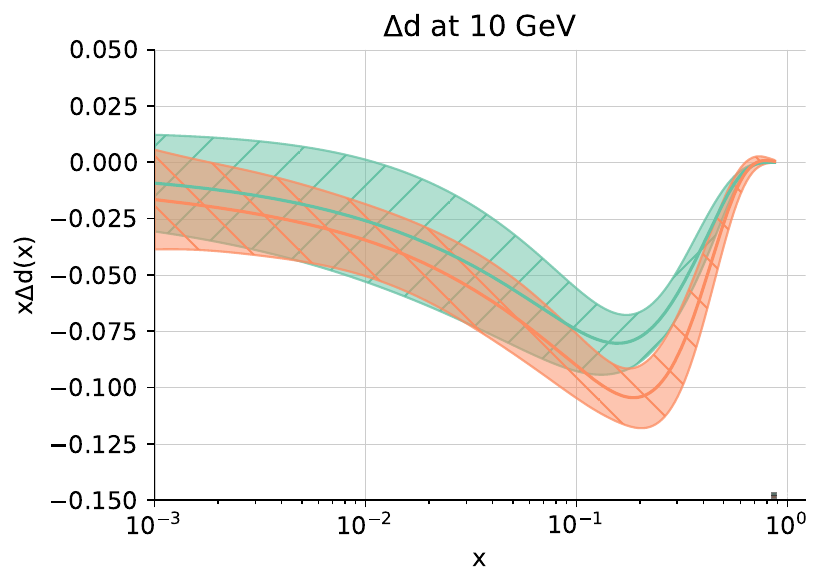}
  \includegraphics[width=0.49\textwidth]{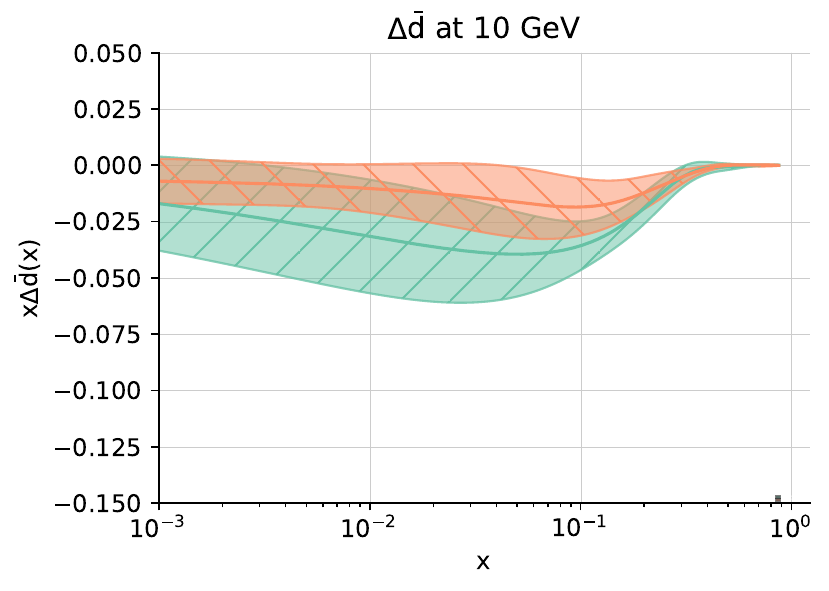}\\
  \includegraphics[width=0.49\textwidth]{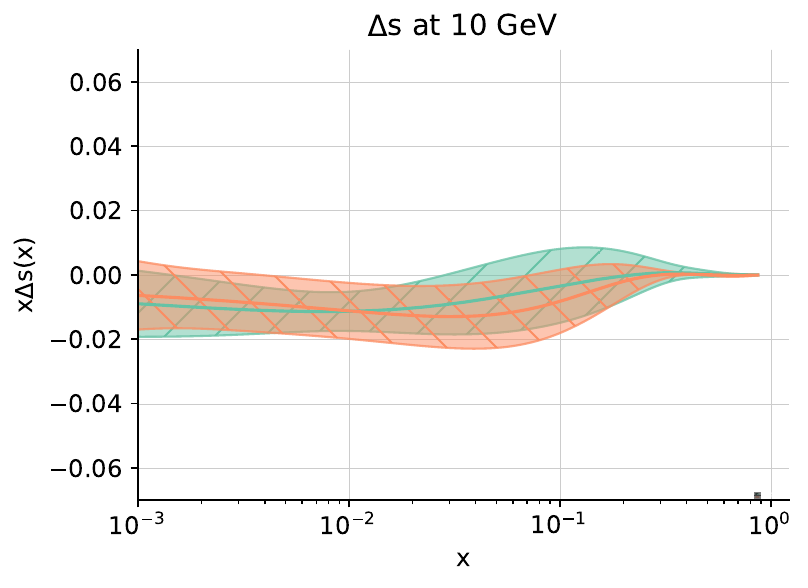}
  \includegraphics[width=0.49\textwidth]{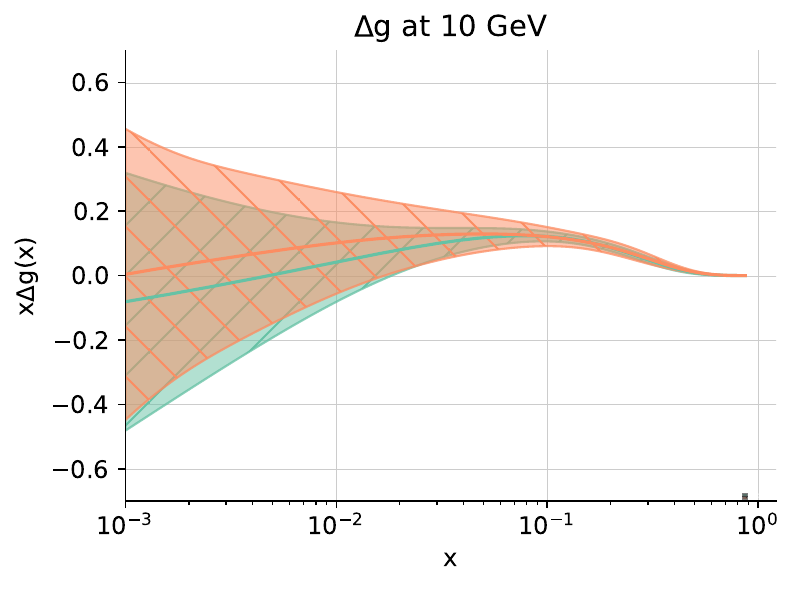}\\
  \caption{Comparison of the NLO {\sc NNPDFpol2.0} and {\sc NNPDFpol1.1} PDFs
    as a function of $x$ at $Q=10$~GeV, in the same format as
    Fig.~\ref{fig:NNPDFpol20_pert_conv}.}
  \label{fig:NNPDFpol20_vs_NNPDFpol12}
\end{figure}

\begin{figure}[!t]
  \centering
  \includegraphics[width=0.49\textwidth]{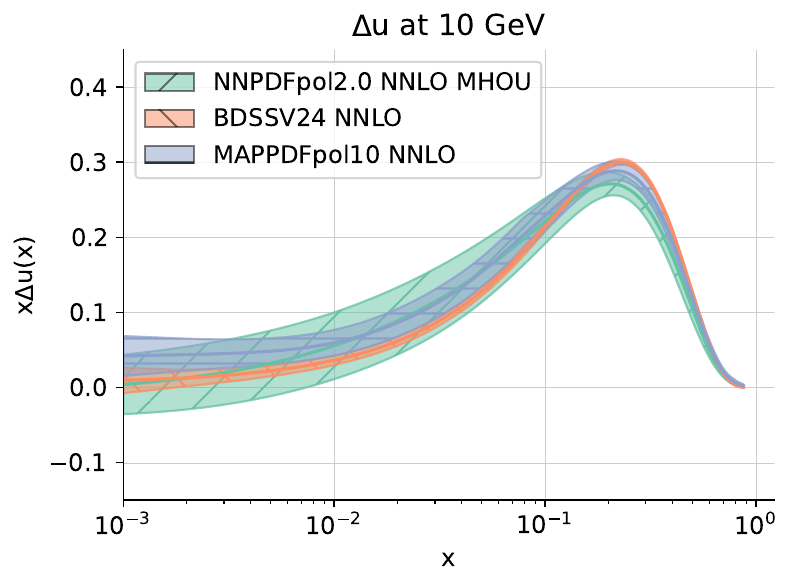}
  \includegraphics[width=0.49\textwidth]{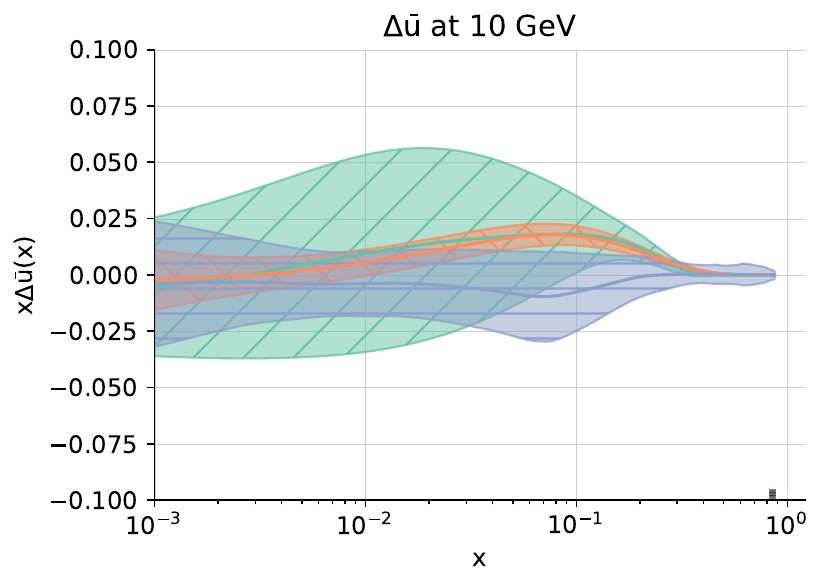}\\
  \includegraphics[width=0.49\textwidth]{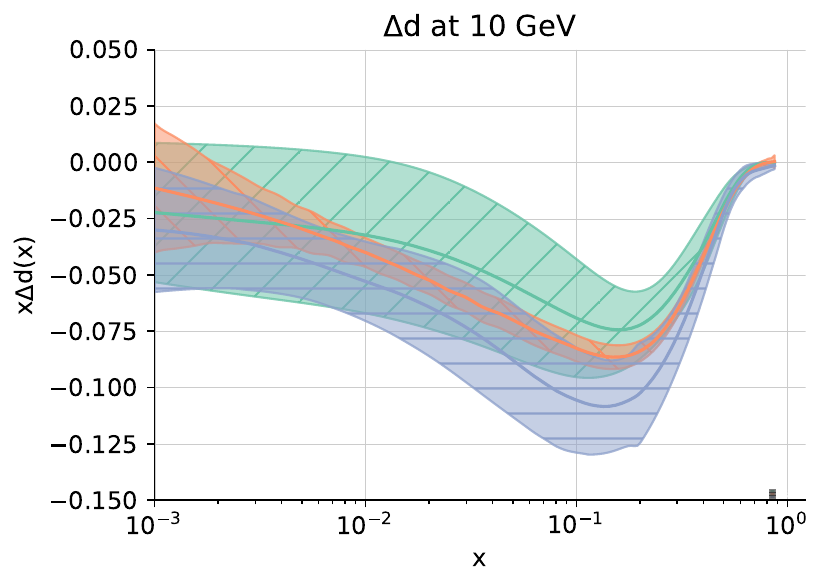}
  \includegraphics[width=0.49\textwidth]{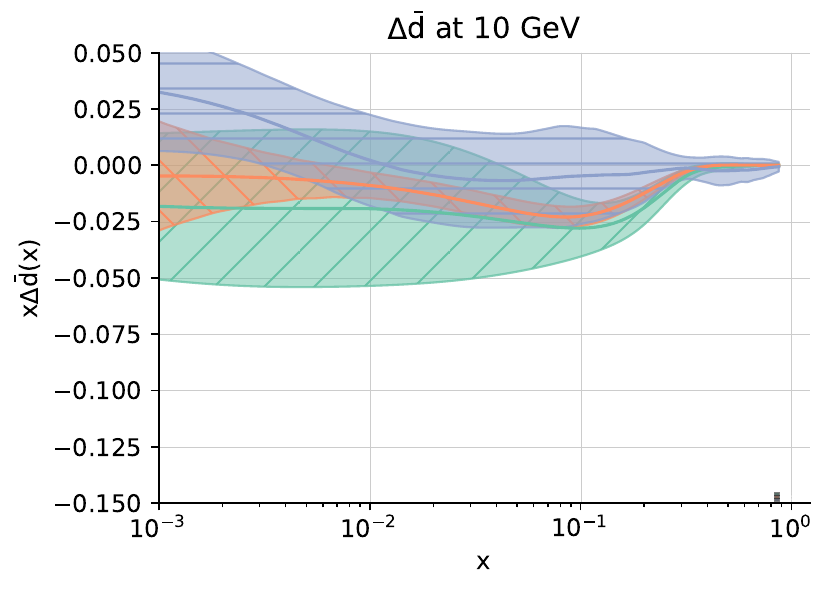}\\
  \includegraphics[width=0.49\textwidth]{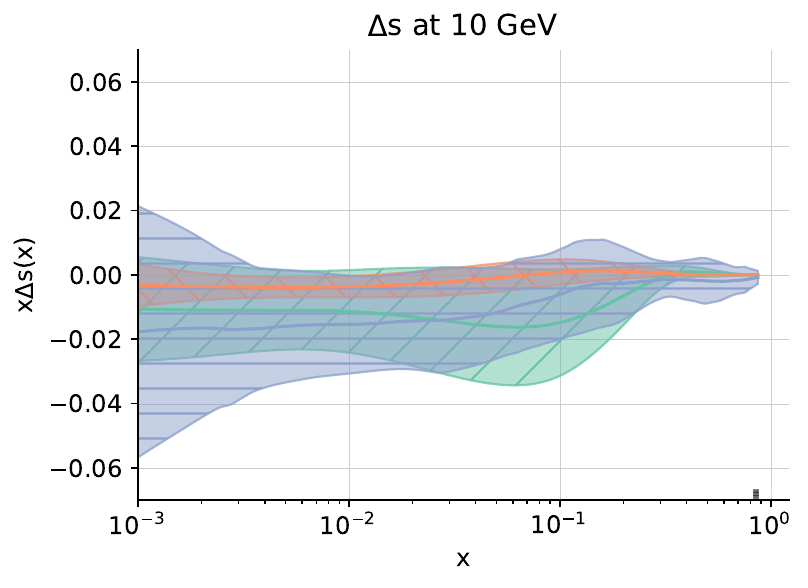}
  \includegraphics[width=0.49\textwidth]{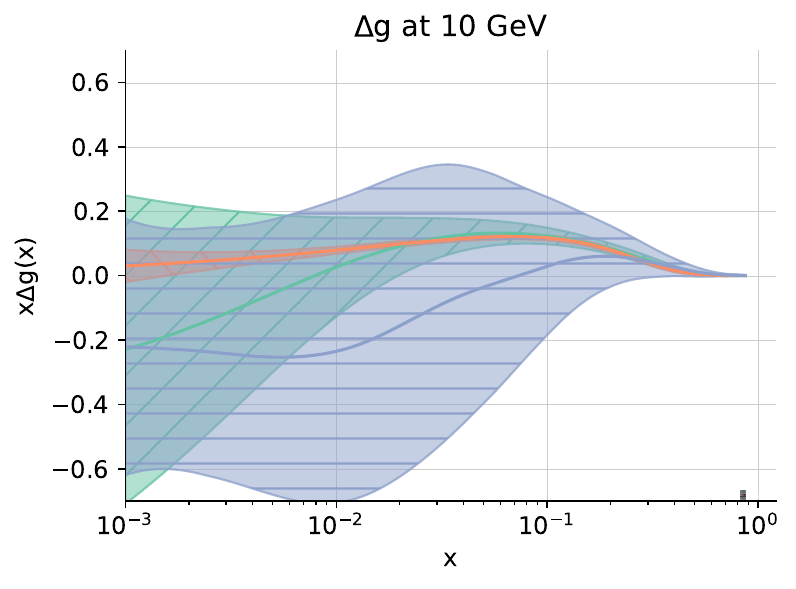}\\
  \caption{Comparison of the NNLO {\sc NNPDFpol2.0} (with MHOUs), BDSSV, and
    MAPPDFpol10 PDFs as a function of $x$ at $Q=10$~GeV, in the same format as
    Fig.~\ref{fig:NNPDFpol20_pert_conv}.}
  \label{fig:NNPDFpol20_vs_others}
\end{figure}

In Figs.~\ref{fig:NNPDFpol20_vs_NNPDFpol12} and~\ref{fig:NNPDFpol20_vs_others}
we compare the {\sc NNPDFpol2.0} polarised PDFs, respectively, to the
{\sc NNPDFpol1.1} PDFs~\cite{Nocera:2014gqa}, and to other recent polarised
PDFs, namely {\sc BDSSV24}~\cite{Borsa:2024mss} and
{\sc MAPPDFpol1.0}~\cite{Bertone:2024taw}. All these PDF sets are provided as
Monte Carlo replica ensembles: therefore, bands correspond to one-sigma
uncertainties computed over the corresponding nominal number of replicas.
When comparing to {\sc NNPDFpol1.1}, we consider the NLO MHOU version of
{\sc NNPDFpol2.0}, given that the highest accuracy of {\sc NNPDFpol1.1} is
NLO. When comparing to {\sc BDSSV24} and {\sc MAPPDFpol1.0}, we consider the
NNLO MHOU version of {\sc NNPDFpol2.0} and the NNLO version of the other sets.
The comparisons in Figs.~\ref{fig:NNPDFpol20_vs_NNPDFpol12}
and~\ref{fig:NNPDFpol20_vs_others} are otherwise in the same format as
Fig.~\ref{fig:NNPDFpol20_pert_conv}.

We first inspect the comparison of {\sc NNPDFpol2.0} to {\sc NNPDFpol1.1}.
We generally observe a fair agreement between the two. For up and down
polarised quarks and anti-quarks, differences between PDF central values
are the largest around $x\sim 0.1$, where they amount to up to slightly more than
$\sqrt{2}$-sigma. For the strange quark and gluon PDF central values, the two
sets are in perfect agreement. Uncertainties are likewise similar, except for
anti-quarks and the gluon PDFs. In the former case, they are smaller in
{\sc NNPDFpol1.1}; in the latter case, they are smaller in {\sc NNPDFpol2.0}
To understand this behaviour, it is useful to remind what the differences are
between the {\sc NNPDFpol1.1} and {\sc NNPDFpol2.0} determinations, both in
terms of experimental data and methodological details of each of them.

Concerning experimental data, as detailed in Sect.~\ref{subsec:data}, both
analyses are based on fixed-target inclusive DIS data, and on collider 
$W$-boson and single-inclusive jet production data. The {\sc NNPDFpol2.0} data
set is however updated and extended with respect to the {\sc NNPDFpol1.2} one.
In particular, more modern DIS measurements for COMPASS and JLab experiments,
the combined $W$-boson RHIC measurement, and a larger array of RHIC
single-inclusive and di-jet measurements have been included in
{\sc NNPDFpol2.0}. Among all of these, the RHIC data play a leading role.
The replacement of the $W$-boson measurement of Ref.~\cite{STAR:2014afm},
included in {\sc NNPDFpol1.1}, with the measurement of Ref.~\cite{STAR:2018fty},
now included in {\sc NNPDFpol2.0}, is responsible for the enhancement of the
anti-up quark and the depletion of the anti-down quark at large $x$. As can be
seen in Ref.~\cite{STAR:2018fty} (see in particular Fig.~5), the newer
measurement, which increases the luminosity of the older by about a factor of
three, is not only more precise, but it also has a preference for values of the
$W^-$ ($W^+$) cross section that are a little higher (lower) than those
predicted by {\sc NNPDFpol2.0}. This means that, roughly, the anti-up quark
should be enhanced and the anti-down quark should be suppressed, as we indeed
observe in Fig.~\ref{fig:NNPDFpol20_vs_NNPDFpol12}. This conclusion aligns
with what the authors of Ref.~\cite{STAR:2018fty} observe when reweighting
{\sc NNPDFpol1.1} with their newer measurement. At the same time, because
the newer DIS data do not alter the sum of quarks and anti-quarks with
respect to {\sc NNPDFpol1.1}, the {\sc NNPDFpol2.0} polarised up (down) PDF is
depleted (enhanced) by an amount equal to that by which the polarised anti-up
(anti-down) PDF is enhanced (depleted). On the other hand, the significant
amount of newer RHIC single-inclusive jet and di-jet production data are
responsible for the reduction of the polarised gluon PDF uncertainty. We
will further comment on the relative impact of single-inclusive jet and
di-jet measurements in Sect.~\ref{subsec:PDF_dependence}.

Concerning methodological details, {\sc NNPDFpol1.1} was obtained by
reweighting~\cite{Ball:2011gg} a certain prior PDF set with collider $W$-boson
and single-inclusive jet production data. The prior was constructed by
performing a fit to inclusive DIS data only, and by enforcing the separation of
up and down quark and anti-quark PDFs --- to which inclusive DIS measurements
are insensitive by construction --- to the result of the {\sc DSSV}
fit~\cite{deFlorian:2008mr,deFlorian:2009vb}. Up and down quark and
anti-quark separation was achieved by fitting SIDIS data in {\sc DSSV}.
Different prior PDF sets were considered, by taking once, twice, three times or
four times the nominal {\sc DSSV} one-sigma uncertainty. It was checked that
the posterior became insensitive to the prior for a sufficiently large
inflation of the {\sc DSSV} uncertainties (three-sigma or more). This
methodology can bias PDF uncertainties, if anything because the available data
cannot constrain the PDFs on the whole range of $x$. For instance, for
$x\lesssim 0.04$ the uncertainties on the polarised anti-up and anti-down
{\sc NNPDFpol1.1} PDFs are completely determined by the assumption
made in the prior. Conversely, the {\sc NNPDFpol2.0} PDFs are determined from a
fit to all the data: because of the very flexible parametrisation, any bias in
the prior is reduced in comparison to {\sc NNPDFpol1.1}. In light of these
considerations, we ascribe the smaller uncertainties of {\sc NNPDFpol1.1}
in comparison to {\sc NNPDFpol2.0} seen for up and down anti-quarks to a bias
in the former. In this respect, {\sc NNPDFpol2.0} is less precise but more
accurate than {\sc NNPDFpol1.1}.

We now turn to compare the NNLO {\sc NNPDFpol2.0} PDFs to the NNLO
{\sc BDSSV} and {\sc MAPPDFpol1.0} PDFs. We generally observe a fair agreement
among the three sets. The shape of the PDFs is generally very similar between
{\sc NNPDFpol2.0} and {\sc BDSSV24}, despite the significantly smaller
uncertainties of the latter. Particularly impressive is the gluon PDF, for
which, in the region constrained by experimental data, $x\gtrsim 0.02$, the
two determination are one spot on top of the other. Somewhat larger
differences, though well within one-sigma uncertainties, are observed
between {\sc NNPDFpol2.0} and {\sc MAPPDFpol1.0}. These differences regard,
in particular, the up and down quark/anti-quark PDFs and the gluon PDF:
the polarised anti-up (anti-down) PDF is positive (negative) in
{\sc NNPDFpol2.0}, whereas it is negative (positive) in {\sc MAPPDFpol1.0};
the polarised gluon PDF is significantly more positive in {\sc NNPDFpol2.0}
than in {\sc MAPPDFpol1.0}. The size of the uncertainties is very similar in
the two PDF sets, except for the gluon PF at large $x$, where it is
significantly smaller in {\sc NNDPFpol2.0}. In order to understand this
behaviour we should again consider the differences due to the data set and
the methodological details of each determination.

Concerning the data set, all the three analyses are based on very similar sets
of inclusive DIS data; the {\sc BDSSV24} and {\sc MAPPDFpol1.0} set also
include semi-inclusive DIS data, from the COMPASS and HERMES experiments;
the {\sc NNPDFpol2.0} and {\sc BDSSV24} sets also include polarised collider
data: $W$-boson and single-inclusive jet production measurements are included
in both, whereas di-jet production measurements are included only in
{\sc NNPDFpol2.0}. These differences explain differences of PDF central values.
Quark/anti-quark separation is driven only by SIDIS data in {\sc MAPPDFpol1.0},
only by $W$-boson production data in {\sc NNPDFpol2.0}, and by an admixture of
the two processes in {\sc BDSSV24}. As we can see from
Fig.~\ref{fig:NNPDFpol20_vs_others}, the {\sc BDSSV24} central values of the
polarised up, anti-up, down, and anti-down PDFs lie in between the
{\sc NNPDFpol2.0} and {\sc MAPPDFpol1.0} central values. This suggests that
SIDIS and $W$-boson production measurements tend to pull polarised quark and
anti-quark PDFs is slightly opposite directions: specifically, SIDIS
measurements tend to suppress (enhance) the polarised anti-up (anti-down) PDF
around $x\sim 0.1$, whereas $W$-boson production measurements tend to do the
opposite. Because the sum of quark and anti-quark PDFs is well constrained by
inclusive DIS data, the effect is that SIDIS measurements enhance (suppress)
the polarised up (down) PDF by a similar amount, whereas $W$-boson production
measurements do the opposite. On the other hand, the polarised gluon PDF is
constrained by higher-order (beyond LO) contributions to the matrix elements of
(semi-)inclusive DIS data and to splitting functions in all the three
analyses; in {\sc NNPDFpol2.0} and in {\sc BDSSV24}, it is further constrained
by LO contributions to the matrix elements of single-inclusive jet (and di-jet)
production measurements. This constraint is clearly dominant with respect to
the one coming from the other data sets, as expected. Be that as it may, all
these differences are well encompassed by PDF uncertainties. This fact suggests
that there are no major tensions across data sets, and that a fit would
possibly be able to accommodate all the data sets with ease, should they all be
included in a fit at the same time.

Concerning the methodological details, all the three analyses are based on a
Monte Carlo PDFs, whereby the data uncertainties are represented into PDF
uncertainties by means of importance sampling. They do however make use of a
different parametrisation of PDFs: whereas a neural network (with a different
architecture) is used by {\sc NNPDFpol2.0} and {\sc MAPPDFpol1.0}, a simpler
polynomial form is used by {\sc BDSSV24}. All the three analyses enforce
positivity of cross sections as in Eq.~\eqref{eq:positivity_pdfs}, and require
that the triplet and octet sum rules in Eq.~\eqref{eq:decay_constants} are
fulfilled. The most relevant methodological difference across methodologies is
therefore the parametrisation, which may be responsible for the observed
differences in PDF uncertainties. These are indeed rather similar between
{\sc NNPDFpol2.0} and {\sc MAPPDFpol2.0}, which use a neural network, while
they are significantly smaller for {\sc BDSSV24}, which use a polynomial.
Conclusive evidence of this fact can be only reached through a benchmark study,
which, despite becoming compelling, will require dedicated future work.

\subsection{Dependence on the positivity constraint and on the data set}
\label{subsec:PDF_dependence}

We now turn to study the stability of the {\sc NNPDFpol2.0} parton set upon
variations of the positivity constraint and of the input data set. Among all the theoretical, methodological, and experimental choices that enter this
analysis, as detailed in Sects.~\ref{sec:data-theory}-\ref{sec:methodology}, we
consider that these two have the largest impact on the polarised PDFs.

\begin{table}[!t]
  \renewcommand{\arraystretch}{1.60}
  \scriptsize
  \centering
  \begin{tabularx}{\textwidth}{Xrccccc}
  \toprule
  Data set
  & $N_{\rm dat}$
  & baseline
  & no pos.
  & no di-jets
  & no s.-i- jets
  & no JLab \\
  \midrule
  DIS NC
  & 704 & 0.93 & 0.91 & 0.92 & 0.91 & 0.88 \\
  DY CC 
  &  12 & 1.17 & 0.98 & 1.05 & 0.98 & 0.66 \\ 
  JETS
  &  97 & 1.03 & 1.00 & 1.02 &  --- & 1.01 \\
  DIJET
  & 138 & 1.07 & 1.07 &  --- & 1.07 & 1.08 \\
  \midrule
  Total
  & 951 & 0.95 & 0.93 & 0.93 & 0.94 & 0.94 \\
  \bottomrule
\end{tabularx}

  \vspace{0.3cm}
  \caption{The number of data points $N_{\rm dat}$ and the $\chi^2$ per data
    point corresponding to the baseline {\sc NNPDFpol2.0} NNLO MHOU fit and to
    variants obtained without the positivity constraint (no pos.), without
    di-jet measurements (no di-jet), without single-inclusive jet measurements
    (no s.-i. jets), and without JLab measurements (no JLab). Numerical values
    are displayed for data sets aggregated according to the process
    categorisation introduced in Sect.~\ref{subsec:th_covmat}, and for the
    total data set.}
  \label{tab:chi2_stability}
\end{table}

\begin{figure}[!t]
  \centering
  \includegraphics[width=0.49\textwidth]{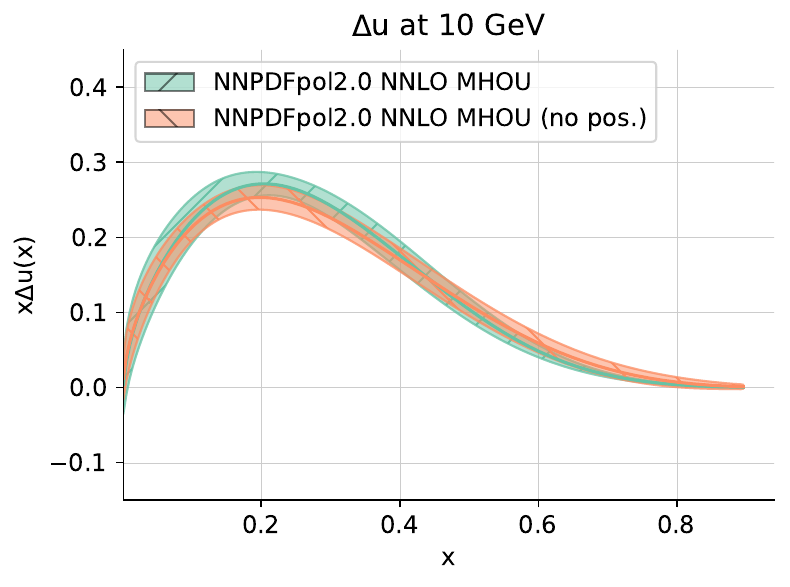}
  \includegraphics[width=0.49\textwidth]{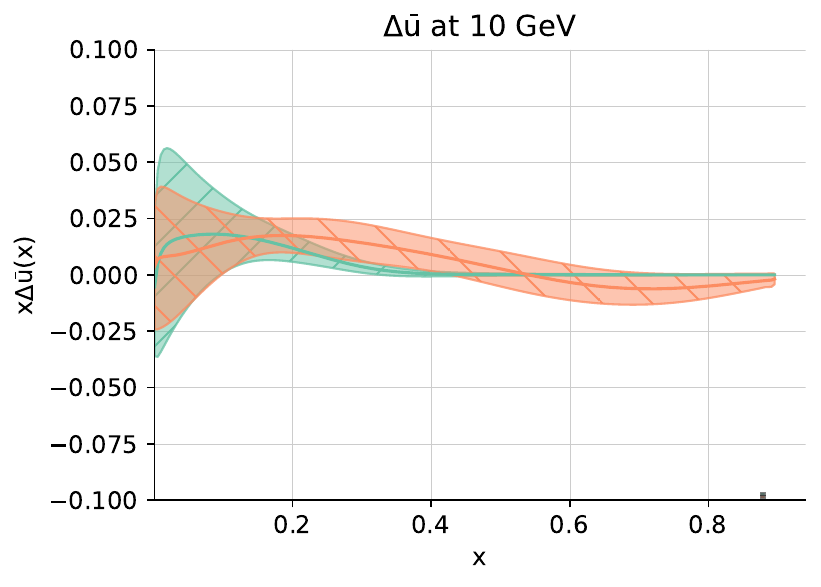}\\
  \includegraphics[width=0.49\textwidth]{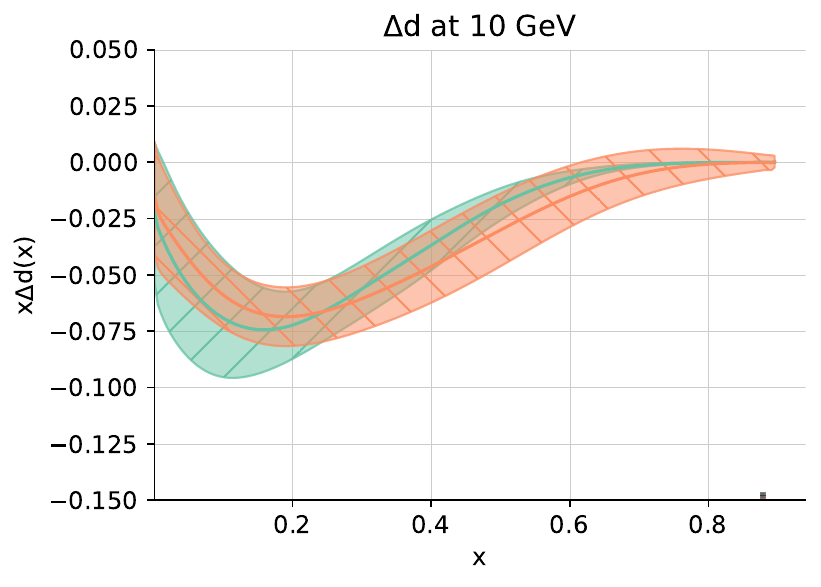}
  \includegraphics[width=0.49\textwidth]{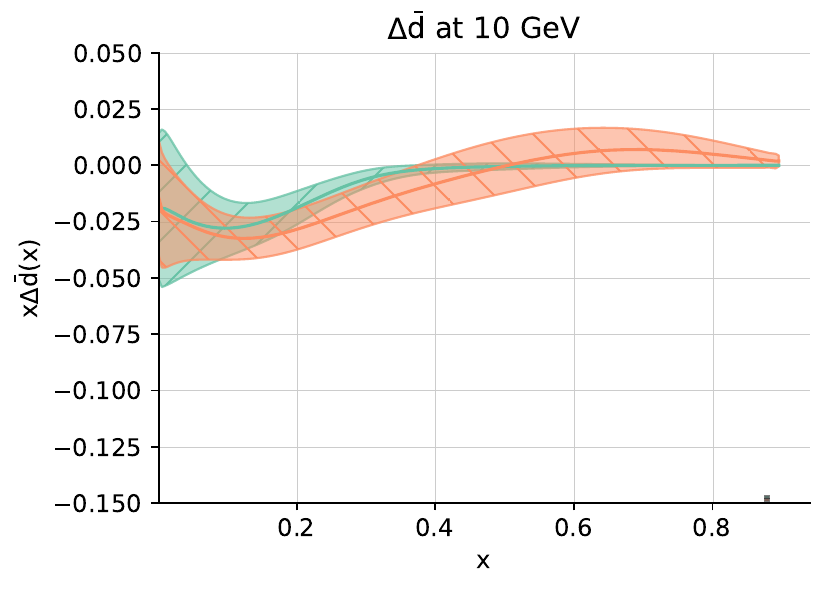}\\
  \includegraphics[width=0.49\textwidth]{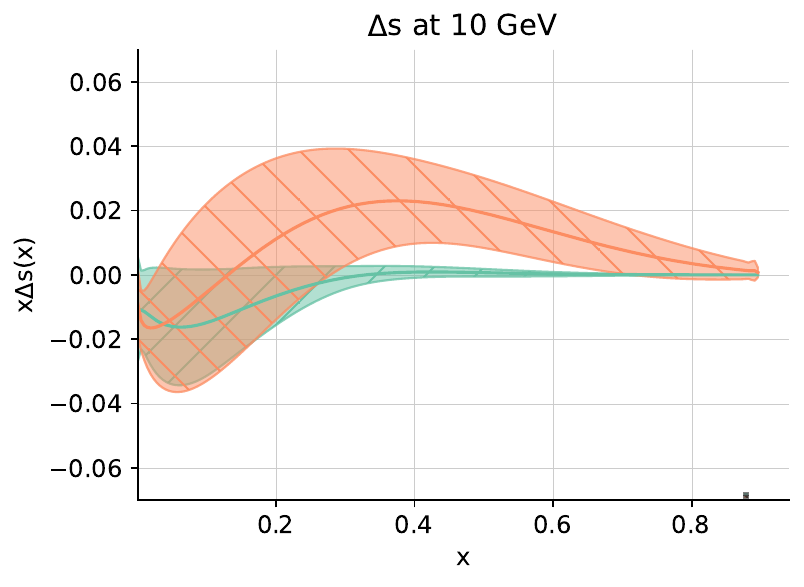}
  \includegraphics[width=0.49\textwidth]{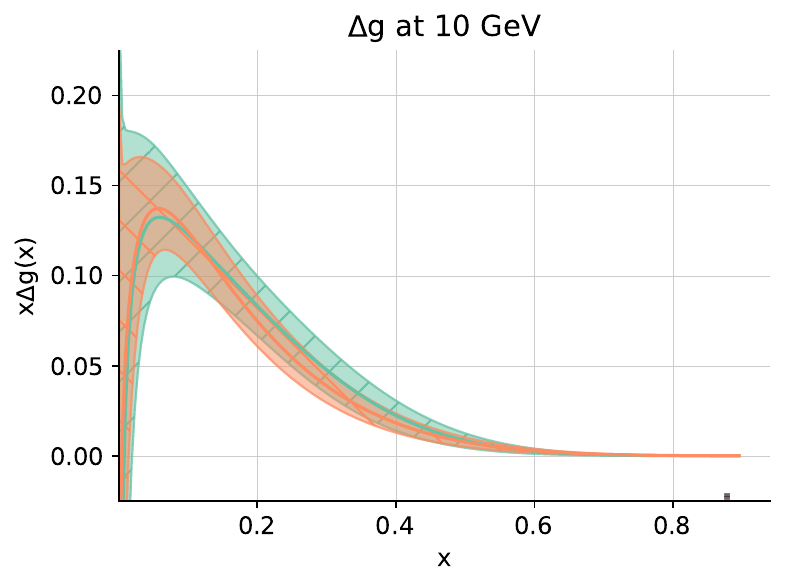}\\
  \caption{Comparison of the baseline {\sc NNPDFpol2.0} NNLO MHOU polarised
    PDFs and PDFs obtained from an equivalent determination without the
    positivity constraint, Eq.~\eqref{eq:positivity_pdfs}. Parton distributions
    are reported as a function of $x$ in linear scale at $Q=10$~GeV, with
    one-sigma uncertainty bands.}
  \label{fig:positivity}
\end{figure}

Concerning the positivity constraint, we recall that we incorporate
Eq.~\eqref{eq:positivity_pdfs} in Eq.~\ref{eq:loss_function} by means of the
penalty term defined in
Eqs.~\eqref{eq:positivity-constraints}-\eqref{eq:c-function}. This positivity
constraint has been deemed essential to stabilise the {\sc NNPDFpol1.0}
fit~\cite{Ball:2013lla}, in particular to drive the behaviour of polarised PDFs
at large values of $x$, due to the scarcity of experimental data in that region.
The phenomenological effect of this positivity constraint and its interplay
with various experimental data have been recently investigated
in~\cite{Zhou:2022wzm,Karpie:2023nyg,Hunt-Smith:2024khs}, in particular
relationship with the sign of the polarised gluon PDF. There it was found that,
if no positivity constraints are imposed in the fit, it is possible to
attain a positive polarised gluon PDF only if one simultaneously accounts for
RHIC measurements of single-inclusive jet production and JLab measurements of
DIS production at very large values of $x$. On the other hand, the validity of
the LO approximation behind Eq.~\eqref{eq:positivity_pdfs} was recently
re-assessed in~\cite{deFlorian:2024utd}. The authors confirmed the results
of~\cite{Altarelli:1998gn}, namely that violations of
Eq.~\eqref{eq:positivity_pdfs} are possible, though they are of the order of
percent at most.

In order to check the impact of the positivity constraint on {\sc NNPDFpol2.0},
we have performed a fit in which we set the maximum value of the Lagrange
multiplier $\Lambda_{\rm pos}$ in Eq.~\eqref{eq:loss_function} to zero. The fit
is otherwise equivalent to the {\sc NNPDFpol2.0} NNLO MHOU baseline
determination. This effectively amounts to removing the positivity constraint
from the fit. The corresponding fit quality, compared to the baseline, is
reported, for data sets aggregated according to the process categorisation
introduced in Sect.~\ref{subsec:th_covmat}, and for the total data set, in
Table~\ref{tab:chi2_stability}. The corresponding PDFs, again compared to the
baseline PDFs, are reported, as a function of $x$ in linear scale at
$Q=10$~GeV, in Fig.~\ref{fig:positivity}. Bands correspond to one-sigma
uncertainties.

As we see from Table~\ref{tab:chi2_stability}, the effect of removing the
positivity constraint is almost immaterial on the fit quality. The improvement
of the value of the $\chi^2$, for groups of data sets and for the total data
set, is less than half a sigma, in units of the standard deviation of the
$\chi^2$ distribution, therefore it is not statistically significant. The effect
of removing the positivity constraint is more noticeable on PDFs, for which we
observe a marked increase of PDF uncertainties, especially at large values of
$x$, see Fig.~\ref{fig:positivity}. Differences are apparent for the sea quark
PDFs, in particular for the polarised strange PDF, which develops a preference
for positive values. We have explicitly checked that this behaviour violates
the more general positivity constraint that relates longitudinal single- and
double-spin asymmetries for gauge-boson production in polarised proton-proton
collisions, see {\it e.g.}~Sect.~4 in~\cite{Kang:2011qz} and Sect.~4.1 
in~\cite{Nocera:2014gqa}. The violation occurs in particular for the production
of a $W^-$ boson, at forward values of its rapidity, $y_W\gtrsim 1.2$, which,
at the RHIC centre-of-mass energy $\sqrt{s}=510$~GeV, correspond to large
values of the proton momentum fraction, $x\gtrsim 0.5$. On the other hand, the
variation of the gluon PDF remains rather limited, and amounts to a moderate
increase of the PDF uncertainty. This fact suggests that the data included in
the fit constrain the gluon PDF very well, as we discuss next.

\begin{figure}[!t]
  \centering
  \includegraphics[width=0.49\textwidth]{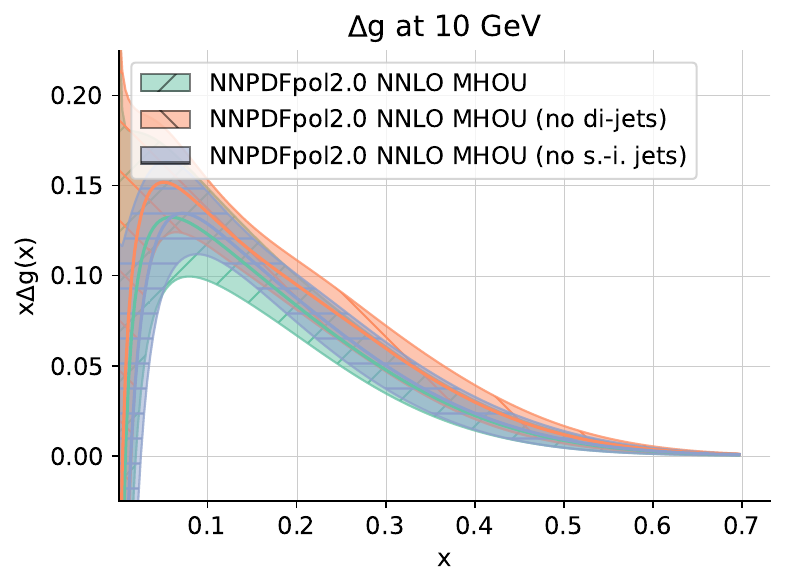}
  \includegraphics[width=0.49\textwidth]{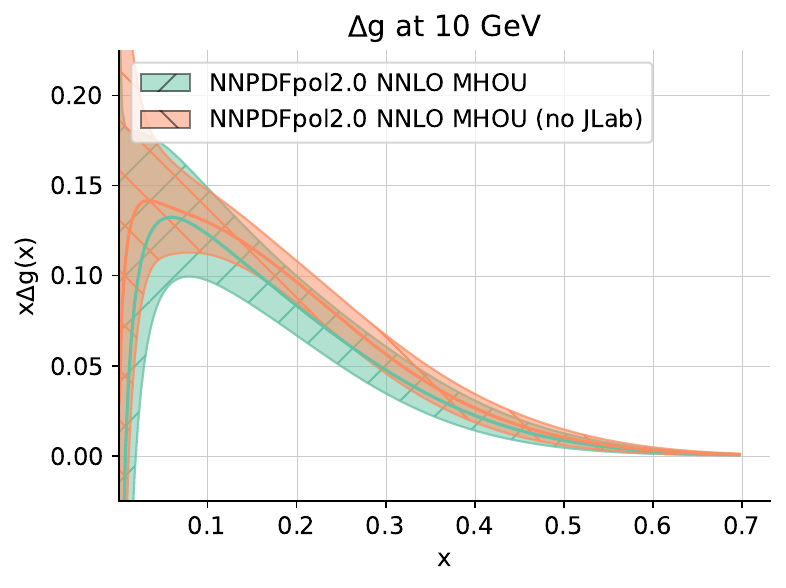}  \\
  \caption{Comparison of the baseline {\sc NNPDFpol2.0} NNLO MHOU polarised
    gluon PDF and the corresponding PDF obtained from equivalent determinations
    without the RHIC di-jet and single-inclusive jet measurements (left) or
    without the JLab DIS measurements (right). The PDF is displayed as a
    function of $x$ in linear scale at $Q=10$~GeV, with one-sigma uncertainty
    bands.}
  \label{fig:data_impact}
\end{figure}

Concerning the data set, we observe that variations are somewhat limited by the
overall composition of the global data set, which does not allow us to remove
measurements without affecting our ability to determine PDFs in a sufficiently
sensible way. We therefore focus on three classes of measurements, which are
collectively sensitive to the polarised gluon PDF at large values of $x$.
These are the spin asymmetries measured by RHIC in di-jet and single-inclusive
jet production and the structure functions measured by JLab in polarised DIS.

In order to check the impact of each of these classes of measurements, we
have performed three additional fits, equivalent to the {\sc NNPDFpol2.0} NNLO
MHOU baseline determination, from which we have removed, respectively, the
di-jet, the single-inclusive jet, and the JLab DIS measurements in turn.
The corresponding fit quality, compared to the baseline, is reported, for
data sets aggregated according to the process categorisation introduced
in Sect.~\ref{subsec:th_covmat}, and for the total data set, in
Table~\ref{tab:chi2_stability}. The corresponding gluon PDF, again compared
to the baseline PDFs, are reported, as a function of $x$ in linear scale at
$Q=10$~GeV, in Fig.~\ref{fig:data_impact}. Bands correspond to one-sigma
uncertainties.

As we see from Table~\ref{tab:chi2_stability}, the effect of removing either
the di-jet or the single-inclusive jet measurements from the fit is immaterial:
changes in fit quality are compatible with statistical fluctuations. The effect
of removing JLab data is small. It leads to an improvement in the overall
description of the DIS and DY data by about one sigma, in units of the standard
deviation of the $\chi^2$ distribution, which is however not noticeable on the
global $\chi^2$. The effect of varying the data set is slightly more visible
on the large-$x$ gluon PDF, see Fig.~\ref{fig:data_impact}. In particular, we
observe that single-inclusive jet measurements have a preference for a slightly
larger gluon PDF in comparison to di-jet measurements. The latter actually
drive the gluon PDF in the global fit, given that this is equivalent to that
obtained in the fit without single-inclusive jet measurements. On the other hand,
the JLab DIS measurements have a preference for a slightly smaller gluon in
comparison to the global data set. All of these affects are however minor, as
they are well encompassed by PDF uncertainties.

\section{Phenomenological applications}
\label{sec:pheno}

In this section, we illustrate two phenomenological applications of the
{\sc NNPDFpol2.0} parton sets. First, we revisit the spin content of the
proton, by computing the lowest moments of relevant polarised PDF combinations.
Second, we compare predictions of single-hadron production in polarised DIS
and in proton-proton collisions with recent experimental measurements.

\subsection{The spin content of the proton revisited}
\label{subsec:proton_spin}

The lowest moments of the gluon and singlet polarised PDFs are proportional to
the proton axial currents, which express the fraction of the proton spin
carried by gluons and quarks. If one defines these moments as
\begin{equation}
  \langle
  \Delta f(Q)
  \rangle ^{[x_{\rm min},x_{\rm max}]}
  =
  \int_{x_{\rm min}}^{x_{\rm max}}\hspace{-17pt} dx\, \Delta f(x,Q)\,,
  \label{eq:truncated_moments}
\end{equation}
the Jaffe-Manohar sum rule~\cite{Jaffe:1989jz} reads, in natural units, as
\begin{equation}
  \frac{1}{2}
  =
  \langle\Delta g(Q) \rangle^{[0,1]}
  +
  \frac{1}{2}\langle\Delta\Sigma (Q)\rangle^{[0,1]}
  +
  \mathcal{L}_g(Q)
  +
  \mathcal{L}_q(Q)\,,
  \label{eq:proton_spindecomposition}
\end{equation}
where $\mathcal{L}_g(Q)$ and $\mathcal{L}_q(Q)$ are the canonical gluon and
quark orbital angular momenta. We therefore seek to compute the first two terms
of \cref{eq:proton_spindecomposition}. However, because the available
piece of experimental information does not allow for an accurate determination
of polarised PDFs over the whole range of $x$~\cite{Adamiak:2025mdy}, we rather consider truncated
quantities, for which $x_{\rm min}>0$ and $x_{\rm max}<1$.

\begin{figure}[!t]
  \centering
  \includegraphics[width=0.7\textwidth]{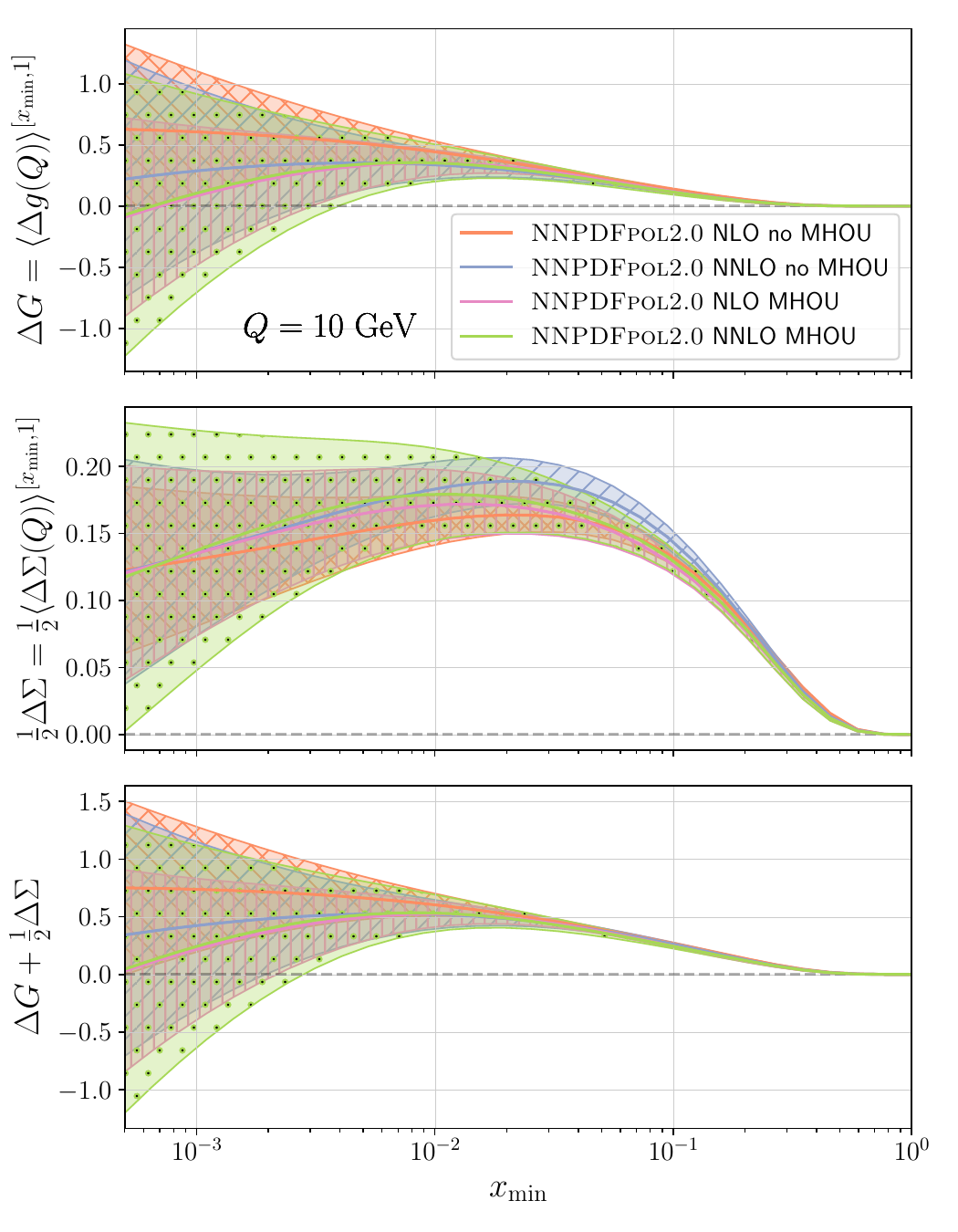}\\
  \caption{The truncated moments, \cref{eq:truncated_moments}, of the
    gluon (top) and singlet (middle) PDFs, together with the sum of the former
    and of half of the latter (bottom). The moments are computed as a function
    of $x_{\rm min}$, with $5\cdot 10^{-4}<x_{\rm min}<1$, with fixed $x_{\rm max}=1$,
    and at $Q=10$~GeV. The NLO and NNLO {\sc NNPDFpol2.0} PDF sets, without
    and with MHOUs are used. Bands correspond to one-sigma PDF uncertainties.}
  \label{fig:spin_frac}
\end{figure}

In \cref{fig:spin_frac} we display the truncated moments,
\cref{eq:truncated_moments}, of the gluon and quark PDFs, together with
the sum of the former and of half of the latter. The moments are computed as a
function of $x_{\rm min}$, with $5\cdot 10^{-4}<x_{\rm min}<1$, with fixed
$x_{\rm max}=1$, and at $Q=10$~GeV. The {\sc NNPDFpol2.0} NLO and NNLO PDF sets,
without and with MHOUs are used. Bands correspond to one-sigma PDF
uncertainties. In \cref{tab:spin_frac} we report the values of the same
quantities computed with the {\sc NNPDFpol1.1} NLO (NNpol1.1), {\sc NNPDFpol2.0}
NNLO MHOU (NNpol2.0), and {\sc BDSSV24} NNLO (BDSSV) PDF sets. Values are
determined at $Q=10$~GeV, in three integration intervals
$[x_{\rm min}, x_{\rm max}]$: [0.001, 1], [0.02, 0.5], and [0.05; 1]. These roughly
correspond to the kinematic ranges covered by the entire data set, by the
single-inclusive jet and di-jet data, and by the large-$x$ DIS data. 

\begin{table}[!t]
  \renewcommand{\arraystretch}{1.60}
  \scriptsize
  \centering
  \begin{tabularx}{\textwidth}{Xccccccccc}
  \toprule
  & \multicolumn{3}{c}{$\langle\Delta g(Q) \rangle^{[\cdot,\cdot]}$}
  & \multicolumn{3}{c}{$\langle\Delta\Sigma(Q)\rangle^{[\cdot,\cdot]}$}
  & \multicolumn{3}{c}{$\langle\Delta g(Q) \rangle^{[\cdot,\cdot]} + 1/2\langle\Delta\Sigma (Q)\rangle^{[\cdot,\cdot]}$}
  \\
  & $[0.001, 1]$ & $[0.02,0.5]$ & $[0.05,1]$
  & $[0.001, 1]$ & $[0.02,0.5]$ & $[0.05,1]$
  & $[0.001, 1]$ & $[0.02,0.5]$ & $[0.05,1]$\\
  \midrule
  NNpol1.1
  & $0.53\pm 0.77$ & $0.32\pm 0.10$ & $0.21\pm 0.04$
  & $0.13\pm 0.07$ & $0.16\pm 0.02$ & $0.16\pm 0.01$
  & $0.66\pm 0.83$ & $0.48\pm 0.10$ & $0.37\pm 0.04$ \\
  NNpol2.0
  & $0.11\pm 0.81$ & $0.31\pm 0.08$ & $0.20\pm 0.04$
  & $0.14\pm 0.09$ & $0.17\pm 0.03$ & $0.16\pm 0.02$
  & $0.24\pm 0.88$ & $0.48\pm 0.08$ & $0.36\pm 0.04$ \\
  BDSSV
  & $0.47\pm 0.07$ & $0.29\pm 0.02$ & $0.19\pm 0.01$
  & $0.18\pm 0.01$ & $0.19\pm 0.01$ & $0.19\pm 0.01$
  & $0.65\pm 0.08$ & $0.48\pm 0.02$ & $0.38\pm 0.01$\\
  \bottomrule
\end{tabularx}

  \vspace{0.3cm}
  \caption{The truncated moments, \cref{eq:truncated_moments}, of the
      gluon, singlet, and the sum of the former and of half of the latter,
      computed with the NNPDFpol1.1 NLO (NNpol1.1), NNPDFpol2.0 NNLO MHOU
      (NNpol2.0), and BDSSV24 NNLO (BDSSV) PDF sets, at $Q=10$~GeV in three
      integration intervals $[x_{\rm min}, x_{\rm max}]$: [0.001, 1], [0.02, 0.5],
      and [0.05, 1]. Errors correspond to one-sigma PDF uncertainties.}
  \label{tab:spin_frac}
\end{table}

From inspection of \cref{fig:spin_frac} and \cref{tab:spin_frac}, we
make the following observations on the determination of the truncated
moments of the gluon and singlet PDFs. These observations also apply to the
residual proton spin fraction not carried by the spin of gluons and quarks.
First, the accuracy of the determination depends very weakly on the
perturbative accuracy of the PDF fit. This result is consistent with the
conclusions on fit quality and perturbative stability reached in
\cref{subsec:pert_stability}. Second, the precision of the determination
strongly depends on the contribution coming from the small-$x$ integration
region, which remains completely unknown. Because
of this fact, the spin of gluons and quarks could account for the total proton
spin budget. Third, in the region covered by experimental data, the three
determinations obtained using different PDF sets are in remarkable agreement
with each other. Moving from {\sc NNPDFpol1.1} to {\sc NNPDFpol2.0}, we
observe a slight reduction of the uncertainty on the truncated moment of the
gluon PDF in the integration region [0.02,0.5] covered by single-inclusive jet
and di-jet measurements; on the other hand, we observe a slight increase of the
uncertainty on the truncated moment of both the gluon and singlet PDFs in the
widest integration region [0.001, 1]. As discussed in
\cref{subsec:comparison}, the first effect is due to the largest array of
RHIC measurements included in {\sc NNPDFpol2.0}, whereas the second effect is
due to an extrapolation bias in {\sc NNPDFpol1.1} associated to the choice of
Bayesian prior. Moving from {\sc NNPDFpol2.0} to {\sc BDSSV24}, we observe
larger uncertainties in the results obtained with the former PDF set,
especially when extending the integration interval to the small-$x$ region,
again consistently with what was observed in \cref{subsec:comparison}.

Finally, we compute the truncated moments corresponding to the charge-even PDF
combinations $\Delta u^+=\Delta u+\Delta\bar u$,
$\Delta d^+=\Delta d+\Delta\bar d$, and $\Delta s^+=\Delta s+\Delta\bar s$,
and to the PDF combination
$x(\Delta u^- - \Delta d^-)=x(\Delta u-\Delta\bar u-\Delta d + \Delta\bar d)$.
These combinations are routinely determined using lattice
QCD~\cite{Lin:2017snn,Constantinou:2020hdm}, therefore these results can serve
as a benchmark for those studies. We use the {\sc NNPDFpol2.0} NNLO MHOU PDF
set, an integration interval [0.001,1], and two energy scales, $Q=2$~GeV and
$Q=10$~GeV, respectively. Our results are collected in \cref{tab:moments}.

\begin{table}[!t]
  \renewcommand{\arraystretch}{1.60}
  \scriptsize
  \centering
  \begin{tabularx}{\textwidth}{Xcccccccc}
  \toprule
  & \multicolumn{2}{c}{$\langle \Delta u^+(Q)\rangle^{[0.001,1]}$} 
  & \multicolumn{2}{c}{$\langle \Delta d^+(Q)\rangle^{[0.001,1]}$} 
  & \multicolumn{2}{c}{$\langle \Delta s^+(Q)\rangle^{[0.001,1]}$} 
  & \multicolumn{2}{c}{$\langle x(\Delta u^- - \Delta d^-)(Q)\rangle^{[0.001,1]}$}
  \\
  & $Q=2$~GeV & $Q=10$~GeV 
  & $Q=2$~GeV & $Q=10$~GeV 
  & $Q=2$~GeV & $Q=10$~GeV 
  & $Q=2$~GeV & $Q=10$~GeV \\
  \midrule
  NNpol2.0
  & $+0.81\pm 0.07$ & $+0.79\pm 0.08$ 
  & $-0.40\pm 0.05$ & $-0.39\pm 0.06$ 
  & $-0.13\pm 0.13$ & $-0.13\pm 0.13$ 
  & $+0.17\pm 0.02$ & $+0.13\pm 0.01$\\
  \bottomrule
\end{tabularx}

  \vspace{0.3cm}
  \caption{The truncated moments corresponding to the charge-even PDF
    combinations $\Delta u^+=\Delta u+\Delta\bar u$,
    $\Delta d^+=\Delta d+\Delta\bar d$, and $\Delta s^+=\Delta s+\Delta\bar s$,
    and to the PDF combination
    $x(\Delta u^- - \Delta d^-)=x(\Delta u-\Delta\bar u-\Delta d
    + \Delta\bar d)$. The moments are computed in the integration interval
    [0.001,1], at $Q=2$~GeV and $Q=10$~GeV, with the {\sc NNPDFpol2.0} NNLO
    MHOU PDF set (NNpol2.0).}
  \label{tab:moments}
\end{table}

\subsection{Single-inclusive particle production}
\label{subsec:predictions}

As an illustration of the predictive power of {\sc NNPDFpol2.0}, we now 
present predictions for the longitudinal double-spin asymmetries $A_{\rm LL}^h$
of single-inclusive particle production in polarised DIS and in proton-proton
collisions. These asymmetries are defined, respectively, in Eq.~(2)
of~\cite{HERMES:2018awh} and in Eq.~(17) of~\cite{Nocera:2014gqa} as ratios of
polarised to unpolarised cross sections with the same final-state particle.
In the case of semi-inclusive DIS, we further approximate the longitudinal
double-spin asymmetry with the virtual-photon-nucleon asymmetry $A_1^h$
(see Eq.~(6) in~\cite{HERMES:2018awh}), which is the ratio between the
polarised and unpolarised SIDIS structure functions.

We compute predictions for these processes at NLO accuracy in perturbative QCD.
For single-hadron production in DIS, NNLO corrections to the matrix elements are
known~\cite{Bonino:2024wgg,Goyal:2024tmo,Goyal:2024emo}, whereas they are not
in proton-proton collisions. Because our aim is solely to demonstrate an
application of {\sc NNPDFpol2.0}, and not to perform a detailed
phenomenological study, we stick to NLO matrix elements in both
cases. Theoretical predictions require knowledge of FFs, in particular for
separate quark and anti-quark flavours and for the gluon. Such knowledge is
lacking in available NNFF sets~\cite{Bertone:2017tyb,Bertone:2018ecm} because
of the limited amount of measurements included in the corresponding
determinations. We therefore select FFs from the {\sc MAPFF1.0}
set~\cite{Khalek:2021gxf,AbdulKhalek:2022laj}. These were determined from a
comprehensive analysis of single-hadron production measurements in
electron-positron annihilation and in DIS with a methodology similar to that
used for {\sc NNPDFpol2.0}. We use the {\sc NNPDF4.0} set for the required
initial-state unpolarised proton PDFs. In the case of single-hadron
production in polarised DIS, we use a piece of software that we have developed
for the purpose of this study, whereas we use the code of~\cite{Jager:2002xm}
in the case of single-hadron production in polarised proton-proton collisions.

\begin{figure}[!t]
  \centering
  \includegraphics[width=0.49\textwidth]{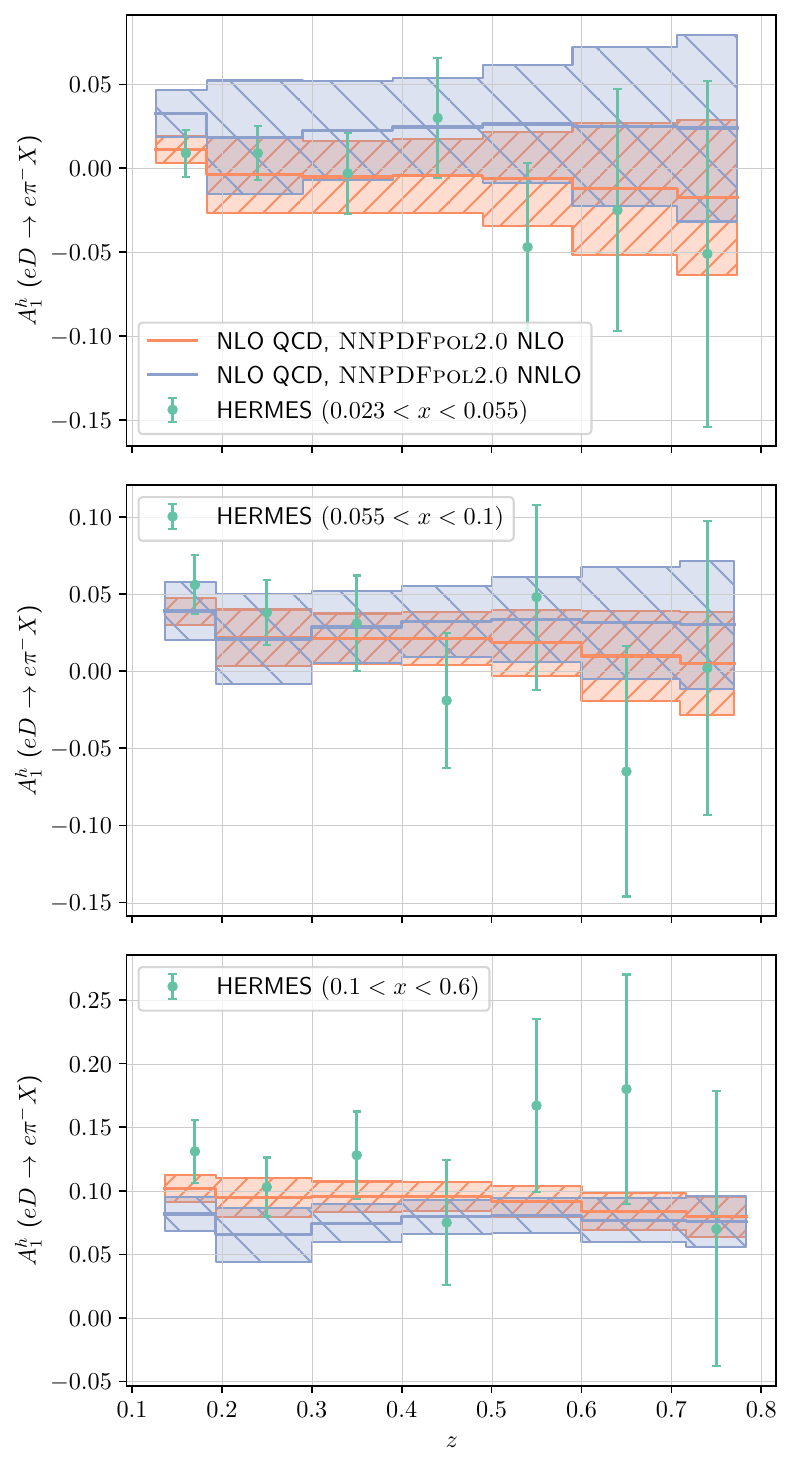}
  \includegraphics[width=0.49\textwidth]{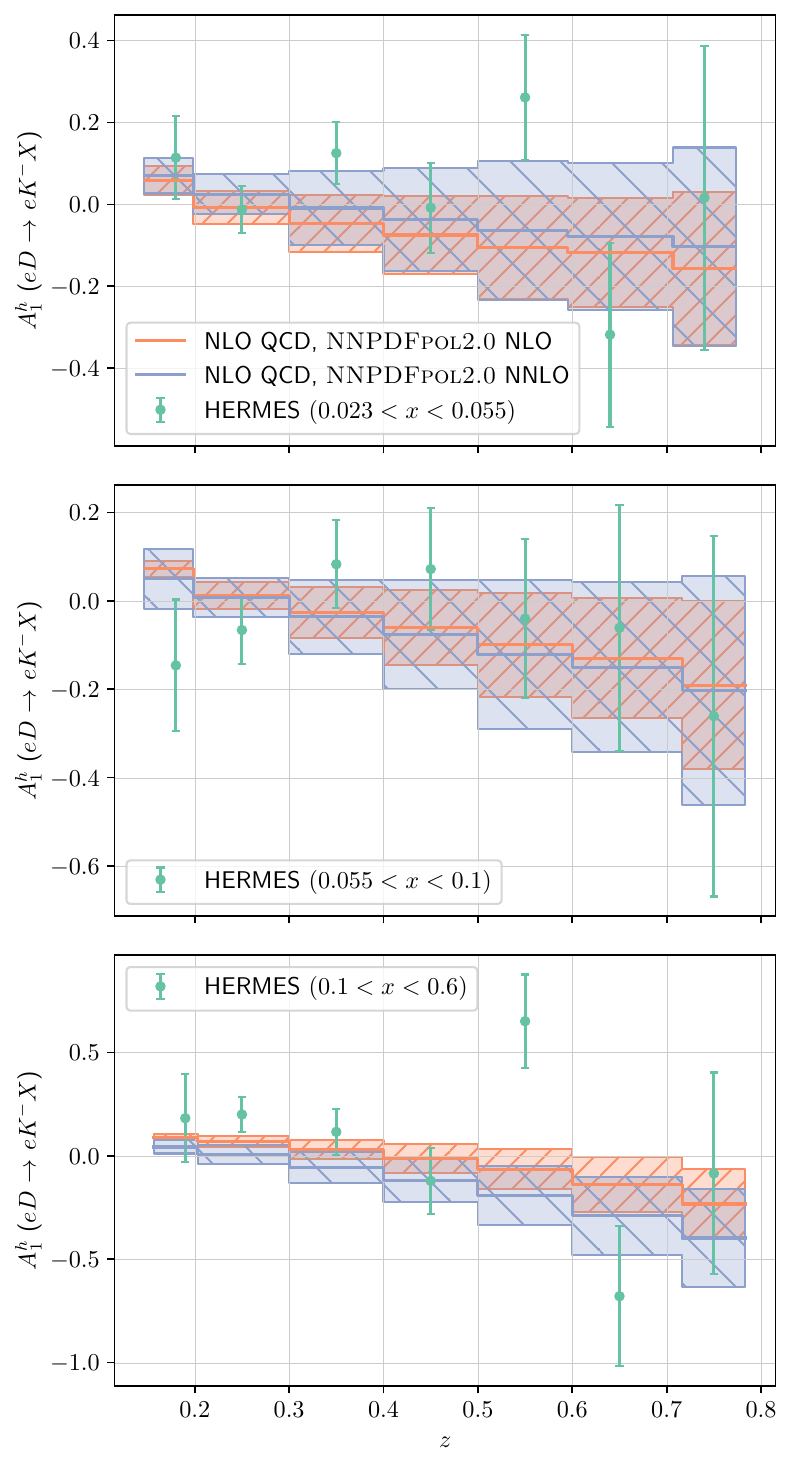}\\
  \caption{Comparison between experimental measurements of the spin asymmetry
    $A_1^h$ for the production of single-inclusive negative pions (left) and
    kaons (right) in polarised semi-inclusive DIS and the corresponding
    theoretical predictions. Experimental data are displayed as a function of
    $z$ for three $x$ bins of the HERMES experiment~\cite{HERMES:2018awh}, see
    text for details. Theoretical predictions are obtained with matrix elements
    accurate to NLO and with {\sc NNPDFpol2.0} NLO or NNLO polarised PDF sets;
    the unpolarised PDF and the FF are fixed to the central value of the
    {\sc NNPDF4.0} and {\sc MAPFF1.0} NLO sets. Therefore, error bands only
    account for one-sigma polarised PDF uncertainties.}
  \label{fig:had_DIS}
\end{figure}

\begin{figure}[!t]
  \centering
  \includegraphics[width=0.49\textwidth]{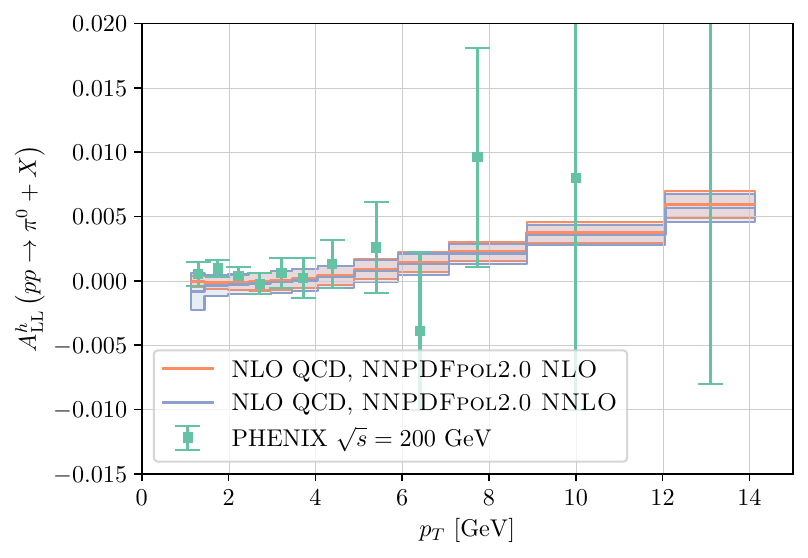}
  \includegraphics[width=0.49\textwidth]{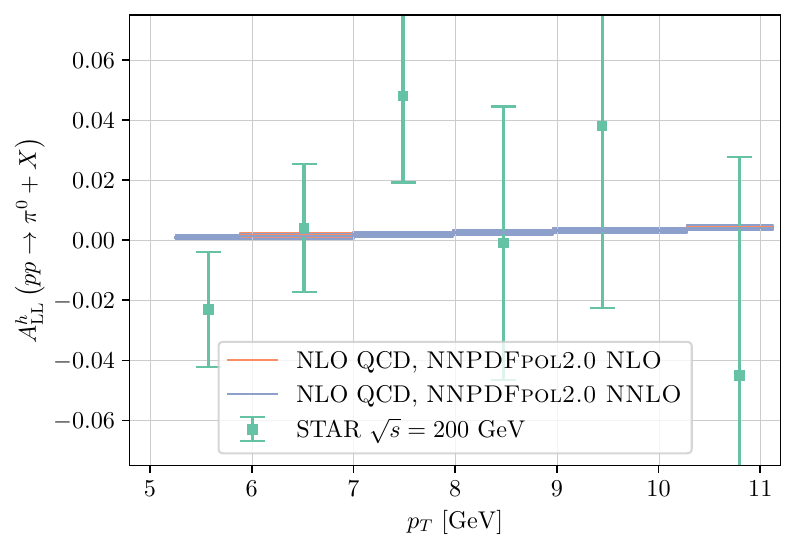}\\
  \caption{Comparison between experimental measurements of the double-spin
    asymmetry $A_{\rm LL}^h$ for the production of single-inclusive neutral
    pions in polarised proton-proton collisions and the corresponding
    theoretical predictions. Experimental data are from the PHENIX
    (left)~\cite{PHENIX:2014gbf} and STAR (right)~\cite{STAR:2018iyz}
    experiments. Theoretical predictions are obtained with matrix elements
    accurate to NLO and with {\sc NNPDFpol2.0} NLO or NNLO polarised PDF sets;
    the unpolarised PDF and the FF are fixed to the central value of the
    {\sc NNPDF4.0} and {\sc MAPFF1.0} NLO sets. Therefore, error bands only
    account for one-sigma polarised PDF uncertainties.}
  \label{fig:had_pp}
\end{figure}

In \cref{fig:had_DIS}, we compare our predictions for negatively charged
pion and kaon production in polarised semi-inclusive electron-deuteron DIS to
measurements performed by the HERMES experiment~\cite{HERMES:2018awh} with
a beam energy of 27.6~GeV. The asymmetry $A_1^h$ is reported as a function of
$z$, the fraction of the virtual-photon energy carried by the observed
ﬁnal-state hadron, in three bins of $x$. In \cref{fig:had_pp}, we compare
our predictions for neutral pion production in polarised proton-proton
collisions by the PHENIX~\cite{PHENIX:2014gbf} and STAR~\cite{STAR:2018iyz}
experiments at a centre-of-mass energy of 200~GeV. The asymmetry $A_{\rm LL}^h$
is reported as a function of the transverse momentum $p_T$ of the neutral pion.
In both \cref{fig:had_DIS} and \cref{fig:had_pp}, matrix elements are
accurate to NLO. Error bands correspond to one-sigma uncertainties computed
from the {\sc NNPDFpol2.0} NLO or NNLO polarised PDF sets; the unpolarised PDF
and the FF are fixed to the central value of the {\sc NNPDF4.0} and
{\sc MAPFF1.0} NLO sets.

Our predictions are always in good agreement with the data within experimental
uncertainties. These remain rather larger than the uncertainty on the
prediction, therefore one may question whether these additional data can
provide any additional constraint on the polarised PDFs. In this respect, we
make two considerations. First, the longitudinal double-spin asymmetry measured
by PHENIX may actually have some impact on the polarised gluon PDF, thanks to
the precision of the small-$p_T$ data points. Second, the uncertainty displayed
in \cref{fig:had_DIS,fig:had_pp}, as mentioned, only represents the
polarised PDF uncertainty. On top of this, there are MHO uncertainties and
uncertainties coming from the PDF and FF. This last uncertainty, in particular,
may be non-negligible. The results displayed in
\cref{fig:had_DIS,fig:had_pp} may therefore have a simultaneous
impact on polarised PDFs and FFs, the assessment of which requires a
determination of FFs beyond the scope of this work.

\section{Summary and delivery}
\label{sec:summary}

In this work, we have presented {\sc NNPDFpol2.0}, a new set of polarised PDFs
of the proton accurate at LO, NLO, and NNLO in the strong coupling expansion
that systematically incorporates, for the first time, MHOUs. This determination
is based on legacy measurements of polarised structure functions in inclusive
DIS and of longitudinally polarised spin asymmetries for the production of $W$
bosons, single-inclusive jets, and di-jet in polarised proton-proton
collisions. These measurements amount to most of all of those that are currently
available, and that are relevant to constrain polarised PDFs. The only notable
exception being measurements whose analysis requires an explicit knowledge of
FFs, namely those for single-hadron production in polarised DIS and
proton-proton collisions. {\sc NNPDFpol2.0} makes use of a
state-of-the-art machine learning methodology, that allows not only for the
reduction of parametrisation bias, through a neural network parametrisation,
but also for the selection of the optimal fitting model, through an automated
hyperparameter scan.

Our results indicate that {\sc NNPDFpol2.0} is a reliable determination
of polarised PDFs. In particular, it displays an excellent fit quality for all
of the considered data sets, a remarkable perturbative stability upon inclusion
of higher-order corrections or MHOUs, and a good consistency upon relaxation of
the positivity constraint or variations of the input data set. Two outcomes
are outstanding. First, the determination exhibits less sensitivity to the
perturbative accuracy than its unpolarised counterpart: the impact of NNLO
perturbative corrections on the fit quality is almost immaterial, and the
impact of MHOUs is moderate at LO and NLO and very limited at NNLO. We therefore
conclude that, with the current data set, theoretical framework, and
methodology, the perturbative expansion has converged. Second, the
determination confirms the relevance of the RHIC data in constraining key
aspects of the polarised PDFs. Measurements of longitudinal spin asymmetries
for single-inclusive jet and di-jet production in polarised proton-proton
collisions are such that the polarised gluon PDF is conclusively positive,
albeit only in the $x$ region constrained by the data, $x\gtrsim 0.02$. This
result is largely independent of the positivity constraint and from the other
data sets included in the fit. Measurements of longitudinal spin asymmetries
for $W$-boson production in polarised proton-proton collisions confirm that the
difference between up and down antiquark PDF is positive, and actually a little
larger than what was found in {\sc NNPDFpol1.1}. The {\sc NNPDFpol2.0} PDFs are
compatible with other recent NNLO polarised PDF determinations, in particular
with {\sc MAPPDFpol1.0} and {\sc BDSSV24}.

From a phenomenological perspective, we have revisited the spin content of the
proton, showing that the fraction of proton spin carried by gluons and quarks
remains consistent with {\sc NNPDFpol1.1}, though it is still subject to large
uncertainties coming from the small-$x$ kinematic region. We have also
demonstrated the applicability of {\sc NNPDFpol2.0} in computing predictions
for single-inclusive hadron production in polarised DIS and proton-proton
collisions, showing good agreement with the available experimental data. Our
work could naturally be extended to incorporate these measurements in a fit.
However, this task would first require to develop a consistent methodological
framework to determine FFs. This is a non-trivial endeavour which requires combining 
PDFs and FFs, with their respective perturbative evolution, and to
account for their possible interplay in physical observables that depend
on both of them. In principle, a simultaneous fit to both should be performed
to quantify this interplay. This exercise will be part of future work.

Together with the {\sc NNPDF4.0} parton
sets~\cite{NNPDF:2021njg,NNPDF:2024dpb,NNPDF:2024nan}, {\sc NNPDFpol2.0}
completes a global set of unpolarised and polarised PDFs determined with a
consistent methodology, including consistent positivity constraints.
These parton sets provide altogether a baseline for the upcoming
physics program of the EIC. The LO, NLO, and NNLO {\sc NNPDFpol2.0} PDF sets
without and with MHOUs are made publicly available in the {\sc LHAPDF}
format~\cite{Buckley:2014ana,LHAPDFurl} as
ensembles of both 1000 and 100 Monte Carlo replicas, the latter being obtained
from compression of the former with the algorithm developed
in~\cite{Carrazza:2021hny}. These sets are also available through the NNPDF
web page~\cite{NNPDFurl}. The open-source NNPDF software~\cite{NNPDF:2021uiq}
has been extended to include the input and tools needed to reproduce the
{\sc NNPDFpol2.0} sets presented here.

The list of available PDF sets, all with a fixed value of the strong coupling
$\alpha_s(m_Z)=0.118$, is as follows.
\begin{itemize}
\item Baseline LO, NLO, and NNLO PDF sets without MHOUs, 1000 Monte Carlo
  replicas
\begin{flushleft}
\tt NNPDFpol20\_lo\_as\_01180\_1000 \\
\tt NNPDFpol20\_nlo\_as\_01180\_1000 \\
\tt NNPDFpol20\_nnlo\_as\_01180\_1000 \\
\end{flushleft}
\item Baseline LO, NLO, and NNLO PDF sets with MHOUs, 1000 Monte Carlo
  replicas
\begin{flushleft}
\tt NNPDFpol20\_lo\_as\_01180\_mhou\_1000 \\
\tt NNPDFpol20\_nlo\_as\_01180\_mhou\_1000 \\
\tt NNPDFpol20\_nnlo\_as\_01180\_mhou\_1000 \\
\end{flushleft}
\item Compressed LO, NLO, and NNLO PDF sets without MHOUs, 100 Monte Carlo
  replicas
\begin{flushleft}
\tt NNPDFpol20\_lo\_as\_01180 \\
\tt NNPDFpol20\_nlo\_as\_01180 \\
\tt NNPDFpol20\_nnlo\_as\_01180 \\
\end{flushleft}
\item Compressed LO, NLO, and NNLO PDF sets with MHOUs, 100 Monte Carlo
  replicas
\begin{flushleft}
\tt NNPDFpol20\_lo\_as\_01180\_mhou \\
\tt NNPDFpol20\_nlo\_as\_01180\_mhou \\
\tt NNPDFpol20\_nnlo\_as\_01180\_mhou \\
\end{flushleft}
\end{itemize}

\section*{Acknowledgements}

Special thanks are due to Christopher Schwan for his continuous and friendly
support with {\sc PineAPPL}. We thank Elke Aschenauer for clarifications on the
usage of the STAR data, Daniel de Florian and Werner Vogelsang for providing
us with the piece of software of~\cite{deFlorian:1998qp,Jager:2004jh},
Frank Petriello and Hai Tao Li for providing us with the modified version of
MCFM developed in~\cite{Boughezal:2021wjw}, and all the members of the NNPDF
collaboration for interesting discussions.

F.H. is supported by the Academy of Finland project 358090 and is funded as a
part of the Center of Excellence in Quark Matter of the Academy of Finland,
project 346326. E.R.N. is supported by the Italian Ministry of University and
Research (MUR) through the ``Rita Levi-Montalcini'' Program. G.M. and J.R. are partially
supported by NWO, the Dutch Research Council. J.R. and T.R.R. are supported by
an ASDI grant from the Netherlands eScience Center (NLeSC).

\appendix
\section{Benchmark of the {\sc PineAPPL} grids}
\label{app:pinebench}

As discussed in \cref{sec:data-theory}, we compute theoretical predictions
corresponding to the data included in the {\sc NNPDFpol2.0} fits by
convolving PDFs with pre-computed PDF-independent fast interpolation grids.
We compute these grids in the {\sc PineAPPL} format~\cite{Carrazza:2020gss},
starting from various pieces of external software, that were typically
developed to compute cross sections of a specific production process.
Making these pieces of software able to generate {\sc PineAPPL} grids
requires nontrivial modifications, therefore one may wonder whether these
introduce a loss of accuracy in the computation of physical observables.
In this appendix, we show, for the specific case of $W$-boson production in
polarised proton-proton collisions, as implemented in the modified version of
{\sc MCFM}~\cite{Boughezal:2021wjw}, that this is not the case.

To this purpose, we perform the following benchmark. We compute the longitudinal
double-spin asymmetry in two ways. First, by running the unmodified version of
the code of~\cite{Boughezal:2021wjw}. In this case, the convolution between
PDFs and matrix elements is performed at every step of the Monte Carlo
integration. Second, by convolving PDFs with the {\sc PineAPPL} grids generated
from a modified version of the same code. In this case, PDFs are effectively
multiplied by the weights stored in the grids only once. The result obtained
with the latter method is equivalent to the result obtained with the
former method, within Monte Carlo and interpolation uncertainties, only if the
required modification to the code in~\cite{Boughezal:2021wjw} are properly
implemented. Given that the accuracy of the integration and of the
interpolation does not depend on the choice of the renormalisation and
factorisation scales, it is sufficient to check the results obtained with the
central scale.

\begin{figure}[!t]
  \centering
  \includegraphics[width=0.495\textwidth]{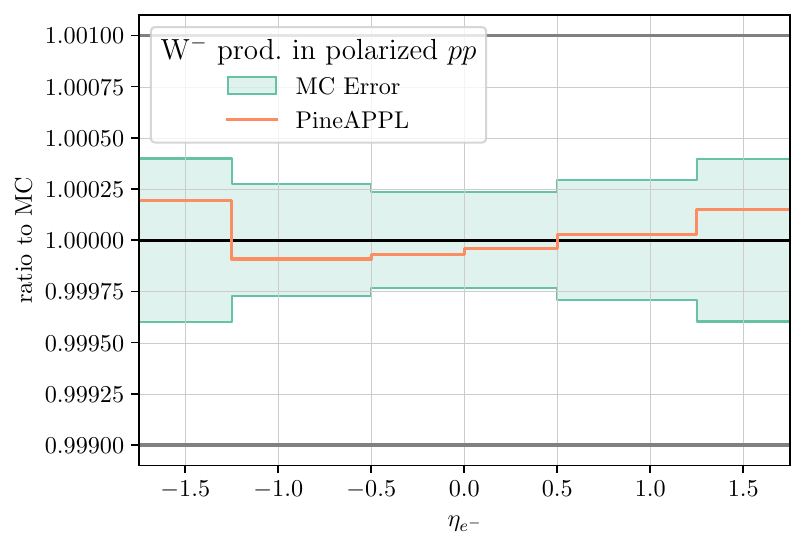}
  \includegraphics[width=0.495\textwidth]{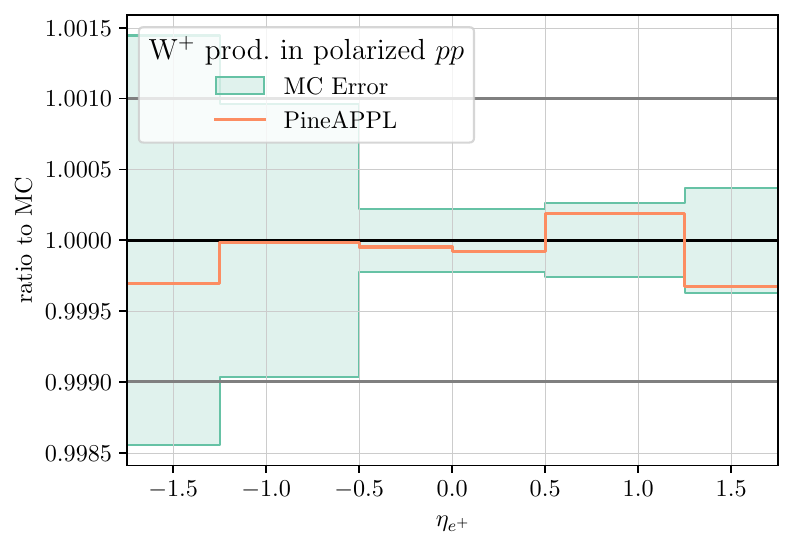}
  \caption{Predictions of the longitudinal spin asymmetry for $W^-$ (left) and
    $W^+$ (right) boson production in polarised proton-proton collisions
    obtained from the code of~\cite{Boughezal:2021wjw} or from convolution
    of PDFs with {\sc PineAPPL} grids generated from a modified version of the
    same code. The results are normalised to the former. The band
    corresponds to Monte Carlo uncertainties due to numerical integration
    performed with the code of~\cite{Boughezal:2021wjw}.}
  \label{fig:pinebench}
\end{figure}

In \cref{fig:pinebench}, we compare the results obtained from the two
aforementioned methods, normalised to the former. Specifically, we compare
the Monte Carlo integration error of the first method with the interpolation
error of the second method. We observe that Monte Carlo errors are below one
permil, except for the leftmost bin in the positron pseudorapidity of the $W^+$
boson, where they are of the order of 1.5\textperthousand. Interpolation
uncertainties are smaller than that: indeed the difference between the two
central values is of the order of 0.5\textperthousand~at most, with very
negligible fluctuations across different pseudorapidity bins. We therefore
conclude that the two methods are equivalent, and, in particular, that the
generation of {\sc PineAPPL} grids does not result in any accuracy loss.


\bibliographystyle{utphys}
\bibliography{nnpdfpol20}

\end{document}